\documentclass[journal]{IEEEtran}
\usepackage[utf8]{inputenc}
\usepackage{amsmath}
\usepackage{amssymb}
\usepackage{amsfonts}
\usepackage{amsthm}
\usepackage{optidef}
\usepackage{multirow}
\usepackage{threeparttable}
\usepackage{todonotes}
\usepackage{array}
\usepackage{algorithm}
\usepackage{algorithmic}
\usepackage{caption} 
\usepackage{subcaption} 
\usepackage{cite}
\usepackage{verbatim} 
\usepackage{harmony} 
\usepackage{scalerel}
\newcommand{\lcirc}[1]{\scaleobj{.6}{\mbox{\Kr{#1}}}}
\usepackage{setspace} 
\usepackage[bookmarks=false]{hyperref}
\hypersetup{colorlinks=true,linkcolor=blue,citecolor=blue}
\allowdisplaybreaks[4] 

\title{Faulty RIS-aided Integrated Sensing and Communication: Modeling and Optimization}
\date{}

\author{Lu Wang, Gui Zhou, Changheng Li, Luis F. Abanto-Leon, Nairy Moghadas Gholian, Matthias Hollick, and Arash Asadi

\thanks{L. Wang, N. Moghadas Gholian, and M. Hollick are with the Department of Computer Science, Technical University of Darmstadt, Darmstadt 64289, Germany (e-mail: lwang@wise.tu-darmstadt.de, ngholian@seemoo.tu-darmstadt.de, mhollick@seemoo.tu-darmstadt.de).}
\thanks{G. Zhou is with the School of Electronic Information and Communication from Huazhong University of Science and Technology (HUST), Wuhan, China (email:
gui\_zhou@hust.edu.cn).}
\thanks{L. F. Abanto-Leon is with Faculty of Electrical Engineering and Information Technology, Ruhr University Bochum, Bochum, Germany (e-mail:luis.abantoleon@ruhr-uni-bochum.de).}
\thanks{C. Li and A. Asadi are with the Embedded Systems Group, Delft University of Technology, 2628 CD Delft, Netherlands (e-mail: C.Li-7@tudelft.nl, a.asadi@tudelft.nl).}
}

\begin{document}
\maketitle

\begin{abstract}

This work investigates a practical reconfigurable intelligent surface (RIS)-aided integrated sensing and communication (ISAC) system, where a subset of RIS elements fail to function properly and reflect incident signals randomly towards unintended directions with attenuation, thereby degrading system performance. To date, no study has addressed such impairments caused by faulty RIS elements in ISAC systems. This work aims to fill the gap. First, to quantify the impact of faulty elements on ISAC performance, we derive the misspecified Cramér-Rao bound (MCRB) for sensing parameter estimation and signal-to-interference-and-noise ratio (SINR) for communication quality. Then, to mitigate the performance loss caused by faulty elements, we jointly design the remaining functional RIS phase shifts and transmit beamforming to minimize the MCRB, subject to the communication SINR and transmit power constraints. The resulting optimization problem is highly non-convex due to the intricate structure of the MCRB expression and constant-modulus constraint imposed on RIS. To address this, we reformulate it into a more tractable form and propose a block coordinate descent (BCD) algorithm that incorporates majorization-minimization (MM), successive convex approximation (SCA), and penalization techniques. Simulation results demonstrate that our proposed approach reduces the performance loss by 21.25\% on average compared to the baseline where the presence of faulty elements is ignored. Furthermore, the performance gain becomes more evident as the number of faulty elements increases.

\end{abstract}



\section{Introduction}
\label{Introduction}
Integrated sensing and communication (ISAC) has been recognized as one of the key features of 6G networks, positioning it as a critical research direction in wireless communications. Within the ISAC framework, sensing and communication are intended to be performed simultaneously using one unified infrastructure, thereby reducing energy consumption and hardware costs~\cite{bFanoverview}. To improve the sensing resolution and communication throughput, ISAC systems are anticipated to operate in the millimeter-wave (mmWave) bands, which offer larger bandwidths. However, mmWave signals are highly susceptible to blockage and significant attenuation due to their short wavelength and high path loss~\cite{bmmWave}. To address these challenges, reconfigurable intelligent surfaces (RISs) can be employed. A RIS comprises passive elements whose phase shifts can be dynamically adjusted to reshape the wireless propagation environment. By incorporating RISs, additional propagation paths can be established to deal with blockages, enhance signal strength, and improve sensing accuracy. Consequently, RIS-aided ISAC architectures hold great promise for substantially enhancing performance in the mmWave spectrum and beyond~\cite{bRISISAC}.

Additionally, practical electronic devices exhibit hardware impairments due to the non-ideal nature of their components, such as phase noise, power amplifier nonlinearity, mutual coupling, etc.~\cite{bHWIall}. Likewise, RISs are inevitably subject to failures due to a variety of reasons, including internal causes (aging after long-term usage) and external factors (manufacturing imperfections, lack of proper maintenance, or natural environmental disaster)~\cite{bfaultyReason1, bfaultyReason2}. There have been works studying the impact of RIS phase noise on the performance of RIS-aided systems, where phase offsets following uniform or Von-Mises distribution are added to the original RIS phases~\cite{bRISHWI1,bRISHWI2}. Note that in such RIS hardware impairment models, all the RIS elements still function under certain phase noise. In contrast, \cite{bNairyRIS} investigated a more severe hardware impairment scenario in which a subset of RIS elements are entirely non-functional, acting as random phase shifters that scatter incident signals in unintended directions with attenuation. Such malfunctioning elements are referred to as \textbf{faulty}. In the following, we provide a more detailed discussion and literature review of RIS-aided systems in the presence of faulty elements.

\subsection{Prior Works and Motivations}
In the presence of faulty elements, RIS phase configurations designed under the assumption of a perfect RIS will no longer be appropriate and may degrade system performance. Consequently, the careful consideration of these faulty elements is necessary but poses a significant challenge to system design. To date, there have been works dealing with faulty element diagnosis to detect the locations and coefficients of faulty elements~\cite{bfaultyDetect1,bfaultyDetect2,bfaultyDetect3}. In~\cite{bfaultyDetect1}, assuming the failure of RIS is sparse and knowing the perfect channel state information (CSI), the authors formulated the faulty element detection and recovery problem based on the compressed sensing technique. With the same assumption of failure sparsity, the work~\cite{bfaultyDetect2} located and retrieved faulty elements for not only the perfect CSI case but also the partial CSI case and no CSI case. Furthermore, the authors in~\cite{bfaultyDetect3} jointly reconstructed sparse signals and diagnosed faulty elements, which does not depend on using pilots. 

Moreover, in real-world applications, the size of RIS can be quite large, often comprising hundreds of elements. In such cases, replacing the entire RIS due to some faulty elements is neither practical nor cost-effective. Instead, the focus should shift to maximizing the use of the remaining functional elements to maintain system performance. So far, there are only a few works that have taken measures to tackle the faulty elements problem existing in RIS-aided systems~\cite{bNairyRIS,bfaultyRIS2,bfaultyRIS3,bfaultyRIS4}.
The work~\cite{bNairyRIS} demonstrated the risk of leaking information to unintended users in the presence of faulty RIS elements. To mitigate this undesired effect, two optimization problems were formulated to maximize the signal-to-leakage-and-noise ratio (SLNR) of the user for both the perfect CSI case and the partial CSI case. The authors in \cite{bfaultyRIS2} studied a faulty RIS-aided satellite communication system, for which a genetic algorithm was applied to optimize the beampattern by designing the operational RIS elements. Except for the RIS-aided communication scenario, the works~\cite{bfaultyRIS3,bfaultyRIS4} focused on RIS-aided localization and jointly investigated faulty element detection and user localization, with~\cite{bfaultyRIS3} based on $l_1$-regularization and hypothesis testing and~\cite{bfaultyRIS4} based on transfer learning, autoencoders, and convolutional neural networks.

However, to the best of our knowledge, no work has studied the impact of faulty RIS elements in the ISAC scenarios. Therefore, two research questions remain unanswered: \textit{(i) how to quantify the impact of faulty elements on both sensing and communication performances? and (ii) how to maximize the effects of the remaining functional RIS elements on both sensing and communication performances?} Note that the presence of faulty elements might not be detected promptly, leading to the assumption of a perfect RIS model. In such cases, a model mismatch (or misspecification) is caused. To account for this effect, the misspecified Cramér-Rao Bound (MCRB) is employed as a performance bound for parameter estimation under model mismatch~\cite{bMCRBorigin}. To date, only one work has examined the model mismatch caused by faulty elements using MCRB in a localization scenario~\cite{bfaultyRIS3}. However, this work assumed sparsity in failures, with faulty elements comprising no more than 2\% of the total. In contrast, we aim to optimize the performance of a faulty RIS-aided ISAC system without imposing any restrictions on the number of faulty elements. Inspired by the application of MCRB, we adopt it as a metric to evaluate sensing performance. For the communication metric, we consider the degraded signal-to-interference-and-noise ratio (SINR) experienced by users due to faulty elements. With MCRB and SINR defined, we optimize the ISAC system performance by fully leveraging the remaining functional RIS elements. Further details addressing these two research questions are provided in the subsequent sections.

\subsection{Contributions}
To fill the research gap identified earlier, this work investigates a RIS-aided ISAC system, where a subset of RIS elements suffers from hardware failures. Assuming that the locations and coefficients of these faulty elements have been accurately identified using the existing detection methods proposed in~\cite{bfaultyDetect1,bfaultyDetect2,bfaultyDetect3}, we focus on optimizing the configuration of the remaining functional elements to mitigate the negative impact of these impairments. To address the first research question, we derive performance metrics that quantify the influence of faulty elements on both sensing and communication. In response to the second research question, we jointly optimize the transmit beamforming and the phase shifts of the operational RIS elements to maintain overall ISAC system performance. The main contributions of this work are summarized as follows.

\begin{itemize}

\item \textbf{First MCRB derivation for faulty RIS-aided ISAC}: We model a RIS-aided ISAC system, where a subset of RIS elements suffer from hardware impairments and reflect incident signals towards random directions with attenuation. The performance degradation due to these faulty elements is evaluated using the SINR for communication and the MCRB for sensing parameter estimation. Unlike the standard Cramér-Rao bound (CRB), the MCRB characterizes the RIS model mismatch between the perfect RIS and faulty RIS models. To the best of the authors' knowledge, this is the first work to derive the MCRB under a faulty RIS-aided ISAC framework and to propose a performance preservation strategy by optimizing the remaining functional elements.

\item \textbf{MCRB-minimization problem formulation to mitigate faulty RIS impact}: To mitigate the adverse impact of faulty RIS elements, we formulate an optimization problem that jointly designs the transmit beamforming and the remaining functional RIS phase shifts. The objective is to minimize the MCRB for estimating the two-dimensional (2D) angle of departure (AoD) of the target, while satisfying both a communication SINR constraint and a total transmit power constraint. This optimization problem is highly non-convex due to the intricate structure of the MCRB expression, which involves inversions and products of nested matrices whose entries depend nonlinearly on system parameters, thereby inducing strong coupling among variables. Such coupling is significantly more complex than that in the conventional CRB expression.

\item \textbf{Tractable reformulation and solution design}: To address this challenge, we first reformulate the original problem into a more tractable form by introducing auxiliary variables and applying the Schur complement for nested matrices. Subsequently, we develop a block coordinate descent (BCD) algorithm that incorporates majorization-minimization (MM), successive convex approximation (SCA), and penalization techniques. These methods enable the decomposition of the non-convex problem into a series of convex semidefinite programming (SDP) subproblems, which can be efficiently solved using standard convex optimization tools such as CVX.

\item \textbf{Performance evaluation and insights analysis}: Simulation results reveal that the proposed MCRB-based optimization significantly narrows the performance gap to the ideal fault-free RIS case, achieving an average performance improvement of 21.25\% over the naive baseline that ignores faulty elements. This confirms that our proposed scheme effectively leverages the functional elements to compensate for performance degradation. Additionally, the inherent trade-off between sensing accuracy and communication quality, as well as the beampatterns under different fault distributions, are analyzed.

\end{itemize}

\subsection{Organization and Notations}
The remainder of this paper is organized as follows. Section \uppercase\expandafter{\romannumeral2} describes the system model for the faulty RIS‑aided ISAC setup. In Section \uppercase\expandafter{\romannumeral3}, we derive the MCRB for estimating the target’s 2D AoD in the presence of faulty RIS elements. Section \uppercase\expandafter{\romannumeral4} formulates the resulting optimization problem, and Section \uppercase\expandafter{\romannumeral5} presents our proposed algorithm along with its convergence and complexity analysis. Simulation results and a comprehensive discussion are provided in Section \uppercase\expandafter{\romannumeral6}. Finally, conclusions are presented in Section \uppercase\expandafter{\romannumeral7}.

\textit{Notation}: Bold-face uppercase and lowercase letters  ${\mathbf{X}}_1$ and ${\mathbf{x}}_1$ symbolize matrices and vectors, respectively. $\mathbb{E}\left\{ \cdot \right\}$ denotes the statistical expectation, while ${\mathcal O}\left(  \cdot  \right)$ denotes the order of computational complexity. For a square matrix ${\bf{X}}_1$, ${\rm{tr}}\left({\bf{X}}_1\right)$ and ${\bf{X}}_1^{-1}$ represent its trace and inverse, respectively. ${{\bf{X}}_1} \succeq 0$ and ${{\bf{X}}_1} \prec 0$ indicate that ${\bf{X}}_1$ is positive semi-definite (PSD) and negative definite, respectively. ${\left(  \cdot  \right)^{\rm{T}}}$ and ${\left(  \cdot  \right)^{\rm{H}}}$ denote the transpose and conjugate transpose of a matrix or vector. ${\left\| \cdot \right\|_{\rm{F}}}$, ${\left\| \cdot \right\|_ * }$,  ${\left\| \cdot \right\|_2}$, and ${\rm{rank}}\left( \cdot \right)$ denote the Frobenius norm, nuclear norm, spectral norm, and rank of a matrix, respectively. ${\rm{diag}}\left(  \cdot  \right)$ forms a diagonal matrix with the elements in $\left(  \cdot  \right)$. ${\mathbb{C}}$ and ${\mathbb{R}}$ denote the set of complex and real numbers, respectively. ${\rm{Re}}\left( \cdot \right)$ denotes the real part of a complex number.

\section{System Model}
\begin{figure}[!t]
    \centering
    \vspace{-0.5cm}
    \includegraphics[width=0.65\linewidth]{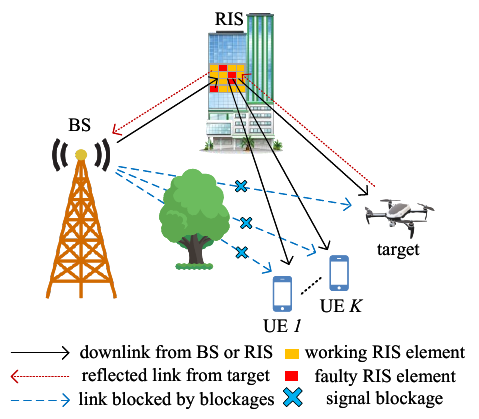}
    \caption{RIS-assisted ISAC system with faulty elements.}
    \label{Sysmol_faultyRIS}
    \vspace{-0.5cm}    
\end{figure}

We consider a RIS-aided ISAC system where direct links between the base station (BS) and intended user equipments (UEs) and targets are obstructed\footnote{To support this assumption practically, we refer to~\cite{b38901}, according to which the attenuation levels of blockages (modeled as stochastic or geometric screens) can reach up to 30 dB depending on the specific blockage size, position, and material. Moreover, Table 7.4.3-1 in~\cite{b38901} reports severe penetration losses for typical building materials, e.g., 125 dB for concrete walls and 8 dB for multi-pane glass at 30 GHz, confirming extreme blockage potential. These practical values emphasize that blockage is not only realistic but also frequent in mmWave systems, thus justifying our assumption. In such scenarios, RIS can be strategically deployed to establish indirect paths to maintain both communication and sensing performance.}, as shown in Fig.~\ref{Sysmol_faultyRIS}. A RIS is deployed in this scenario to enable both sensing and communication. Equipped with a $N_t$-antenna transmitting uniform linear array (ULA) and a $N_r$-antenna receiving ULA, the BS communicates with $K$ UEs and senses one target simultaneously. The RIS is configured with a uniform planar array (UPA) with $R_y$ elements in the horizontal direction and $R_z$ elements in the vertical direction. The total number of elements is $R=R_y  R_z$. In practice, due to the reasons described in Section~\ref{Introduction}, the RIS elements may experience failures.

\subsection{Signal Model}
To simultaneously realize communication and sensing functionalities, the dual-functional transmit signal at time slot $t$ is expressed as
\begin{equation}\label{Xtrans}
{\bf{x}}\left( t \right) = \sum\limits_{k = 1}^K {{{\bf{w}}_k}} {c_k}\left( t \right) + {\bf{s}}\left( t \right) = {\bf{Wc}}\left( t \right) + {\bf{s}}\left( t \right),
\end{equation}
where ${\bf{c}}\left( t \right) = {\left[ {{c_1}\left( t \right),...,{c_K}\left( t \right)} \right]^{\rm{T}}}\in {\mathbb{C}^{K \times 1}}$ denotes the information symbols delivered to $K$ UEs at time slot $t$, modeled as independently and identically distributed random variables with zero mean and unit variance, satisfying $\mathbb{E} \{ {{\bf{c}}\left( t \right){\bf{c}}{{\left( t \right)}^{\rm{H}}}} \} = {{\bf{I}}_K}$. In addition, ${\bf{W}} = \left[ {{{\bf{w}}_1},...,{{\bf{w}}_K}} \right] \in {\mathbb{C}^{N_t \times K}}$ represents the transmit beamforming matrix for all $K$ UEs, where the $k$-th column ${{\bf{w}}_k}\in {\mathbb{C}^{N_t \times 1}}$ corresponds to the beamforming vector for the $k$-th UE. The dedicated sensing signal ${\bf{s}}\left( t \right)\in {\mathbb{C}^{N_t \times 1}}$ is introduced to achieve full degree of freedom (DoF) for sensing, which is generated by pseudo-random coding satisfying $\mathbb{E} \{ {{\bf{c}}\left( t \right){\bf{s}}{{\left( t \right)}^{\rm{H}}}} \} = {{\bf{0}}_{K \times {N_t}}}$ \cite{XiangLiu}. When the total transmitting time slots $T$ is sufficiently large, the sample covariance matrix of the transmitted signal ${\bf{x}}\left( t \right)$ is equivalent to its statistical covariance matrix, namely
\begin{equation}\label{XcovaM}
{{\bf{R}}_x} \buildrel \Delta \over = \frac{1}{T}\sum\limits_{t = 1}^T {{\bf{x}}\left( t \right){{\bf{x}}\left( t \right)^{\rm{H}}}}  \approx \mathbb{E}\left\{ {{\bf{x}}\left( t \right){{\bf{x}}\left( t \right)^{\rm{H}}}} \right\} \!=\! {\bf{W}}{{\bf{W}}^{\rm{H}}} \!+\! {{\bf{R}}_s},
\end{equation}
where ${{\bf{R}}_s}$ is the sensing covariance matrix, satisfying
\begin{equation}\label{scovaM}
{{\bf{R}}_s} = \frac{1}{T}\sum\limits_{t = 1}^T {{\bf{s}}\left( t \right){{\bf{s}}\left( t \right)^{\rm{H}}}} \succeq {\bf{0}}.
\end{equation}
Moreover, the total power allocated to communication and sensing must not exceed the maximum transmit power $P_{\max}$ of the BS, enforced via
\begin{equation}\label{Ptran}
{\rm{tr}}({{\bf{R}}_x}) = {\rm{tr}}({\bf{W}}{{\bf{W}}^{\rm{H}}}) + {\rm{tr}}({{\bf{R}}_s}) \le {P_{\max }}.
\end{equation}

\subsection{RIS Model and Channel Model}
\label{FaultyRISChannel}
In practice, the presence of faulty elements may not be detected in a timely manner, resulting in the assumption of a perfect RIS model. Hence, there exists a mismatch in the RIS model. Next, we introduce the perfect RIS model and the faulty RIS model.

\subsubsection{Perfect RIS Case}
\label{RISassumed}
When the presence of faulty elements is not noticed promptly, the ideal passive RIS model with perfectly functional elements is assumed. Let ${\boldsymbol{\Theta }} = {\rm{diag}}\left( {{\beta _1}{e^{j{\theta _1}}},...,{\beta _r}{e^{j{\theta _r}}},...,{\beta _R}{e^{j{\theta _R}}}} \right) \in {\mathbb{C}^{R \times R}}$ denote the reflection coefficient matrix of the RIS, where ${\beta _r}$ and ${\theta _r}$ represent the amplitude and phase shift of the $r$-th reflective element. Letting ${\bf{v}} = {\left[ {{\beta _1}{e^{j{\theta _1}}},...,{\beta _R}{e^{j{\theta _R}}}} \right]^{\rm{H}}} \in \mathbb{C}^{ R \times 1}$ denote the coefficient vector, the cascaded communication channel from the BS to UE $k$ is given by
\begin{equation}\label{HkVassumed}
{{{\bf{\tilde h}}}_k} = {\bf{h}}_k^{\rm{H}}{\bf{\Theta H}}_{BR}^{\rm{H}} = {{\bf{v}}^{\rm{H}}}{{\bf{H}}_k},
\end{equation}
where ${{\bf{H}}_{BR}} \in {\mathbb{C}^{N_t \times R}}$ and ${{\bf{h}}_k} \in {\mathbb{C}^{R \times 1}}$ represent the channels from BS to RIS and from RIS to the $k$-th UE, respectively. Additionally, ${{\bf{H}}_k} = {( {{\rm{diag}}( {{\bf{h}}_k^{\rm{H}}} ){\bf{H}}_{BR}^{\rm{H}}} )^{\rm{H}}} \in \mathbb{C}^{ {N_t \times R}}$.

For sensing, the target's parameter estimation is the focus of this work. We denote the 2D AoD from the RIS to the target as ${\boldsymbol{\phi}} = {\left[ {{\phi _e},{\phi _a}} \right]^{\rm{T}}}$, where ${\phi _e}$ and ${\phi _a}$ are the elevation and azimuth AoD, respectively. The cascaded sensing channel under the perfect RIS model is expressed as follows
\begin{equation}\label{GsensingAssumed}
{\bf{\tilde G}} = \alpha {{\bf{H}}_{BR}}{{\bf{\Theta }}^{\rm{H}}}{\bf{a}}\left( {{\phi _e},{\phi _a}} \right){{\bf{a}}^{\rm{H}}}\left( {{\phi _e},{\phi _a}} \right){\bf{\Theta H}}_{BR}^{\rm{H}},
\end{equation}
where $\alpha$ is the round-trip channel attenuation between the RIS and the target\footnote{$\alpha$ is intended to model the round-trip channel attenuation between the RIS and the target, rather than the cascaded attenuation between the BS and the target. This is because the BS-RIS channel ${{\bf{H}}_{BR}}$ already incorporates the path loss from the BS to the RIS.}, accounting for physical and geometric properties. It models how strongly a sensing signal is reflected from the target and received back by the receiver, including the signal strength and phase shift\cite{bRadarB}. $\alpha$ is mathematically expressed as $|\alpha {|^2} = \frac{{{\lambda ^2}{\sigma _{RCS}}}}{{{{\left( {4\pi } \right)}^3}{{d_{RT}}^4}}}$, where $\lambda$ denotes the carrier wavelength, $\sigma_{\rm{RCS}}$ means the radar cross section, and $d_{RT}$ refers to the distance between the RIS and the target. Meanwhile, ${\bf{a}}\left( {{\phi _e},{\phi _a}} \right) \in {\mathbb{C}^{R \times 1}}$ denotes the steering vector with respect to RIS deployed in the y-z plane, given by

\begin{equation}\label{SVa}
{\bf{a}}\left( {{\phi _e},{\phi _a}} \right) = \left[ {{a}{{\left( {{\phi _e},{\phi _a}}  \right)}_1},...,{a}{{\left( {{\phi _e},{\phi _a}} \right)}_R}} \right]^{\rm{T}}.
\end{equation}
Each element is expressed as
\begin{equation}\label{SVa_r}
{a}{\left( {{\phi _e},{\phi _a}} \right)_r} = {e^{-j\kappa \left[ {{d_y}\cos \left( {{\phi _e}} \right)\sin \left( {{\phi _a}} \right){r_y} + {d_z}\sin \left( {{\phi _e}} \right){r_z}} \right]}}.
\end{equation}
where $\kappa  = {{2\pi } \mathord{\left/ {\vphantom {{2\pi } \lambda }} \right. \kern-\nulldelimiterspace} \lambda }$ is the wave number, while $d_y$ and $d_z$ are the spacing between adjacent RIS elements along y-axis and z-axis, which are indexed by ${r_y} \in \left\{ {0,1,...,{R_y} - 1} \right\}$ and ${r_z} \in \left\{ {0,1,...,{R_z} - 1} \right\}$, respectively.

\subsubsection{Faulty RIS Case} \label{RIStrue}
To model the faulty RIS, we denote the number of faulty and working RIS elements as $F$ and $W$, respectively, which satisfy $F+W=R$. For faulty elements, we define ${{\bf{v}}_F} \in \mathbb{C}^{ F \times 1}$, collecting the reflection coefficients of $F$ faulty elements. Each element of ${{\bf{v}}_F}$ is expressed as ${v_f} = {\beta _f}{e^{j{\theta _f}}}$, with amplitudes and phase shifts modeled as ${\beta _f} \sim U\left( {0,1} \right)$ and ${\theta_f} \sim U\left( {0,2 \pi} \right)$~\cite{bNairyRIS}. This means that the signals incident on faulty elements experience random attenuation and scattering, causing them to reflect in unintended directions. Meanwhile, we assume the faulty elements are randomly distributed over the total $R$ elements, which is a widely adopted assumption to model unpredictable hardware impairments. In practice, the specific values of ${\beta _f}$ and ${\theta_f}$, as well as the locations of faulty elements, can be obtained using the fault diagnosis methods proposed in~\cite{bfaultyDetect1,bfaultyDetect2,bfaultyDetect3}, and are assumed to be known at the BS\footnote{The specific values of ${\beta _f}$ and ${\theta_f}$ are sampled from their respective distributions during each Monte Carlo simulation.}. We denote the set indexing faulty elements ${v_f}$ as ${\mathcal S_F}$, namely the index  $f \in {\mathcal S_F}$. The remaining functional elements are grouped into a new vector ${{\bf{v}}_W}\in \mathbb{C}^{ W \times 1}$, representing the reflection coefficients of working elements. Each working element is expressed as ${v_w} = {\beta _w}{e^{j{\theta _w}}}$, where the amplitude and phase shift satisfy ${\beta _w} = 1$ and ${\theta _w} \in \left[ {0,2\pi } \right)$. The index of working elements is denoted by ${\mathcal S_W}$.
Additionally, the reflection coefficient matrices of faulty elements and working elements are modeled as ${\boldsymbol{\Theta }_F} = {\rm{diag}}\left( {{\bf{v}}_F} \right) \in {\mathbb{C}^{F \times F}}$ and ${\boldsymbol{\Theta }_W} = {\rm{diag}}\left( {{\bf{v}}_W} \right) \in {\mathbb{C}^{W \times W}}$, respectively.

Likewise, we define ${{\bf{H}}_{W,k}} \in \mathbb{C}^{N_t \times W}$ and ${{\bf{H}}_{F,k}} \in \mathbb{C}^{N_t \times F}$, which group UE $k$'s channel from ${\bf{H}}_{k}$ into the separate channels assisted by the working and faulty elements, respectively. Thus, under the faulty RIS model, the cascaded communication channel from BS to RIS to the $k$-th UE, denoted by ${{\bf{\bar h}}_k} \in \mathbb{C}^{ 1 \times N_t}$, can be expressed as 
\begin{equation}\label{Hk:faulty}
{{\bf{\bar h}}_k} = \left[ {{\bf{v}}_W^{\rm{H}}\,{\bf{v}}_F^{\rm{H}}} \right]{\left[ {{{\bf{H}}_{W,k}}\,{{\bf{H}}_{F,k}}} \right]^{\rm{T}}} = {\bf{v}}_W^{\rm{H}}{{\bf{H}}_{W,k}} + {\bf{v}}_F^{\rm{H}}{{\bf{H}}_{F,k}},
\end{equation}
where ${{{\bf{\bar h}}}_{F,k}} = {\bf{v}}_F^{\rm{H}}{{\bf{H}}_{F,k}} \in \mathbb{C}^{ 1 \times N_t}$ corresponds to UE $k$'s cascaded communication channel affected by faulty elements.
In sensing applications, the presence of scatters within the mainlobe directions of faulty elements can lead to echo interference at the receiver. To simplify the analysis, we model this interference by assuming the existence of a single scatter in addition to the target. Moreover, to separately characterize the channels between the working RIS elements and the target, and between the faulty elements and the scatter, we define distinct target response matrices for the working and faulty elements, respectively, as follows
\begin{align}
& {{\bf{H}}_{{\rm{TRM}},W}} = {{\bf{a}}_W}\left( {{\phi_e},{\phi_a}} \right){\bf{a}}_W^{\rm{H}}\left( {{\phi_e},{\phi_a}} \right),\label{H_TRMW}\\
& {{\bf{H}}_{{\rm{TRM}},F}} = {{\bf{a}}_F}\left( {{{\phi}_e'},{{\phi}_a'}} \right){\bf{a}}_F^{\rm{H}}\left( {{{\phi}_e'},{{\phi}_a'}} \right)\label{H_TRMF},
\end{align}
where ${{\bf{a}}_W}\left( {{\phi_e},{\phi_a}} \right) \in \mathbb{C}^{W \times 1}$ and ${{\bf{a}}_F}\left( {{{\phi}_e'},{{\phi}_a'}} \right) \in \mathbb{C}^{F \times 1}$ are the steering vectors corresponding to working and faulty elements, respectively. ${{\phi}_e'}$ and ${{\phi}_a'}$ are the elevation and azimuth angles from the RIS toward the scatter. \eqref{H_TRMW} and \eqref{H_TRMF} are obtained by excluding $F$ faulty phase delays (indexed by ${\mathcal S_F}$) and $W$ functional phase delays (indexed by ${\mathcal S_W}$), respectively, from the total $R$ phase delays in ${\bf{a}}\left( {{\phi_e},{\phi_a}} \right)$, as defined in \eqref{SVa}.
The cascaded sensing channel through working and faulty elements can be written as follows
\begin{align}
& {{{\bf{\bar G}}}_W} = \alpha {{\bf{H}}_{BR,W}}{\bf{\Theta }}_W^{\rm{H}}{{\bf{H}}_{{\rm{TRM}},W}}{{\bf{\Theta }}_W}{\bf{H}}_{BR,W}^{\rm{H}}, \label{Gw}\\
& {{{\bf{\bar G}}}_F} = \alpha' {{\bf{H}}_{BR,F}}{\bf{\Theta }}_F^{\rm{H}}{{\bf{H}}_{{\rm{TRM}},F}}{{\bf{\Theta }}_F}{\bf{H}}_{BR,F}^{\rm{H}},\label{Gf}
\end{align}
where ${{{\bf{H}}_{BR,W}}}\in {\mathbb{C}^{N_t \times W}}$ and ${{{\bf{H}}_{BR,F}}}\in {\mathbb{C}^{N_t \times F}}$ are the channels from BS to RIS corresponding to working and faulty elements, respectively. $\alpha'$ is the channel attenuation under the reflection toward the scatter. 
The overall cascaded sensing channel ${\bf{\bar G}} \in \mathbb{C}^{N_t \times N_t}$ can be obtained by adding up (\ref{Gw}) and (\ref{Gf}), namely ${\bf{\bar G}} = {{{\bf{\bar G}}}_W} + {{{\bf{\bar G}}}_F}$.

\textit{Remark 1: In the context of sensing, signals traverse the RIS twice: initially during transmission from the BS, and subsequently from echoes reflected by either the target or the scatter. These signals interact with both working and faulty elements. Due to the attenuation caused by faulty elements, the echoes received by them are significantly weaker than those received by working elements. Signal paths involving emission from working elements and reception by faulty elements, or vice versa, are also considerably attenuated. As these contributions are negligible relative to the dominant signal components, \eqref{Gw} considers only the signals that are both transmitted and received by working elements.
Additionally, since the scatter is assumed to be far from the target, the echoes reflected by the scatter and subsequently received by working elements exhibit low power and are similarly neglected in \eqref{Gw}. This simplification is justified by the fact that the working elements are designed to form a beam directed toward the target rather than the scatter. In contrast, echoes both transmitted and received by faulty elements are more diffusely distributed in space. For tractability, \eqref{Gf} only accounts for the echoes that leak toward the direction of the scatter.
}

\subsection{Communication Model}
\label{COMSINR}
Given the transmitted signal ${\bf{x}}\left( t \right)$, the received signal of the $k$-th UE at time slot $t$ is expressed as follows
\begin{equation}\label{Yck}
\begin{array}{*{20}{l}}
{{y_{c,k}}\left( t \right) = {{{\bf{\bar h}}}_k}{\bf{x}}\left( t \right) = \underbrace {{{{\bf{\bar h}}}_k}{{\bf{w}}_k}{c_k}\left( t \right)}_{{\rm{desired}}\;{\rm{signal}}} + }\\
{\underbrace {\sum\limits_{i = 1,i \ne k}^K {{{{\bf{\bar h}}}_k}{{\bf{w}}_i}{c_i}\left( t \right)} }_{{\rm{inter - user}}\;{\rm{interference}}} + \underbrace {{{{\bf{\bar h}}}_k}{\bf{s}}\left( t \right)}_{{\rm{sensing}}\;{\rm{interference}}} + {n_k}(t),}
\end{array}
\end{equation}
where ${n_k}(t) \sim \mathcal{C}\mathcal{N}\left( {0,\sigma _k^2} \right)$ represents the complex Gaussian white noise experienced by the $k$-th UE. Signals incident on faulty RIS elements suffer from attenuation and get unwanted phase changes, thus being reflected towards unintended directions. Consequently, each UE may receive signals intended for other UEs, causing inter-user interference and potential information leakage. To address this, we fully exploit the remaining functional RIS elements to preserve communication performance. The SINR is employed to evaluate each UE's performance under RIS faults, aiming to enhance desired signal power while suppressing interference from other UEs and the target. The SINR is expressed as follows
\begin{equation}\label{CSINR}
{\gamma _k} = \frac{{{{\left| {{{{\bf{\bar h}}}_k}{{\bf{w}}_k}} \right|}^2}}}{{\sum\limits_{i = 1,i \ne k}^K {{{\left| {{{{\bf{\bar h}}}_k}{{\bf{w}}_i}} \right|}^2}}  + {{{\bf{\bar h}}}_k}{{\bf{R}}_s}{\bf{\bar h}}_k^{\rm{H}} + \sigma _k^2}}.
\end{equation}

\subsection{Sensing Model}
For target parameter estimation, the received echo signals at the BS are analyzed over a coherent time block consisting of $T$ time slots, namely
\begin{equation}\label{YsT}
{{\bf{Y}}_s} = {\bf{\bar GX}} + {{\bf{N}}_s},
\end{equation}
where ${\bf{X}} = \left[ {{\bf{x}}\left( 1 \right),...,{\bf{x}}\left( T \right)} \right] \in {\mathbb{C}^{N_t \times T}}$. ${{\bf{N}}_s} \in {\mathbb{C}^{N_t \times T}}$ is the noise with each entry obeying the complex Gaussian distribution with zero mean and variance $\sigma_s^2$. In this work, we focus on the estimation of the 2D AoD from the RIS towards the target. To measure the accuracy of AoD estimation, CRB can be utilized to provide a lower bound of the mean square error for unbiased estimation. However, in the considered faulty RIS scenario, CRB becomes inappropriate as it does not account for the RIS model mismatch caused by faulty elements. To solve this issue, we resort to MCRB and lower bound (LB) to capture the RIS model mismatch, which will be introduced and derived in the next section.
\section{MCRB Derivation and Analysis}

In this section, we analyze the performance degradation of AoD estimation for sensing, caused by the model mismatch between the faulty RIS model and the perfect RIS model. To quantify this degradation, we adopt MCRB and {\color{blue}LB} as the performance metrics. We begin by introducing the definition of MCRB and  {\color{blue}LB}, followed by the derivation of MCRB under the faulty RIS-aided ISAC scenario.

\subsection{Solution to the Pseudo-true Parameter}
Under the faulty RIS model, the unknown parameters to estimate are denoted by ${\boldsymbol{\bar \eta }} = {\left[ {{{\bar \phi }_e},{{\bar \phi }_a},{\rm{Re}}\left( {\bar \alpha } \right),{\rm{Im}}\left( {\bar \alpha } \right)} \right]^{\rm{T}}} \in {\mathbb{R}^{4\times 1}}$. To derive the MCRB, we first vectorize~(\ref{YsT}) as follows
\begin{equation}\label{YsVector}
{{\bf{y}}_s} = {\rm{vec}}\left( {{{\bf{Y}}_s}} \right) = {\rm{vec}}\left( {{\bf{\bar GX}}} \right) + {\rm{vec}}\left( {{{\bf{N}}_s}} \right) = {\boldsymbol{\bar \mu }} + {{\bf{n}}_s},
\end{equation}
where ${\boldsymbol{\bar \mu }} = {\rm{vec}}\left( {{\bf{\bar GX}}} \right) \in {\mathbb{C}^{{N_t}T\times 1}}$ means the noise-free observation of the received echo. Then, we can obtain the probability density function (PDF) of the true\footnote{\textit{True} refers to the realistic scenario with faulty RIS elements.} received echo signals under faulty RIS elements, namely
\begin{equation}\label{PDFtrue}
p\left( {{{\bf{y}}_s}} \right) = \frac{1}{{\sqrt {{{\left( {2\pi \sigma _s^2} \right)}^{{N_t}T}}} }}\exp \left( { - \frac{{{{\left\| {{{\bf{y}}_s} - {\boldsymbol{\bar \mu }}} \right\|}^2}}}{{2\sigma _s^2}}} \right).
\end{equation}
In contrast, the parameters to estimate in the perfect RIS model are denoted by ${\boldsymbol{\eta }} = {\left[ {{\phi _e},{\phi _a},{\rm{Re}}\left( \alpha  \right),{\rm{Im}}\left( \alpha  \right)} \right]^{\rm{T}}} \in {\mathbb{R}^{4\times 1}}$. Correspondingly, the noise-free observation of the received echo from (\ref{GsensingAssumed}) is denoted by ${\boldsymbol{\tilde \mu }} = {\rm{vec}}( {{\bf{\tilde GX}}} ) \in {\mathbb{C}^{{N_t}T\times 1}}$. The misspecified\footnote{\textit{Misspecified} corresponds to the idealized assumption of a perfect RIS model, which is mismatched/misspecified to the true case.} PDF of the received echo signals for $\boldsymbol{\eta}$ under the perfect RIS model is expressed as
\begin{equation}\label{PDFmisspec}
\tilde p\left( {{{\bf{y}}_s}\left| {\boldsymbol{\eta }} \right.} \right) = \frac{1}{{\sqrt {{{\left( {2\pi \sigma _s^2} \right)}^{{N_t}T}}} }}\exp \left( { - \frac{{{{\left\| {{{\bf{y}}_s} - {\boldsymbol{\tilde \mu }}\left( {\boldsymbol{\eta }} \right)} \right\|}^2}}}{{2\sigma _s^2}}} \right).
\end{equation}
Then, we introduce the pseudo-true parameter ${{\boldsymbol{\eta }}_0}\in {\mathbb{R}^{4\times 1}}$ to derive the MCRB. ${\boldsymbol{\eta }}_0$ is defined to minimize the Kullback-Leibler (KL) divergence between the true and misspecified PDF in~(\ref{PDFtrue}) and~(\ref{PDFmisspec}), given by
\begin{equation}\label{PseudoP}
{{\boldsymbol{\eta }}_0} = \mathop {\arg \min }\limits_{\boldsymbol{\eta }} {D_{KL}}\left( {p\left( {{{\bf{y}}_s}} \right)\left\| {\tilde p\left( {{{\bf{y}}_s}\left| {\boldsymbol{\eta }} \right.} \right)} \right.} \right),
\end{equation}
where ${D_{KL}}$ stands for the KL divergence between ${p\left( {{{\bf{y}}_s}} \right)}$ and ${\tilde p\left( {{{\bf{y}}_s}\left| {\boldsymbol{\eta }} \right.} \right)}$. ${{\boldsymbol{\eta }}_0}$ corresponds to the parameter set in the perfect RIS model whose PDF is closest, in the KL divergence sense, to the true PDF under the faulty RIS model\cite{bMCRBorigin}. We solve (\ref{PseudoP}) based on the method proposed in~\cite{bMCRB} (see Lemma 1 and equations (23) and (24) in~\cite{bMCRB}) to obtain ${{\boldsymbol{\eta }}_0}$.

\subsection{MCRB and LB Derivation}
\label{MCRBsensing}
We denote the misspecified-unbiased (MS-unbiased) estimator of the true parameters ${\boldsymbol{\bar \eta }}$ as ${\boldsymbol{\hat \eta }}\left( {{{\bf{y}}_s}} \right)$. Thus, the mean of the estimator ${\boldsymbol{\hat \eta }}\left( {{{\bf{y}}_s}} \right)$ under the faulty RIS model is the pseudo-true parameter ${{\boldsymbol{\eta }}_0}$. The MCRB is defined as the lower bound of the error covariance matrix of any MS-unbiased estimator~\cite{bMCRB}, where the error is the deviation between the estimator and the pseudo-true parameters ${{\boldsymbol{\eta }}_0}$. Thus, we have
\begin{equation}\label{E_MCRB}
\mathbb{E}\left\{ {\left( {{\boldsymbol{\hat \eta }}\left( {{{\bf{y}}_s}} \right) - {{\boldsymbol{\eta }}_0}} \right){{\left( {{\boldsymbol{\hat \eta }}\left( {{{\bf{y}}_s}} \right) - {{\boldsymbol{\eta }}_0}} \right)}^{\rm{T}}}} \right\} \succeq \rm{MCRB}\left( {{{\boldsymbol{\eta }}_0}} \right).
\end{equation}
The MCRB is further obtained as
\begin{equation}\label{MCRB}
\rm{MCRB} \buildrel \Delta \over = {\bf{A}}_{{{\boldsymbol{\eta }}_0}}^{ - 1}{{\bf{B}}_{{{\boldsymbol{\eta }}_0}}}{\bf{A}}_{{{\boldsymbol{\eta }}_0}}^{ - 1}.
\end{equation}
The $\left( {i,j} \right)$-th entry of ${{\bf{A}}_{{{\boldsymbol{\eta }}_0}}}\in {\mathbb{R}^{4\times 4}}$ and ${{\bf{B}}_{{{\boldsymbol{\eta }}_0}}}\in {\mathbb{R}^{4\times 4}}$ are given in (\ref{MCRBA}) and (\ref{MCRBB})~\cite{bModelMisAB}, respectively, shown at the top of next page, where ${\boldsymbol{\varepsilon }}\left( {\boldsymbol{\eta }} \right) \buildrel \Delta \over = {\boldsymbol{\bar \mu }} - {\boldsymbol{\tilde \mu }}\left( {\boldsymbol{\eta }} \right) \in {\mathbb{C}^{{N_t}T\times 1}}$.
\begin{figure*}[!t]
\begin{equation}\label{MCRBA}
{\left[ {{{\bf{A}}_{{{\boldsymbol{\eta }}_0}}}} \right]_{ij}} = \mathbb{E} \left\{ {{{\left. {\frac{{{\partial ^2}}}{{\partial {\eta _i}\partial {\eta _j}}}\log \tilde p\left( {{{\bf{y}}_s}\left| {\boldsymbol{\eta }} \right.} \right)} \right|}_{{\boldsymbol{\eta }} = {{\boldsymbol{\eta }}_0}}}} \right\} = {\left. {\frac{2}{{\sigma _s^2}}{\rm{Re}}\left[ {{\boldsymbol{\varepsilon }}{{\left( {\boldsymbol{\eta }} \right)}^{\rm{H}}}\frac{{{\partial ^2}{\boldsymbol{\tilde \mu }}\left( {\boldsymbol{\eta }} \right)}}{{\partial {\eta _i}\partial {\eta _j}}} - {{\left( {\frac{{\partial {\boldsymbol{\tilde \mu }}\left( {\boldsymbol{\eta }} \right)}}{{\partial {\eta _i}}}} \right)}^{\rm{H}}}\frac{{\partial {\boldsymbol{\tilde \mu }}\left( {\boldsymbol{\eta }} \right)}}{{\partial {\eta _j}}}} \right]} \right|_{{\boldsymbol{\eta }} = {{\boldsymbol{\eta }}_0}}}.
\end{equation}
\vspace{-0.7cm}
\end{figure*}
\begin{figure*}[!t]
\begin{equation}\label{MCRBB}
{\left[ {{{\bf{B}}_{{{\boldsymbol{\eta }}_0}}}} \right]_{ij}} = {\left. {\left[ \mathbb{E}{\left\{ {\frac{{\partial \log \tilde p\left( {{{\bf{y}}_s}\left| {\boldsymbol{\eta }} \right.} \right)}}{{\partial {\eta _i}}}\frac{{\partial \log \tilde p\left( {{{\bf{y}}_s}\left| {\boldsymbol{\eta }} \right.} \right)}}{{\partial {\eta _j}}}} \right\} - \frac{{\partial {D_{KL}}}}{{\partial {\eta _i}}}\frac{{\partial {D_{KL}}}}{{\partial {\eta _j}}}} \right]} \right|_{{\boldsymbol{\eta }} = {{\boldsymbol{\eta }}_0}}} = \frac{2}{{\sigma _s^2}}{\left. {{\rm{Re}}\left\{ {{{\left( {\frac{{\partial {\boldsymbol{\tilde \mu }}\left( {\boldsymbol{\eta }} \right)}}{{\partial {\eta _i}}}} \right)}^{\rm{H}}}\frac{{\partial {\boldsymbol{\tilde \mu }}\left( {\boldsymbol{\eta }} \right)}}{{\partial {\eta _j}}}} \right\}} \right|_{{\boldsymbol{\eta }} = {{\boldsymbol{\eta }}_0}}}.
\end{equation}
\vspace{-0.6cm}
\end{figure*}
Moreover, the lower bound of the covariance matrix of any MS-unbiased estimator for the true parameters ${\boldsymbol{\bar \eta }}$ is denoted as LB, namely
\begin{equation}\label{E_LB}
\mathbb{E}\left\{ {\left( {{\boldsymbol{\hat \eta }}\left( {{{\bf{y}}_s}} \right) - {\boldsymbol{\bar \eta }}} \right){{\left( {{\boldsymbol{\hat \eta }}\left( {{{\bf{y}}_s}} \right) - {\boldsymbol{\bar \eta }}} \right)}^{\rm{T}}}} \right\} \succeq \rm{LB}\left( {{{\boldsymbol{\eta }}_0}} \right).
\end{equation}
The LB is further calculated as follows
\begin{equation}\label{LB}
{\rm{LB}}\left( {{{\boldsymbol{\eta }}_0}} \right) \buildrel \Delta \over = \underbrace {{\bf{A}}_{{{\boldsymbol{\eta }}_0}}^{ - 1}{{\bf{B}}_{{{\boldsymbol{\eta }}_0}}}{\bf{A}}_{{{\boldsymbol{\eta }}_0}}^{ - 1}}_{{\rm{MCRB}}\left( {{{\boldsymbol{\eta }}_0}} \right)} + \underbrace {\left( {{\boldsymbol{\bar \eta }} - {{\boldsymbol{\eta }}_0}} \right){{\left( {{\boldsymbol{\bar \eta }} - {{\boldsymbol{\eta }}_0}} \right)}^{\rm{T}}}}_{{\rm{bias}}\left( {{{\boldsymbol{\eta }}_0}} \right)},
\end{equation}
where the second term in (\ref{LB}) is a bias term. Note that the bias term in LB is unaffected by the specific transmit beamforming and RIS phase shift design. With an obtained ${{\boldsymbol{\eta }}_0}$ from (\ref{PseudoP}), optimizing LB is equivalent to optimizing MCRB. Thus, to simplify the problem formulation, we leverage MCRB as the metric to capture the lower bound of mean square error for AoD estimation in the presence of faulty RIS elements. In this work, we focus on the target's 2D AoD estimation with respect to the RIS, including the elevation AoD ${\phi _e}$ and azimuth AoD ${\phi _a}$. The specific derivation of MCRB with respect to ${\boldsymbol{\phi}} = {\left[ {{\phi _e},{\phi _a}} \right]^{\rm{T}}}$, denoted by ${{\rm{MCRB}}_{\boldsymbol{\phi}} }\in \mathbb{R}^{2 \times 2}$, is given as follows.

According to \eqref{MCRB}, ${\rm{MCRB}}_{\boldsymbol{\phi}}$ depends on ${{\bf{A}}_{{{\boldsymbol{\eta }}_0}}}$ and ${{\bf{B}}_{{{\boldsymbol{\eta }}_0}}}$, which consist of block matrices with respect to the parameters $\boldsymbol{\phi}$ and $\alpha$, namely ${{\bf{A}}_{\boldsymbol{\phi} \boldsymbol{\phi} }},{{\bf{A}}_{\alpha \boldsymbol{\phi} }},{{\bf{A}}_{\boldsymbol{\phi} \alpha }},{\bf{A}}_{\alpha \alpha }$, ${{\bf{B}}_{\boldsymbol{\phi} \boldsymbol{\phi} }},{{\bf{B}}_{\alpha \boldsymbol{\phi} }},{{\bf{B}}_{\boldsymbol{\phi} \alpha }},{\bf{B}}_{\alpha \alpha }$. The detailed derivations are given in Appendix~\ref{AppCalcuAB}. For a block matrix ${{\bf{A}}_{{{\boldsymbol{\eta }}_0}}}$ shown in (\ref{ABblock}), its inverse is
\begin{equation}\label{Ainv}
{\bf{A}}_{{{\boldsymbol{\eta }}_0}}^{ - 1} \!=\! \left[ \! {\begin{array}{*{20}{c}}
{{{\bf{Z}}^{ - 1}}}&{ - {{\bf{Z}}^{ - 1}}{{\bf{A}}_{\alpha \phi }}{\bf{A}}_{\alpha \alpha }^{ - 1}}\\
{ - {\bf{A}}_{\alpha \alpha }^{ - 1}{{\bf{A}}_{\phi \alpha }}{{\bf{Z}}^{ - 1}}}&{{\bf{A}}_{\alpha \alpha }^{ - 1} + {\bf{A}}_{\alpha \alpha }^{ - 1}{{\bf{A}}_{\phi \alpha }}{{\bf{Z}}^{ - 1}}{{\bf{A}}_{\alpha \phi }}{\bf{A}}_{\alpha \alpha }^{ - 1}}
\end{array}} \! \right],
\end{equation}
where ${\bf{Z}} = {{\bf{A}}_{\boldsymbol{\phi} \boldsymbol{\phi} }} - {{\bf{A}}_{\alpha \boldsymbol{\phi} }}{\bf{A}}_{\alpha \alpha }^{ - 1}{{\bf{A}}_{\boldsymbol{\phi} \alpha }}\in \mathbb{R}^{2 \times 2}$. According to (\ref{MCRB}), we can obtain MCRB by substituting the specific values of ${\bf{A}}_{{{\boldsymbol{\eta }}_0}}^{ - 1}$ and ${\bf{B}}_{{{\boldsymbol{\eta }}_0}}$ into it, namely substituting (\ref{Ainv}) and (\ref{ABblock}) into (\ref{MCRB}). The first diagonal block matrix of MCRB is namely ${{\rm{MCRB}}_{\boldsymbol{\phi}} }$, with the final derived result given in the equation \lcirc{1} of (\ref{MCRB_AoD}). In the following sections, we optimize the performance of the ISAC system based on the derived ${{\rm{MCRB}}_{\boldsymbol{\phi}} }$.

\begin{figure*}[!t]
\begin{equation}\label{MCRB_AoD}
\begin{array}{*{20}{l}}
{{\rm{MCRB}_\phi}\mathop  = \limits^{\lcirc{1}} {{\bf{Z}}^{ - 1}}\left( {{{\bf{B}}_{\phi \phi }} + {{\bf{A}}_{\alpha \phi }}{\bf{A}}_{\alpha \alpha }^{ - 1}{{\bf{B}}_{\alpha \alpha }}{\bf{A}}_{\alpha \alpha }^{ - 1}{{\bf{A}}_{\phi \alpha }} - {{\bf{A}}_{\alpha \phi }}{\bf{A}}_{\alpha \alpha }^{ - 1}{{\bf{B}}_{\phi \alpha }} - {{\bf{B}}_{\alpha \phi }}{\bf{A}}_{\alpha \alpha }^{ - 1}{{\bf{A}}_{\phi \alpha }}} \right){{\bf{Z}}^{ - 1}}}\\
{\mathop  = \limits^{\lcirc{2}} {{\bf{Z}}^{ - 1}}\left( {{{\bf{B}}_{\phi \phi }} - {{\bf{A}}_{\alpha \phi }}{\bf{A}}_{\alpha \alpha }^{ - 1}{{\bf{A}}_{\phi \alpha }} - \left( {{{\bf{A}}_{\alpha \phi }}{\bf{A}}_{\alpha \alpha }^{ - 1}{{\bf{B}}_{\phi \alpha }} + {{\bf{B}}_{\alpha \phi }}{\bf{A}}_{\alpha \alpha }^{ - 1}{{\bf{A}}_{\phi \alpha }}} \right)} \right){{\bf{Z}}^{ - 1}}}\\
{ = {{\bf{Z}}^{ - 1}}\left( {{{\bf{B}}_{\phi \phi }} - {{\bf{A}}_{\alpha \phi }}{\bf{A}}_{\alpha \alpha }^{ - 1}{{\bf{A}}_{\phi \alpha }} + {{\bf{A}}_{\alpha \phi }}{\bf{A}}_{\alpha \alpha }^{ - 1}{{\bf{A}}_{\phi \alpha }} + {{\bf{B}}_{\alpha \phi }}{\bf{A}}_{\alpha \alpha }^{ - 1}{{\bf{B}}_{\phi \alpha }} - \left( {{{\bf{A}}_{\alpha \phi }} + {{\bf{B}}_{\alpha \phi }}} \right){\bf{A}}_{\alpha \alpha }^{ - 1}\left( {{{\bf{A}}_{\phi \alpha }} + {{\bf{B}}_{\phi \alpha }}} \right)} \right){{\bf{Z}}^{ - 1}}}\\
{ = {{\bf{Z}}^{ - 1}}\left( {{{\bf{B}}_{\phi \phi }} + {{\bf{B}}_{\alpha \phi }}{\bf{A}}_{\alpha \alpha }^{ - 1}{{\bf{B}}_{\phi \alpha }} - \left( {{{\bf{A}}_{\alpha \phi }} + {{\bf{B}}_{\alpha \phi }}} \right){\bf{A}}_{\alpha \alpha }^{ - 1}\left( {{{\bf{A}}_{\phi \alpha }} + {{\bf{B}}_{\phi \alpha }}} \right)} \right){{\bf{Z}}^{ - 1}}}\\
{ = {{\bf{Z}}^{ - 1}}\left( {{{\bf{B}}_{\phi \phi }} - {{\bf{B}}_{\alpha \phi }}{\bf{B}}_{\alpha \alpha }^{ - 1}{{\bf{B}}_{\phi \alpha }} + \left( {{{\bf{A}}_{\alpha \phi }} + {{\bf{B}}_{\alpha \phi }}} \right){\bf{B}}_{\alpha \alpha }^{ - 1}\left( {{{\bf{A}}_{\phi \alpha }} + {{\bf{B}}_{\phi \alpha }}} \right)} \right){{\bf{Z}}^{ - 1}}}\\
{ = {{\bf{Z}}^{ - 1}}{\bf{U}}{{\bf{Z}}^{ - 1}}.}
\end{array}
\end{equation}
\vspace{-1cm}
\end{figure*}
\section{Problem Formulation}
In Section~\ref{COMSINR} and \ref{MCRBsensing}, we have derived the metrics SINR for communication and MCRB for sensing to measure the impact of faulty RIS elements on the RIS-aided ISAC system. In this section, we aim to mitigate the negative impact of faulty elements on both sensing and communication performances. To realize this goal, we jointly design the transmit beamforming, sensing covariance matrix, and the remaining working RIS phase shifts to optimize the performance of the considered ISAC system. An optimization problem is formulated to minimize the MCRB of ${\phi _e}$ and ${\phi _a}$, subject to the communication SINR requirements, transmit power, and RIS phase shift constraints, mathematically written as follows
\begin{subequations}\label{OptP1}
\begin{align}
& ({\rm P}1) \mathop {\min }\limits_{\{ {\bf{W}},{{\bf{R}}_s},{{\bf{v}}_W}\} } {\rm{tr}}\left( {{\rm{MCRB}}_{\boldsymbol{\phi}} } \right) \label{P1Obj}\\
\mbox{s.t.}&
\frac{{{{\left| {{{{\bf{\bar h}}}_k}{{\bf{w}}_k}} \right|}^2}}}{{\sum\limits_{i = 1,i \ne k}^K {{{\left| {{{{\bf{\bar h}}}_k}{{\bf{w}}_i}} \right|}^2}}  + {{{\bf{\bar h}}}_k}{{\bf{R}}_s}{\bf{\bar h}}_k^{\rm{H}} + \sigma _k^2}} \ge {\gamma _{th}},\forall k \in {\mathcal K}, \label{const_COM} \\
& {\rm{tr}}({\bf{W}}{{\bf{W}}^{\rm{H}}}) + {\rm{tr}}({{\bf{R}}_s}) \le {P_{\max },}\label{const_Pwr} \\
& {{\bf{R}}_s} \succeq 0 , \label{const_Rs} \\
& \left| {{v_w}} \right| = 1,\forall w \in {\mathcal S_W}. \label{const_RIS}
\end{align}
\end{subequations}
Constraint (\ref{const_COM}) ensures communication performance by guaranteeing each UE's SINR is not smaller than a predefined threshold ${\gamma _{th}}$. Constraint (\ref{const_Pwr}) means the power used for sensing and communication should not exceed the maximum power $P_{\max }$ by the BS. Constraint (\ref{const_Rs}) guarantees the property of the sensing covariance matrix to be PSD. Constraint (\ref{const_RIS}) is the unit modulus constraint for the working RIS elements. The formulated problem $\rm{P}$1 is highly non-convex due to the complex form of the objective ${\rm{MCRB}}_{\boldsymbol{\phi}}$, the coupling among variables, and nonlinear constraints, posing significant challenges for optimization. In the following section, we present a tailored solution to address problem $\rm{P}$1.
\section{Solution}
In this section, we solve the formulated problem $\rm{P}$1 by jointly optimizing the transmit beamforming and the remaining functional RIS phase shifts. Due to the highly complex expression of ${{\rm{MCRB}}_{\boldsymbol{\phi}}}$, which is challenging to solve directly, we first transform problem $\rm{P}$1 into a more tractable form. Subsequently, we propose a BCD-based algorithm that incorporates the MM and SCA techniques and the penalty method, with more details provided below.

The structure of ${{\rm{MCRB}}_{\boldsymbol{\phi}}}$ is reformulated through the transformation shown in equation \lcirc{2} of (\ref{MCRB_AoD}), where the last step is due to ${{\bf{A}}_{\alpha \alpha }} =  - {{\bf{B}}_{\alpha \alpha }}$, provided in (\ref{Aalphaphi}) in Appendix~\ref{AppCalcuAB}. Then, we prove that ${\bf{U}}$ and ${\bf{Z}}$ in (\ref{MCRB_AoD}) are PSD and negative definite, respectively. Note that for ${{\bf{B}}_{{{\boldsymbol{\eta }}_0}}}$ defined in (\ref{MCRBB}), we have the following (\ref{BPSD}) for any column vector ${\bf{m}}$,
\begin{equation}\label{BPSD}
{{\bf{m}}^{\rm{H}}}{\bf{Bm}} = \frac{2}{{\sigma _s^2}}{{\bf{m}}^{\rm{H}}}{\rm{Re}}\left\{ {{{\left( {\frac{{\partial {\boldsymbol{\tilde \mu }}\left( {\boldsymbol{\eta }} \right)}}{{\partial {\bf{\eta }}}}} \right)}^{\rm{H}}}\frac{{\partial {\boldsymbol{\tilde \mu }}\left( {\boldsymbol{\eta }} \right)}}{{\partial {\bf{\eta }}}}} \right\}{\bf{m}} \ge 0,
\end{equation}
infering that ${{\bf{B}}_{{{\boldsymbol{\eta }}_0}}}$ and its diagonal block matrices, namely ${\bf{B}}_{{\boldsymbol{\phi}} {\boldsymbol{\phi}} }$ and ${{\bf{B}}_{\alpha \alpha }}$, are all PSD matrices. Thus, the Schur complement of ${{\bf{B}}_{\alpha \alpha }}$ is also PSD, namely ${{{\bf{B}}_{{\boldsymbol{\phi}} {\boldsymbol{\phi}} }} - {{\bf{B}}_{\alpha {\boldsymbol{\phi}} }}{\bf{B}}_{\alpha \alpha }^{ - 1}{{\bf{B}}_{{\boldsymbol{\phi}} \alpha }}} \succeq 0$~\cite{bPSD}. Meanwhile, (\ref{AplusBPSD}) is also satisfied for any column vector ${\bf{m}}$,
\begin{equation}\label{AplusBPSD}
{{\bf{m}}^{\rm{H}}}\left( {{{\bf{A}}_{\alpha {\boldsymbol{\phi}} }} + {{\bf{B}}_{\alpha {\boldsymbol{\phi}} }}} \right){\bf{B}}_{\alpha \alpha }^{ - 1}\left( {{{\bf{A}}_{{\boldsymbol{\phi}} \alpha }} + {{\bf{B}}_{{\boldsymbol{\phi}} \alpha }}} \right){\bf{m}} \ge 0,
\end{equation}
which infers ${\left( {{{\bf{A}}_{\alpha {\boldsymbol{\phi}} }} + {{\bf{B}}_{\alpha {\boldsymbol{\phi}} }}} \right){\bf{B}}_{\alpha \alpha }^{ - 1}\left( {{{\bf{A}}_{{\boldsymbol{\phi}} \alpha }} + {{\bf{B}}_{{\boldsymbol{\phi}} \alpha }}} \right)}$ is PSD. Since the addition of two PSD matrices is also PSD, ${\bf{U}}$ in (\ref{MCRB_AoD}) is PSD.

Note that the Fisher information matrix (FIM) is expressed as $\mathbb{E} \left\{ {\frac{{\partial \log \tilde p\left( {{{\bf{y}}_s}\left| {\boldsymbol{\eta }} \right.} \right)}}{{\partial {\eta _i}}}\frac{{\partial \log \tilde p\left( {{{\bf{y}}_s}\left| {\boldsymbol{\eta }} \right.} \right)}}{{\partial {\eta _j}}}} \right\} =  - \mathbb{E} \left\{ {\frac{{{\partial ^2}\log \tilde p\left( {{{\bf{y}}_s}\left| {\boldsymbol{\eta }} \right.} \right)}}{{\partial {\eta _i}\partial {\eta _j}}}} \right\}$~\cite{CRBKay}, which is PSD due to its left-hand expression of the first-order derivative. We can observe that for ${{\bf{A}}_{{{\boldsymbol{\eta }}_0}}}$ defined in (\ref{MCRBA}), $\left\{ {\frac{{{\partial ^2}}}{{\partial {\eta _i}\partial {\eta _j}}}\log \tilde p\left( {{{\bf{y}}_s}\left| {\boldsymbol{\eta}} \right.} \right)} \right\}$ is equal to the negative FIM. Therefore, ${\bf{A}}_{{{\boldsymbol{\eta }}_0}}$ and ${\bf{A}}_{\alpha \alpha }$ are negative definite. Hence, we can infer that ${\bf{Z}} = {{\bf{A}}_{\boldsymbol{\phi} \boldsymbol{\phi} }} - {{\bf{A}}_{\alpha \boldsymbol{\phi} }}{\bf{A}}_{\alpha \alpha }^{ - 1}{{\bf{A}}_{\boldsymbol{\phi} \alpha }} \prec 0$ is negative definite. Thus, ${{\bf{Z}}^{ - 1}}$ is negative definite, while ${{\bf{Z}}^{ - 1}}{{\bf{Z}}^{ - 1}} = {\left( {{\bf{ZZ}}} \right)^{ - 1}} = {\left( {{{\bf{Z}}^{\rm{T}}}{\bf{Z}}} \right)^{ - 1}}$ is positive definite, proved in Appendix~\ref{AppZZ}.

Due to the multiplicative relationship among the terms in (\ref{MCRB_AoD}), the variables exhibit a high degree of coupling. To deal with the variable coupling and make problem $\rm{P}$1 more tractable, we introduce two auxiliary matrices ${\bf{\tilde C}}\in \mathbb{R}^{2 \times 2}$ and ${\bf{\tilde D}}\in \mathbb{R}^{2 \times 2}$ to obtain an upper bound of the original problem $\rm{P}$1. To realize this goal, we first introduce a PSD matrix ${{\bf{\tilde C}}} \succeq 0$ for ${\bf{U}}$ in (\ref{MCRB_AoD}), and let $( {{{\bf{B}}_{{\boldsymbol{\phi}} {\boldsymbol{\phi}} }} + \left( {{{\bf{A}}_{\alpha {\boldsymbol{\phi}} }} + {{\bf{B}}_{\alpha {\boldsymbol{\phi}} }}} \right){\bf{B}}_{\alpha \alpha }^{ - 1}\left( {{{\bf{A}}_{{\boldsymbol{\phi}} \alpha }} + {{\bf{B}}_{{\boldsymbol{\phi}} \alpha }}} \right)} ) \le {\bf{\tilde C}}$ in PSD space, which can be rewritten as $( {\bf{\tilde C}} -  {\bf{B}}_{{\boldsymbol{\phi}} {\boldsymbol{\phi}} } ) - \left( {{{\bf{A}}_{\alpha {\boldsymbol{\phi}} }} + {{\bf{B}}_{\alpha {\boldsymbol{\phi}} }}} \right){\bf{B}}_{\alpha \alpha }^{ - 1}\left( {{{\bf{A}}_{{\boldsymbol{\phi}} \alpha }} + {{\bf{B}}_{{\boldsymbol{\phi}} \alpha }}} \right) \succeq 0$. Based on the Schur complement, it can be transformed into constraint
\begin{equation}\label{const_AuxilC1}
\left[ {\begin{array}{*{20}{c}}
{{\bf{\tilde C}} -  {{\bf{B}}_{{\boldsymbol{\phi}} {\boldsymbol{\phi}}}}}&{{{\bf{A}}_{\alpha {\boldsymbol{\phi}} }} + {{\bf{B}}_{\alpha {\boldsymbol{\phi}} }}}\\
{{{\bf{A}}_{{\boldsymbol{\phi}} \alpha }} + {{\bf{B}}_{{\boldsymbol{\phi}} \alpha }}}&{{{\bf{B}}_{\alpha \alpha }}}
\end{array}} \right] \succeq 0.
\end{equation}
Therefore, on the condition of \eqref{const_AuxilC1}, we obtain the relation ${\rm{tr}}( {{{\bf{Z}}^{ - 1}}{\bf{U}}{{\bf{Z}}^{ - 1}}})$ $\le$ ${\rm{tr}}( {{{\bf{Z}}^{ - 1}}( {{\bf{\tilde C}} - {{\bf{B}}_{\alpha {\boldsymbol{\phi}}}}{\bf{B}}_{\alpha \alpha }^{ - 1}{{\bf{B}}_{{\boldsymbol{\phi}} \alpha }}}){{\bf{Z}}^{ - 1}}} )$.
Next, we introduce a negative definite matrix ${{\bf{\tilde D}}} \prec 0$ and let ${{\bf{Z}}^{ - 1}} = {( {{{\bf{A}}_{{\boldsymbol{\phi}} {\boldsymbol{\phi}} }} - {{\bf{A}}_{\alpha {\boldsymbol{\phi}} }}{\bf{A}}_{\alpha \alpha }^{ - 1}{{\bf{A}}_{{\boldsymbol{\phi}} \alpha }}})^{ - 1}} \ge {\bf{\tilde D}}$, which is equivalent to (\ref{DAschur}) based on the Schur complement, with the proof given in Appendix~\ref{DAproof}.
\begin{equation}\label{DAschur}
- \left[ {\begin{array}{*{20}{c}}
{{\bf{\tilde D}}}&{{{\bf{I}}_2}}&{\bf{0}}\\
{{{\bf{I}}_2}}&{{{\bf{A}}_{{\boldsymbol{\phi}} {\boldsymbol{\phi}} }}}&{{{\bf{A}}_{\alpha {\boldsymbol{\phi}} }}}\\
{\bf{0}}&{{{\bf{A}}_{{\boldsymbol{\phi}} \alpha }}}&{{{\bf{A}}_{\alpha \alpha }}}
\end{array}} \right] \succeq 0 .
\end{equation}
Likewise, we further obtain ${\rm{tr}}( {{{\bf{Z}}^{ - 1}}( {{\bf{\tilde C}} - {{\bf{B}}_{\alpha {\boldsymbol{\phi}}}}{\bf{B}}_{\alpha \alpha }^{ - 1}{{\bf{B}}_{{\boldsymbol{\phi}}\alpha }}}){{\bf{Z}}^{ - 1}}} )$ $\le$ ${\rm{tr}}( {{{\bf{\tilde D}}}( {{\bf{\tilde C}} - {{\bf{B}}_{\alpha {\boldsymbol{\phi}}}}{\bf{B}}_{\alpha \alpha }^{ - 1}{{\bf{B}}_{{\boldsymbol{\phi}} \alpha }}}){{\bf{\tilde D}}}} )$, on the condition of \eqref{const_AuxilC1} and \eqref{DAschur}. In ${\rm{tr}}( {{{\bf{\tilde D}}}( {{\bf{\tilde C}} - {{\bf{B}}_{\alpha {\boldsymbol{\phi}}}}{\bf{B}}_{\alpha \alpha }^{ - 1}{{\bf{B}}_{{\boldsymbol{\phi}} \alpha }}}){{\bf{\tilde D}}}} )$, ${\bf{\tilde C}} - {{\bf{B}}_{\alpha {\boldsymbol{\phi}}}}{\bf{B}}_{\alpha \alpha }^{ - 1}{{\bf{B}}_{{\boldsymbol{\phi}} \alpha }}$ and ${\bf{\tilde D\tilde D}}$ are PSD and positive definite (the proof is similar to that provided in Appendix \ref{AppZZ}), respectively, based on which we have the following condition for their product.
\begin{equation}\label{CmBDDRelax}
\begin{array}{l}
{\rm{tr}}\left( {{{\bf{Z}}^{ - 1}}{\bf{U}}{{\bf{Z}}^{ - 1}}} \right) \le {\rm{ tr}}\left( {{\bf{\tilde D}}\left( {{\bf{\tilde C}} - {{\bf{B}}_{\alpha {\boldsymbol{\phi}}}}{\bf{B}}_{\alpha \alpha }^{ - 1}{{\bf{B}}_{{\boldsymbol{\phi}} \alpha }}} \right){\bf{\tilde D}}} \right)\\
\le {\rm{tr}}\left( {{\bf{\tilde C}} - {{\bf{B}}_{\alpha {\boldsymbol{\phi}}}}{\bf{B}}_{\alpha \alpha }^{ - 1}{{\bf{B}}_{{\boldsymbol{\phi}} \alpha }}} \right){\rm{tr}}\left( {{\bf{\tilde D\tilde D}}} \right).
\end{array}
\end{equation}
Thus, subject to \eqref{const_AuxilC1} and \eqref{DAschur}, ${\rm{tr}}( {{\bf{\tilde C}} - {{\bf{B}}_{\alpha {\boldsymbol{\phi}}}}{\bf{B}}_{\alpha \alpha }^{ - 1}{{\bf{B}}_{{\boldsymbol{\phi}}\alpha }}}){\rm{tr}}( {{\bf{\tilde D\tilde D}}})$ is an upper bound of the original objective ${\rm{tr}}({{{\bf{Z}}^{ - 1}}{\bf{U}}{{\bf{Z}}^{ - 1}}})$. With the above transformations, problem $\rm{P}$1 is relaxed into problem $\rm{P}$1.1, defined below
\begin{subequations}\label{OptP11}
\begin{align}
({\rm P}1.1) & \mathop {\min }\limits_{\{ {\bf{W}},{{\bf{R}}_s},{{\bf{v}}_W},{\bf{\tilde C}},{\bf{\tilde D}}\} } {\rm{tr}}\left( {{\bf{\tilde C}} - {{\bf{B}}_{\alpha {\boldsymbol{\phi}} }}{\bf{B}}_{\alpha \alpha }^{ - 1}{{\bf{B}}_{{\boldsymbol{\phi}} \alpha }}} \right){\rm{tr}}\left( {{{\bf{\tilde D}}}{{\bf{\tilde D}}}} \right)\label{P11Obj}\\
\mbox{s.t.}&
\left[ {\begin{array}{*{20}{c}}
{{\bf{\tilde C}} - {{\bf{B}}_{{\boldsymbol{\phi}} {\boldsymbol{\phi}} }}}&{{{\bf{A}}_{\alpha {\boldsymbol{\phi}} }} + {{\bf{B}}_{\alpha {\boldsymbol{\phi}} }}}\\
{{{\bf{A}}_{{\boldsymbol{\phi}} \alpha }} + {{\bf{B}}_{{\boldsymbol{\phi}} \alpha }}}&{{{\bf{B}}_{\alpha \alpha }}}
\end{array}} \right] \succeq 0, \label{const_AuxilC}\\
& - \left[ {\begin{array}{*{20}{c}}
{{\bf{\tilde D}}}&{{{\bf{I}}_2}}&{\bf{0}}\\
{{{\bf{I}}_2}}&{{{\bf{A}}_{{\boldsymbol{\phi}} {\boldsymbol{\phi}} }}}&{{{\bf{A}}_{\alpha {\boldsymbol{\phi}} }}}\\
{\bf{0}}&{{{\bf{A}}_{{\boldsymbol{\phi}} \alpha }}}&{{{\bf{A}}_{\alpha \alpha }}}
\end{array}} \right] \succeq 0, \label{const_AuxilD}\\
& {{\bf{\tilde C}}} \succeq 0, {{\bf{\tilde D}}} \prec 0,\label{const_CDPSD}\\
&\eqref{const_COM},\eqref{const_Pwr},\eqref{const_Rs}, \eqref{const_RIS}.
\end{align}
\end{subequations}
We can observe that in problem $\rm{P}$1.1, the degree of coupling in the objective function is highly reduced by splitting the product of the block matrices ${{\bf{A}}_{\boldsymbol{\phi} \boldsymbol{\phi} }},{{\bf{A}}_{\alpha \boldsymbol{\phi} }},{{\bf{A}}_{\boldsymbol{\phi} \alpha }},{\bf{A}}_{\alpha \alpha }$, ${{\bf{B}}_{\boldsymbol{\phi} \boldsymbol{\phi} }},{{\bf{B}}_{\alpha \boldsymbol{\phi} }},{{\bf{B}}_{\boldsymbol{\phi} \alpha }},{\bf{B}}_{\alpha \alpha }$ in ${\bf{Z}}$ and ${\bf{U}}$, compared to the original ${{\rm{MCRB}}_{\boldsymbol{\phi}}}$ expression in (\ref{MCRB_AoD}). However, $\{ {\bf{W}},{{\bf{R}}_s},{{\bf{v}}_W}\}$ are still coupled within these block matrices, as evidenced in the objective function (\ref{P11Obj}) and constraints (\ref{const_COM}), (\ref{const_AuxilC}), (\ref{const_AuxilD}). Meanwhile, $\{{\bf{\tilde C}},{\bf{\tilde D}}\}$ are still coupled in the objective function (\ref{P11Obj}). To address this, we divide these variables into two blocks, namely $\{ {\bf{W}},{{\bf{R}}_s},{\bf{\tilde C}}\}$ and $\{{{\bf{v}}_W},{\bf{\tilde D}}\}$, and then apply the BCD framework to iteratively update each block while fixing the variables in the other block until convergence. The detailed solutions to each block are given in the following.

\subsection{Sub-problem with Respect to $\{ {\bf{W}},{{\bf{R}}_s},{\bf{\tilde C}}\}$}
Given $\{{{\bf{v}}_W},{\bf{\tilde D}}\}$, the sub-problem for $\{ {\bf{W}},{{\bf{R}}_s},{\bf{\tilde C}}\}$ is written as follows, after omitting the unrelated term.
\begin{subequations}\label{OptP12}
\begin{align}
({\rm P}1.2) & \mathop {\min }\limits_{\{ {\bf{W}},{{\bf{R}}_s},{\bf{\tilde C}}\} } {\rm{tr}}\left( {{\bf{\tilde C}} - {{\bf{B}}_{\alpha {\boldsymbol{\phi}} }}{\bf{B}}_{\alpha \alpha }^{ - 1}{{\bf{B}}_{{\boldsymbol{\phi}} \alpha }}} \right)\label{P12Obj}\\
\mbox{s.t.}&
\eqref{const_AuxilC},\eqref{const_AuxilD},\eqref{const_CDPSD}, \eqref{const_COM},\eqref{const_Pwr},\eqref{const_Rs}.
\end{align}
\end{subequations}
We define ${{\bf{W}}_k} = {{\bf{w}}_k}{\bf{w}}_k^{\rm{H}} \in {\mathbb{C}^{N_t \times N_t}}$, satisfying ${{\bf{W}}_k} \succeq 0$ and ${\rm{rank}}\left( {\bf{W}}_k \right) = 1$, for UE $k\in {\mathcal K}$. We also define ${{{\bf{\bar H}}}_k} = {\bf{\bar h}}_k^{\rm{H}}{{{\bf{\bar h}}}_k} \in {\mathbb{C}^{N_t \times N_t}}$. The SINR constraint (\ref{const_COM}) for UE $k$ can be transformed into the following form.
\begin{equation}\label{SINRtransW}
\left( {\frac{1}{{{\gamma _{th}}}} \!+\! 1} \right){\rm{tr}}\left( {{{{\bf{\bar H}}}_k}{{\bf{W}}_k}} \right) \!-\! {\rm{tr}}\left( {{{{\bf{\bar H}}}_k}\left( {\sum\limits_{k = 1}^K {{{\bf{W}}_k} \!+\! {{\bf{R}}_s}} } \right)} \right) \!\ge\! \sigma _k^2.
\end{equation}
Notice that the variables in the product term ${{{\bf{B}}_{\alpha {\boldsymbol{\phi}} }}{\bf{B}}_{\alpha \alpha }^{ - 1}{{\bf{B}}_{{\boldsymbol{\phi}} \alpha }}}$ from \eqref{P12Obj} are still coupled. To address this problem, we apply the MM technique, where we linearize the trace of this product term by its first-order Taylor approximation and obtain its lower bound ${\rm{tr}}\left( {{\bf{B'}}} \right)$~\cite{MM} as follows.
\begin{equation}\label{BBBlinear}
\begin{array}{l}
{\rm{tr}}\left( {{{\bf{B}}_{\alpha {\boldsymbol{\phi}} }}{\bf{B}}_{\alpha \alpha }^{ - 1}{{\bf{B}}_{{\boldsymbol{\phi}} \alpha }}} \right) \ge {\rm{tr}}\left( {{\bf{B'}}} \right) = \\
2{\rm{tr}}\left( {\mathring{{\bf{B}}_{\alpha {\boldsymbol{\phi}} }} {{\mathring{\left( {{\bf{B}}_{\alpha \alpha }^{ - 1}} \right)}} }{{\bf{B}}_{{\boldsymbol{\phi}} \alpha }}} \right) \!-\! {\rm{tr}}\left( {{{\mathring{\left( {{\bf{B}}_{\alpha \alpha }^{ - 1}} \right)}}}{\mathring{\bf{B}}_{{\boldsymbol{\phi}} \alpha }} {\mathring{\bf{B}}_{\alpha {\boldsymbol{\phi}} }} {\mathring{{\left( {{\bf{B}}_{\alpha \alpha }^{ - 1}} \right)}}}} \right).
\end{array}
\end{equation}
The values with $\mathring {}$ are the respective updated values from the previous iteration. Therefore, we can obtain the upper bound of ${\rm{tr}}( {{\bf{\tilde C}} - {{\bf{B}}_{\alpha {\boldsymbol{\phi}} }}{\bf{B}}_{\alpha \alpha }^{ - 1}{{\bf{B}}_{{\boldsymbol{\phi}} \alpha }}} )$ as ${\rm{tr}}({{\bf{\tilde C}}}) - {\rm{tr}}({{\bf{B'}}})$. By equivalently rewriting ${{\bf{R}}_x} = {\bf{W}}{{\bf{W}}^{\rm{H}}} + {{\bf{R}}_s}$ as ${{\bf{R}}_x} = {\sum\nolimits_{k = 1}^K {{{\bf{W}}_k} + {{\bf{R}}_s}} }$ in ${{\bf{A}}_{\boldsymbol{\phi} \boldsymbol{\phi} }},{{\bf{A}}_{\alpha \boldsymbol{\phi} }},{{\bf{A}}_{\boldsymbol{\phi} \alpha }},{\bf{A}}_{\alpha \alpha }$, ${{\bf{B}}_{\boldsymbol{\phi} \boldsymbol{\phi} }},{{\bf{B}}_{\alpha \boldsymbol{\phi} }},{{\bf{B}}_{\boldsymbol{\phi} \alpha }}$, and ${\bf{B}}_{\alpha \alpha }$, the problem to be solved at each MM iteration is given as follows.
\begin{subequations}\label{OptP13}
\begin{align}
&({\rm P}1.3) \mathop {\min }\limits_{\{ {\bf{W}}_k,{{\bf{R}}_s},{\bf{\tilde C}}\} } \left( {{\rm{tr}}\left( {{\bf{\tilde C}}} \right) - {\rm{tr}}\left( {{\bf{B'}}({{\bf{R}}_x})} \right)} \right) \label{P13Obj}\\
\mbox{s.t.}&\!
\left[ {\begin{array}{*{20}{c}}\!
{{\bf{\tilde C}} \!-\! {{\bf{B}}_{{\boldsymbol{\phi}} {\boldsymbol{\phi}} }}({{\bf{R}}_x})}&\!{{{\bf{A}}_{\alpha {\boldsymbol{\phi}} }}({{\bf{R}}_x}) \!+\! {{\bf{B}}_{\alpha {\boldsymbol{\phi}} }}({{\bf{R}}_x})}\\
{{{\bf{A}}_{{\boldsymbol{\phi}} \alpha }}({{\bf{R}}_x}) \!+\! {{\bf{B}}_{{\boldsymbol{\phi}} \alpha }}({{\bf{R}}_x})}&{{{\bf{B}}_{\alpha \alpha }}\left( {{{\bf{R}}_x}} \right)}
\end{array}} \right] \succeq 0, \label{const_AuxilC2} \\
& - \left[ {\begin{array}{*{20}{c}}
{{\bf{\tilde D}}}&{{{\bf{I}}_2}}&{\bf{0}}\\
{{{\bf{I}}_2}}&{{{\bf{A}}_{{\boldsymbol{\phi}} {\boldsymbol{\phi}} }}({{{\bf{R}}}_x})}&{{{\bf{A}}_{\alpha {\boldsymbol{\phi}} }}({{{\bf{R}}}_x})}\\
{\bf{0}}&{{{\bf{A}}_{{\boldsymbol{\phi}} \alpha }}({{{\bf{R}}}_x})}&{{{\bf{A}}_{\alpha \alpha }}({{{\bf{R}}}_x})}
\end{array}} \right] \succeq 0, \label{const_AuxilD2} \\
& {{\bf{W}}_k} \succeq 0, {\rm{rank}}\left( {\bf{W}}_k \right) = 1, \forall k \in {\mathcal K}, \label{const_WkPSD} \\
&\eqref{const_Pwr},\eqref{const_Rs}, \eqref{const_CDPSD},\eqref{SINRtransW}.
\end{align}
\end{subequations}
Problem $\rm{P}$1.3 is non-convex due to the existence of the rank-one constraint. To address this, we apply the semi-definite relaxation (SDR) technique. By dropping the rank-one constraint on ${\bf{W}}_k$, $\rm{P}$1.3 becomes an SDP problem, denoted as problem ($\rm{P}$1.3 SDP), which can be solved by the off-the-shelf tool CVX. Note that the optimal ${\bf{W}}_k$ obtained from problem ($\rm{P}$1.3 SDP) may have a higher rank. However, with Proposition 1, we can obtain the optimal solution of problem $\rm{P}$1.3 satisfying the rank-one constraint (\ref{const_WkPSD}).

\textbf{Proposition 1}: Let ${{{\bf{\hat W}}}_k}$ and ${{{\bf{\hat R}}}_s}$ denote the obtained optimal values from problem ($\rm{P}$1.3 SDP). The optimal values to problem $\rm{P}$1.3 can be constructed as follows.
\begin{equation}\label{Wopt}
{\bf{w}}_k^* = {\left( {{{{\bf{\bar h}}}_k}{{{\bf{\hat W}}}_k}{\bf{\bar h}}_k^{\rm{H}}} \right)^{ - {1 \mathord{\left/
 {\vphantom {1 2}} \right.
 \kern-\nulldelimiterspace} 2}}}{{{\bf{\hat W}}}_k}{\bf{\bar h}}_k^{\rm{H}}, \forall k \in {\mathcal K},
\end{equation}
\begin{equation}\label{Rsopt}
{\bf{R}}_s^* = {{{\bf{\hat R}}}_s} + \sum\limits_{k = 1}^K {{{{\bf{\hat W}}}_k}}  - \sum\limits_{k = 1}^K {{\bf{w}}_k^*{{\left( {{\bf{w}}_k^*} \right)}^{\rm{H}}}} .
\end{equation}
The proof can be found in~\cite{XiangLiu,XianxinSongCRB}.
The algorithm to solve problem $\rm{P}$1.2 is summarized as Algorithm~\ref{AlgoP12}.
\begin{algorithm}[htbp]
\caption{MM-based algorithm for solving problem ${\rm P}1.2$}
\label{AlgoP12}
\begin{algorithmic}[1]
\STATE \textbf{Initialization}: Initialize ${{\bf{v}}_W}$, ${\bf{\tilde D}}$, ${\mathring{\bf{B}}_{\alpha {\boldsymbol{\phi}} }}$, ${\mathring{\bf{B}}_{{\boldsymbol{\phi}} \alpha }}$, and ${\mathring{ {{\bf{B}}_{\alpha \alpha }^{ - 1}}}}$ with feasible values.
\REPEAT 
\STATE Given ${{\bf{v}}_W}$ and ${\bf{\tilde D}}$, solve problem (${\rm P}1.3$ SDP) by CVX to obtain $\{ {{{\bf{\hat W}}}_k},{{{\bf{\hat R}}}_s},{\bf{\tilde C}}\}$.
\STATE Update ${\mathring{\bf{B}}_{\alpha {\boldsymbol{\phi}} }}$, ${\mathring{\bf{B}}_{{\boldsymbol{\phi}} \alpha }}$, and ${\mathring{ {{\bf{B}}_{\alpha \alpha }^{ - 1}}}}$ based on the solution obtained from step 3.
\UNTIL the reduction of the objective function (\ref{P13Obj}) is smaller than the threshold $\varepsilon$. \\
\STATE Obtain ${\bf{w}}_k^*$ and ${\bf{R}}_s^*$ by (\ref{Wopt}) and (\ref{Rsopt}).
\STATE \textbf{Output}: $\{ {\bf{W}},{\bf{R}}_s,{\bf{\tilde C}}\}$. \\
\end{algorithmic}
\end{algorithm}

\subsection{Sub-problem with Respect to $\{{{\bf{v}}_W},{\bf{\tilde D}}\}$}
Given $\{ {\bf{W}},{{\bf{R}}_s},{\bf{\tilde C}}\}$, the sub-problem with respect to $\{{{\bf{v}}_W},{\bf{\tilde D}}\}$, omitting unrelated terms, is written as follows.
\begin{subequations}\label{OptP14}
\begin{align}
({\rm P}1.4)\; & \mathop {\min }\limits_{\{ {{\bf{v}}_W},{\bf{\tilde D}}\} }{\rm{tr}}\left( {{\bf{\tilde D\tilde D}}} \right) \label{P14Obj}\\
\mbox{s.t.}& \;
(\ref{const_COM}), (\ref{const_RIS}),(\ref{const_AuxilC}), (\ref{const_AuxilD}).
\end{align}
\end{subequations}
In (\ref{Hk:faulty}), we have derived ${{{\bf{\bar h}}}_k} = {\bf{v}}_W^{\rm{H}}{{\bf{H}}_{W,k}} + {{{\bf{\bar h}}}_{F,k}}$, based on which the SINR constraint (\ref{const_COM}) is transformed into ${\bf{\tilde v}}_W^{\rm{H}}{{\bf{Q}}_k}{{{\bf{\tilde v}}}_W} \ge {\gamma _{th}}\sigma _k^2, \forall k \in {\mathcal K}$, where ${{{\bf{\tilde v}}}_W}$ and ${{\bf{Q}}_k}$ are given by
\begin{equation}\label{vWtilde}
{{{\bf{\tilde v}}}_W} = {\left[ {{\bf{v}}_W^{\rm{H}} \;1} \right]^{\rm{H}}},
\end{equation}
\begin{equation}\label{Qk}
\begin{array}{l}
{{\bf{Q}}_k} = \\
\!\left[\! {\begin{array}{*{20}{c}}\!
{{{\bf{H}}_{W,k}}}\!\\
{{{{\bf{\bar h}}}_{F,k}}}\!
\end{array}} \!\right]\left( {{{\bf{W}}_k} \!-\! {\gamma _{th}}\!\left(\! {\sum\limits_{i = 1,i \ne k}^K {{{\bf{W}}_i}}  \!+\! {{\bf{R}}_s}} \right)} \!\right)\!\left[ \!{{\bf{H}}_{W,k}^{\rm{H}}\;{\bf{\bar h}}_{F,k}^{\rm{H}}} \right].
\end{array}
\end{equation}
Defining ${{{\bf{\tilde V}}}_W} = {{{\bf{\tilde v}}}_W}{\bf{\tilde v}}_W^{\rm{H}} \in {\mathbb{C}^{(W+1) \times (W+1)}}$, which satisfies ${{{\bf{\tilde V}}}_W} \succeq 0$ and ${\rm{rank}}( {{{\bf{\tilde V}}}_W} ) = 1$, the SINR constraint with respect to ${{{\bf{\tilde V}}}_W}$ can be further converted into
\begin{equation}\label{SINRtransV}
{\rm{tr}}\left( {{{\bf{Q}}_k}{{{\bf{\tilde V}}}_W}} \right) \ge {\gamma _{th}}\sigma _k^2,\forall k \in {\cal K}.
\end{equation}

On the other hand, the sensing related constraints involving the working RIS variable ${{\bf{\Theta }}_W} = {\rm{diag}}\left( {{{\bf{v}}_W}} \right)$ are in (\ref{const_AuxilC}) and (\ref{const_AuxilD}), where the variable ${{\bf{\Theta }}_W}$ affects all block matrices in ${\bf{A}}_{{{\boldsymbol{\eta }}_0}}$, namely ${{\bf{A}}_{{\boldsymbol{\phi}} {\boldsymbol{\phi}} }}$, ${{\bf{A}}_{\alpha {\boldsymbol{\phi}} }}$, ${\bf{A}}_{\alpha \alpha }$, and ${{\bf{A}}_{{\boldsymbol{\phi}} \alpha }}$. We can observe that the working RIS variable ${{\bf{\Theta }}_W} = {\rm{diag}}\left( {{{\bf{v}}_W}} \right)$ appears twice but separately in the cascaded sensing channel ${{{\bf{\bar G}}}_W}$ shown in (\ref{Gw}), or ${{{\bf{\bar \Omega }}}_W}$ defined in (\ref{OmegaWbar}). To facilitate solving the RIS phase shift sub-problem, we transform ${{{\bf{\bar \Omega }}}_W}$ with respect to ${{\bf{\Theta }}_W}$ into an equivalent form with respect to ${{\bf{v}}_W}$, allowing the variable ${{\bf{v}}_W}$ to appear in adjacent positions rather than separately, as shown below.
\begin{equation}\label{OmegaWbarTrans}
\begin{array}{l}
{{{\bf{\bar \Omega }}}_W} = {{\bf{H}}_{BR,W}}{\bf{\Theta }}_W^{\rm{H}}{{\bf{H}}_{{\rm{TRM}},W}}{{\bf{\Theta }}_W}{\bf{H}}_{BR,W}^{\rm{H}}\\
 = {{\bf{H}}_{BR,W}}{\rm{diag}}\left( {{{\bf{a}}_W}} \right){{\bf{v}}_W}{\bf{v}}_W^{\rm{H}}{\rm{diag}}\left( {{\bf{a}}_W^{\rm{H}}} \right){\bf{H}}_{BR,W}^{\rm{H}}\\
 = {\bf{G}}_W^{\rm{H}}{{\bf{v}}_W}{\bf{v}}_W^{\rm{H}}{{\bf{G}}_W}.
\end{array}
\end{equation}
where ${{\bf{a}}_W}$ represents ${{\bf{a}}_W}\left( {{\phi_e},{\phi_a}} \right)$ (see definition in (\ref{H_TRMW})) and ${{\bf{G}}_W} = {\rm{diag}}\left( {{\bf{a}}_W^{\rm{H}}} \right){\bf{H}}_{BR,W}^{\rm{H}}$.
After this transformation, we can introduce ${{\bf{V}}_W} = {{\bf{v}}_W}{\bf{v}}_W^{\rm{H}}\in {\mathbb{C}^{W \times W}}$, satisfying ${{\bf{V}}_W} \succeq 0$ and ${\rm{rank}}\left( {{\bf{V}}_W} \right) = 1$, to replace ${{\bf{v}}_W}{\bf{v}}_W^{\rm{H}}$. In fact, ${{\bf{V}}_W}$ is the first $W \times W$ elements in ${{{\bf{\tilde V}}}_W}$, which is introduced earlier in the constraint (\ref{SINRtransV}). Therefore, ${{{\bf{\tilde V}}}_W}$ and ${{{\bf{V}}}_W}$ can be considered as one variable. When ${{{\bf{\tilde V}}}_W}$ is solved, we can simply take its first $W \times W$ elements to obtain ${{{\bf{V}}}_W}$. With these reformulations, the sub-problem with respect to $\{{{\bf{v}}_W},{\bf{\tilde D}}\}$ can be rewritten as problem $\rm{P}$1.5.
\begin{subequations}\label{OptP15}
\begin{align}
&({\rm P}1.5) \mathop {\min }\limits_{\{ {\bf{\tilde V}}_W, {\bf{\tilde D}}\} } {\rm{tr}}\left( {{\bf{\tilde D\tilde D}}} \right) \label{P15Obj}\\
\mbox{s.t.}&
\left[ {\begin{array}{*{20}{c}}
{{\bf{\tilde C}} - {{\bf{B}}_{{\boldsymbol{\phi}} {\boldsymbol{\phi}} }}}&{{{\bf{A}}_{\alpha {\boldsymbol{\phi}} }}({{\bf{V}}_W}) + {{\bf{B}}_{\alpha {\boldsymbol{\phi}} }}}\\
{{{\bf{A}}_{{\boldsymbol{\phi}} \alpha }}({{\bf{V}}_W}) + {{\bf{B}}_{{\boldsymbol{\phi}} \alpha }}}&{{{\bf{B}}_{\alpha \alpha }}}
\end{array}} \right] \succeq 0, \label{const_AuxilC4} \\
& - \left[ {\begin{array}{*{20}{c}}
{{\bf{\tilde D}}}&{{{\bf{I}}_2}}&{\bf{0}}\\
{{{\bf{I}}_2}}&{{{\bf{A}}_{{\boldsymbol{\phi}} {\boldsymbol{\phi}} }}({{\bf{V}}_W})}&{{{\bf{A}}_{\alpha {\boldsymbol{\phi}} }}({{\bf{V}}_W})}\\
{\bf{0}}&{{{\bf{A}}_{{\boldsymbol{\phi}} \alpha }}({{\bf{V}}_W})}&{{\bf{A}}_{\alpha \alpha }}
\end{array}} \right] \succeq 0 , \label{const_AuxilD4} \\
&{\rm{tr}}\left( {{{\bf{Q}}_k}{{{\bf{\tilde V}}}_W}} \right) \ge {\gamma _{th}}\sigma _k^2,\forall k \in {\cal K}, \label{const_SINRtransV} \\
&{{{\bf{\tilde V}}}_W} \succeq 0, \label{const_VPSD}\\
&{{{\bf{\tilde V}}}_{w,w}} = 1, \forall w \in \left\{ {1,...,W + 1} \right\}, \label{const_Vw1}\\
&{\rm{rank}}\left( {{{\bf{\tilde V}}}_W} \right) = 1. \label{const_Vrank1}
\end{align}
\end{subequations}
where ${{{\bf{\tilde V}}}_{w,w}}$ is the element in the $w$-row and $w$-column.

In problem $\rm{P}$1.5, the auxiliary real variable $\bf{\tilde D}$ is symmetric. Thus, we have
\begin{equation}\label{DDnorm}
{\rm{tr}}\left( {{\bf{\tilde D\tilde D}}} \right) = {\rm{tr}}\left( {{{{\bf{\tilde D}}}^{\rm{T}}}{\bf{\tilde D}}} \right){\rm{ = }}\left\| {{\bf{\tilde D}}} \right\|_F^2.
\end{equation}
Note that the squared Frobenius norm is a sum of the squares of all the entries in matrix ${\bf{\tilde D}}$. Thus, (\ref{DDnorm}) is convex with respect to ${\bf{\tilde D}}$. In addition, among all the constraints in problem $\rm{P}$1.5, the rank-one constraint (\ref{const_Vrank1}) is the only constraint that makes problem $\rm{P}$1.5 nonconvex. A common way to deal with the rank-one constraint is to apply the SDR technique and then recover the rank-one solution using the Gaussian randomization technique~\cite{bGaussAppro}. However, in our problem, multiple constraints involve ${{{\bf{v}}}_W}$. In such cases, it would be extremely difficult to find a feasible ${{{\bf{v}}}_W}$ that satisfies all the constraints using Gaussian randomization. Thus, the solution obtained by this technique is probably not a good solution to our problem. 

In order to obtain a performance-guaranteed solution, we tackle the rank-one constraint (\ref{const_Vrank1}) by exploiting the penalty-based method~\cite{Rank1}. The rank-one constraint (\ref{const_Vrank1}) is equivalent to the following constraint
\begin{equation}\label{const_Vrank1equ}
{\left\| {{{{\bf{\tilde V}}}_W}} \right\|_ * } - {\left\| {{{{\bf{\tilde V}}}_W}} \right\|_2} \le 0,
\end{equation}
where ${\left\| {{{{\bf{\tilde V}}}_W}} \right\|_ * } = \sum\nolimits_i^W {{\lambda _i}}$ is the nuclear norm of ${{\bf{\tilde V}}}_W$. $\lambda _i$ is the $i$-th eigenvalue of ${{\bf{\tilde V}}}_W$. ${\left\| {{{{\bf{\tilde V}}}_W}} \right\|_2} = \mathop {\max }\limits_i \left\{ {{\lambda _i}} \right\}$ is the spectral norm of ${{\bf{\tilde V}}}_W$. Note that (\ref{const_Vrank1equ}) holds if and only if ${{\bf{\tilde V}}}_W$ is rank-one. Therefore, (\ref{const_Vrank1}) and (\ref{const_Vrank1equ}) are equivalent.
Next, (\ref{const_Vrank1equ}) is penalized into the objective function, namely
\begin{subequations}\label{OptP16}
\begin{align}
&({\rm P}1.6) \mathop {\min }\limits_{\{ {\bf{\tilde V}}_W, {\bf{\tilde D}}\} } {\rm{tr}}\left( {{\bf{\tilde D\tilde D}}} \right) + \rho \left( {{{\left\| {{{{\bf{\tilde V}}}_W}} \right\|}_ * } \!-\! {{\left\| {{{{\bf{\tilde V}}}_W}} \right\|}_2}} \right) \label{P16Obj}\\
\mbox{s.t.}&
(\ref{const_AuxilC4}),(\ref{const_AuxilD4}), \ref{const_SINRtransV}),(\ref{const_VPSD}),(\ref{const_Vw1}).
\end{align}
\end{subequations}
where $\rho>0$ is the penalty parameter. Problems $\rm{P}$1.5 and $\rm{P}$1.6 are equivalent, with the proof shown in~\cite{Rank1}. We observe that the objective function (\ref{P16Obj}) is a form of difference of convex (DC). Thus, the SCA technique can provide Karush-Kuhn-Tucker (KKT) conditions for problem $\rm{P}$1.6~\cite{dcSCA}. At each iteration of SCA, the lower bound of ${\left\| {{{{\bf{\tilde V}}}_W}} \right\|_2}$ is obtained by its first-order Taylor approximation, namely 
\begin{equation}\label{VspecNormApp}
\begin{array}{l}
{\left\| {{{{\bf{\tilde V}}}_W}} \right\|_2} \ge {\left\| {{{{\bf{\tilde V'}}}_W}} \right\|_2} + \\
{\rm{tr}}\left( {{{\bf{u}}_{\max }}\left( {{{{\bf{\tilde V'}}}_W}} \right) \!\times\! {\bf{u}}_{\max }^{\rm{H}}\left( {{{{\bf{\tilde V'}}}_W}} \right) \!\times\! \left( {{{{\bf{\tilde V}}}_W} \!-\! {{{\bf{\tilde V'}}}_W}} \right)} \right),
\end{array}
\end{equation}
where ${{{\bf{u}}_{\max }}( {{{\bf{\tilde V'}}}_W} )}$ is the eigenvector corresponding to the largest eigenvalue of ${{{\bf{\tilde V'}}}_W}$. ${{{{\bf{\tilde V'}}}_W}}$ is the obtained ${\bf{\tilde V}}_W$ from the previous iteration. At each iteration of SCA, the problem to be solved is shown as follows
\begin{subequations}\label{OptP17}
\begin{align}
& \mathop {\mathop {\min }\limits_{\{ {{{\bf{\tilde V}}}_W},{\bf{\tilde D}}\} } }\limits^{({\rm{P}}1.7)} \begin{array}{l}
{\rm{tr}}\left( {{\bf{\tilde D\tilde D}}} \right) + \rho \left\{ {{{\left\| {{{{\bf{\tilde V}}}_W}} \right\|}_ * } - } \right.{\left\| {{{{\bf{\tilde V'}}}_W}} \right\|_2} - \\
\left. {{\rm{tr}}\!\left[ {{{\bf{u}}_{\max }}\left( {{{{\bf{\tilde V'}}}_W}} \right) \!\times\! {\bf{u}}_{\max }^{\rm{H}}\left( {{{{\bf{\tilde V'}}}_W}} \right) \!\times\! \left( {{{{\bf{\tilde V}}}_W} \!-\! {{{\bf{\tilde V'}}}_W}} \right)} \!\right]} \!\right\}
\end{array} \label{P17Obj}\\
& \mbox{s.t.}
(\ref{const_AuxilC4}),(\ref{const_AuxilD4}), \ref{const_SINRtransV}),(\ref{const_VPSD}),(\ref{const_Vw1}).
\end{align}
\end{subequations}
Problem $\rm{P}$1.7 is convex with respect to ${\bf{\tilde V}}_W$ and ${\bf{\tilde D}}$, which can be solved using tools such as CVX. After problem $\rm{P}$1.7 is solved at each SCA iteration, the penalty parameter $\rho$ is increased to enforce satisfying the rank-one constraint (\ref{const_Vrank1}). As the penalty parameter gradually increases, the rank-one constraint (\ref{const_Vrank1}) is ultimately satisfied, which means problem $\rm{P}$1.5 is solved. The procedure to solve problem $\rm{P}$1.4 is summarized as Algorithm~\ref{AlgoP14}.

\begin{algorithm}[htbp]
\caption{SCA-based algorithm for solving problem ${\rm P}1.4$}
\label{AlgoP14}
\begin{algorithmic}[1]
\STATE \textbf{Initialization}: Initialize ${{\bf{\tilde V'}}}_W$ with feasible values.
\REPEAT 
\STATE Solve problem $\rm{P}$1.7 by CVX to obtain $\{{{\bf{\tilde V}}_W},{\bf{\tilde D}}\}$.
\STATE Update ${{\bf{\tilde V'}}}_W$ according to the solution of step 3.
\STATE Increase the penalty parameter $\rho$.
\UNTIL the reduction of the objective function (\ref{P17Obj}) is smaller than the threshold $\varepsilon$. \\
\STATE \textbf{Output}: $\{ {{\bf{v}}_W},{\bf{\tilde D}}\}$. \\
\end{algorithmic}
\end{algorithm}

\subsection{Summarization and Convergence Analysis}
\label{Converanalysis}
The complete algorithm to solve problem $\rm{P}$1 is shown in Algorithm~\ref{AlgoP1}, where problem $\rm{P}$1 is decomposed into two sub-problems and solved iteratively in an alternating manner until convergence based on the BCD framework. On the one hand, during each MM iteration in the first sub-problem with respect to $\{ {\bf{W}},{{\bf{R}}_s},{\bf{\tilde C}}\}$ (problem $\rm{P}$1.2), the obtained objective value (\ref{P13Obj}) serves as an upper bound of (\ref{P12Obj}). Since the problem ($\rm{P}$1.3 SDP) is convex and can be solved optimally at each iteration, this upper bound is monotonically tightened as the MM iterations progress. Moreover, the recovery of $\{ {\bf{W}},{{\bf{R}}_s}\}$ does not affect the optimal values based on Proposition 1. Therefore, the objective value (\ref{P13Obj}) gradually converges to a stationary point of problem $\rm{P}$1.2. 

On the other hand, for the second sub-problem with respect to $\{{{\bf{v}}_W},{\bf{\tilde D}}\}$ (problem $\rm{P}$1.4), the optimal objective value (\ref{P17Obj}) obtained at each SCA iteration is an upper bound of the objective (\ref{P16Obj}) in problem $\rm{P}$1.6. With the penalty parameter increment and SCA iterations, this upper bound can be monotonically tightened, after which the solution of problem $\rm{P}$1.7 gradually reaches the KKT point of problem $\rm{P}$1.4.

Finally, we can observe that both problem $\rm{P}$1.2 and problem $\rm{P}$1.4 converge to a stationary point. After each problem is alternately iterated, the objective value of problem $\rm{P}$1.1 (\ref{P11Obj}) forms a monotonically non-increasing sequence and converges to a stationary point. As mentioned earlier, the objective of problem $\rm{P}$1.1 (\ref{P11Obj}) is a relaxed upper bound of the original objective ${\rm{tr}}\left( {{\rm{MCRB}}_{\boldsymbol{\phi}} } \right)$ (\ref{P1Obj}). Note that (\ref{P1Obj}) is strongly constrained by (\ref{const_AuxilC}) and (\ref{const_AuxilD}). Thus, both the original and relaxed objective functions can theoretically converge to a stationary point, which will also be numerically proved in Section~\ref{NumerialConver}. Furthermore, a local optimal solution is reached.

\begin{algorithm}[htbp]
\caption{BCD-based algorithm for solving problem ${\rm P}1$}
\label{AlgoP1}
\begin{algorithmic}[1]
\STATE \textbf{Initialization}: Initialize ${{\bf{v}}_W}$, ${\bf{\tilde D}}$, ${\mathring{\bf{B}}_{\alpha {\boldsymbol{\phi}} }}$, ${\mathring{\bf{B}}_{{\boldsymbol{\phi}} \alpha }}$, and ${\mathring{ {{\bf{B}}_{\alpha \alpha }^{ - 1}}}}$ with feasible values.
\REPEAT 
\STATE Solve problem ${\rm P}1.2$ to obtain $\{ {\bf{W}},{{\bf{R}}_s},{\bf{\tilde C}}\}$.
\STATE Solve problem ${\rm P}1.4$ to obtain $\{{{\bf{v}}_W},{\bf{\tilde D}}\}$.
\UNTIL the reduction of the objective function \eqref{P11Obj} is smaller than the threshold $\varepsilon $. \\
\STATE \textbf{Output}: $\{ {\bf{W}},{{\bf{R}}_s},{{\bf{v}}_W},{\bf{\tilde C}},{\bf{\tilde D}}\}$. \\
\end{algorithmic}
\end{algorithm}

\subsection{Complexity Analysis}
In this sub-section, we analyze the computational complexity of our proposed algorithm. In Algorithm~\ref{AlgoP1}, the main computational complexity depends on the update of $\{ {\bf{W}},{{\bf{R}}_s},{\bf{\tilde C}}\}$ and $\{{{\bf{v}}_W},{\bf{\tilde D}}\}$ in solving problems ${\rm P}1.2$ and ${\rm P}1.4$, respectively, which are both transformed into SDP problems. We assume that the interior point method is applied by the CVX solver to solve an SDP problem, which has the complexity of ${\mathcal{O}\left( N^3 \right)}$~\cite{bCInteriorpoint}, where $N$ denotes the variable dimension. Thus, the complexity of problems ${\rm P}1.2$ and ${\rm P}1.4$ depends on their variable dimensions.

In problem ${\rm P}1.2$, the variables are with the dimensions ${\bf{W}}_k \in {\mathbb{C}^{N_t \times N_t}}$ for UE $k$, ${{\bf{R}}_s} \in {\mathbb{C}^{N_t \times N_t}}$, ${\bf{\tilde C}} \in {\mathbb{R}^{2 \times 2}}$. Since ${\bf{W}}_k$, ${{\bf{R}}_s}$, and ${\bf{\tilde C}}$ are all symmetric matrices, the total variable dimensions can be reduced to $K \times \frac{{{N_t}\left( {{N_t} + 1} \right)}}{2} + \frac{{{N_t}\left( {{N_t} + 1} \right)}}{2} + \frac{{2\left( {2 + 1} \right)}}{2} = \frac{{\left( {K + 1} \right){N_t}\left( {{N_t} + 1} \right)}}{2} + 3$. When $K$ and $N_t$ grow, the leading dimension increasing the complexity is the term ${\mathcal{O}\left( KN_t^2 \right)}$. Applying the interior-point method, the complexity of solving the problem ${\rm P}1.2$ is ${\mathcal{O}\left( {{K^3}N_t^6{I_{{\rm{MM}}}}} \right)}$, where ${I_{{\rm{MM}}}}$ is the number of iterations required by the MM algorithm utilized in problem ${\rm P}1.2$. Likewise, the variables in problem ${\rm P}1.4$ are with the dimensions ${{\bf{\tilde V}}_W} \in {\mathbb{C}^{(W+1) \times (W+1)}}$ and ${\bf{\tilde D}}\in {\mathbb{R}^{2 \times 2}}$, which are symmetric as well. Thus, the total variable dimensions in problem ${\rm P}1.4$ are calculated as $\frac{{\left( {W + 2} \right)\left( {W + 1} \right)}}{2} + \frac{{2\left( {2 + 1} \right)}}{2}$, with the leading dimension ${\mathcal{O}\left( W^2 \right)}$. Therefore, the complexity of solving problem ${\rm P}1.4$ based on the interior-point method is ${\mathcal{O}\left( {{W^{6}}{I_{{\rm{SCA}}}}} \right)}$, where ${I_{{\rm{SCA}}}}$ is the number of iterations required by the SCA technique applied to problem ${\rm P}1.4$. Since the BCD algorithm updates problem ${\rm P}1.2$ and problem ${\rm P}1.4$ in an alternating iterative manner, the complexity of solving problem ${\rm P}1$ is $\mathcal{O}( {\left( {{K^3}N_t^6{I_{{\rm{MM}}}} + {W^{6}}{I_{{\rm{SCA}}}}} \right){I_{{\rm{BCD}}}}} )$, where ${I_{{\rm{BCD}}}}$ is the number of iterations of the BCD algorithm.

\subsection{Initialization}
The proposed algorithm is updated in an alternating iterative manner, where the choice of the initial points plays a crucial role in the convergence and quality of the solution.
Since we first update the block $\{ {\bf{W}},{{\bf{R}}_s},{\bf{\tilde C}}\}$, as shown in Algorithm~\ref{AlgoP1}, the initial values of the other block $\{{{\bf{v}}_W},{\bf{\tilde D}}\}$, ${\mathring{\bf{B}}_{\alpha {\boldsymbol{\phi}} }}$, ${\mathring{\bf{B}}_{{\boldsymbol{\phi}} \alpha }}$, and ${\mathring{ {{\bf{B}}_{\alpha \alpha }^{ - 1}}}}$ are required. Instead of relying on random initializations, we aim to search for their feasible initial values. Specifically, to quickly determine the initial values of $\bf{\tilde D}$, ${\mathring{\bf{B}}_{\alpha {\boldsymbol{\phi}} }}$, ${\mathring{\bf{B}}_{{\boldsymbol{\phi}} \alpha }}$, and ${\mathring{ {{\bf{B}}_{\alpha \alpha }^{ - 1}}}}$ without having to iterate, we solve problem ${\rm P}1.2$ but replacing its objective with the objective of problem ${\rm P}1.4$, namely ${\rm{tr}}({{\bf{\tilde D\tilde D}}})$. Such a convex problem can be solved by CVX, after which the initial values of $\bf{\tilde D}$, ${\mathring{\bf{B}}_{\alpha {\boldsymbol{\phi}} }}$, ${\mathring{\bf{B}}_{{\boldsymbol{\phi}} \alpha }}$, and ${\mathring{ {{\bf{B}}_{\alpha \alpha }^{ - 1}}}}$ are obtained. Meanwhile, assuming all RIS elements are fully functional, a problem analogous to problem ${\rm P}1$, with the CRB as the sensing metric, can be solved to determine the complete set of phase shifts. Subsequently, only the phase shifts corresponding to the actually functional elements are used as the initial values of ${\bf{v}}_W$.
\section{Performance Evaluation}
In this section, we evaluate the performance of our proposed algorithm through Monte Carlo simulations, focusing on five key aspects. First, the convergence of the proposed algorithm is validated under varying numbers of faulty elements. Then, the 2D beampattern generated by the RIS towards the target is visualized. Next, the impact of crucial parameters on system performance is investigated, including the number of faulty RIS elements, the RIS size, and the transmit power. Furthermore, the trade-off between sensing accuracy and communication quality is explored under varying numbers of UEs and SINR thresholds. Finally, the performance comparisons under different fault distributions are provided. The performance of the proposed algorithm is compared against three benchmarks:
\begin{itemize}
    \item \textbf{Perfect RIS - upper bound (UB)}: In this scheme, all RIS elements function perfectly without any faults. To solve this case, we formulate an optimization problem similar to ${\rm P}1$ by substituting the MCRB with the conventional CRB as the sensing metric. Problem ${\rm P}3$ in work~\cite{XianxinSongCRB} addressed such a problem. However, their approach was limited to estimating only one-dimensional azimuth AoD. Here, we extend the estimation framework to 2D AoD for both azimuth and elevation angles, which is solved in a similar alternating iterative manner. The resulting performance can be interpreted as an upper bound on the achievable performance for practical RIS models with faulty elements.
    
    \item \textbf{Faulty RIS - naive}: The realistic RIS model with faulty elements is accounted for in this scheme. However, the presence of these faulty elements is naively ignored during optimization. The RIS phase shifts are designed under the incorrect assumption of fully functional elements, with the objective of minimizing the CRB under communication SINR and transmit power constraints.
    
    \item \textbf{Faulty RIS - random}: Faulty RIS elements are also considered here. To assess the performance improvement achieved by optimizing the functional elements in our proposed approach, these functional elements are configured with random phase shifts. Meanwhile, the transmit beamforming and sensing covariance matrix remain consistent with the proposed design.
\end{itemize}

\begin{figure}[!t]
    \centering
    \vspace{-0.1cm}
    \includegraphics[width=0.8\linewidth]{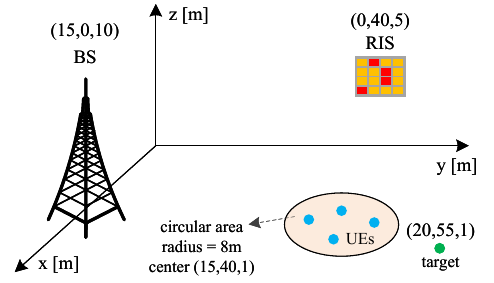}
    \caption{Simulation setup.}
    \vspace{-0.4cm}
    \label{SimuTopo}
\end{figure}

The simulation is conducted in a three-dimensional Cartesian coordinate system. As shown in Fig.~\ref{SimuTopo}, the BS is equipped with a ULA aligning with the x-axis and centered at [15, 0, 10] m. The RIS is configured with a UPA and is deployed in the z-y plane, centered at [0, 40, 5] m. UEs are uniformly distributed in a circular area with center at [15, 40, 1] m and radius of 8 m. The target is located at [20, 55, 1] m which corresponds to its 2D AoD with respect to the RIS ${\boldsymbol{\phi}} = {\left[ {{\phi _e},{\phi _a}} \right]^{\rm{T}}} = {\left[{-9.09^ \circ},{36.87^ \circ} \right]^{\rm{T}}}$. Additionally, the location of scatter is randomly generated such that its azimuth AoD ${\phi}_a'$ and elevation AoD ${\phi}_e'$ towards the RIS lie within $\left[{-90^ \circ},{90^ \circ} \right]$.
The small-scale fading characteristics of both the BS-RIS and RIS-UE channels are modeled according to Rician fading distributions, with respective K-factors of 10 and 1. The path loss is modeled as $PL(d) = {K_0}{\left( {{d \mathord{\left/ {\vphantom {d {{d_0}}}} \right. \kern-\nulldelimiterspace} {{d_0}}}} \right)^{ - {\mu}}}$, where ${K_0} = -30$ dB is the path loss at the reference distance $d_0 = 1$ m, and $d$ is the distance between transmitter and receiver. ${\mu} = (2, 2.2)$ is the path loss exponent for BS-RIS and RIS-UE channels, respectively~\cite{UPAVahid}. The target's channel was introduced in section~\ref{FaultyRISChannel} from (\ref{SVa}) to (\ref{Gf}). The penalty parameter $\rho$ increases fivefold at each iteration from 0.1 to 10000. Unless otherwise specified, all other simulation parameters are set according to Table~\ref{Tab1}.

\begin{table}[!t]
\caption{Simulation parameters}
\label{Tab1}
\setlength{\tabcolsep}{4pt} 
\centering
\begin{tabular}{ccc}
\hline
\textbf{Parameter} & \textbf{Value} & \textbf{Description} \\
\hline
$N_t$ & 32 & number of BS transmit antennas \\
$N_r$ & 32 & number of BS receive antennas \\
$R_z$ & 10 & number of RIS elements in z-axis \\
$R_y$ & 15 & number of RIS elements in y-axis \\
$F$ & 40 & number of faulty RIS elements \\
$K$ & 4 & number of UEs \\
$f_c$ & 28 GHz & carrier frequency \\
$P_{\max }$ & 33 dBm & transmit power \\
$\sigma _{\rm{RCS}} $ & 1 dBsm & radar cross section \\
$\sigma _s^2$ & -$110$ dBm & sensing noise power \\
$\sigma _c^2$ & -$110$ dBm & communication noise power\\
${\gamma _{th}}$ & 10 dB & SINR threshold \\
$\varepsilon $ & ${10^{ - 5}}$ &convergence tolerance \\
\hline
\end{tabular}
\vspace{-0.1cm}
\end{table}

\subsection{Convergence Performance}
\label{NumerialConver}
\begin{figure}[!t]
    \centering
    \includegraphics[width=0.8\linewidth]{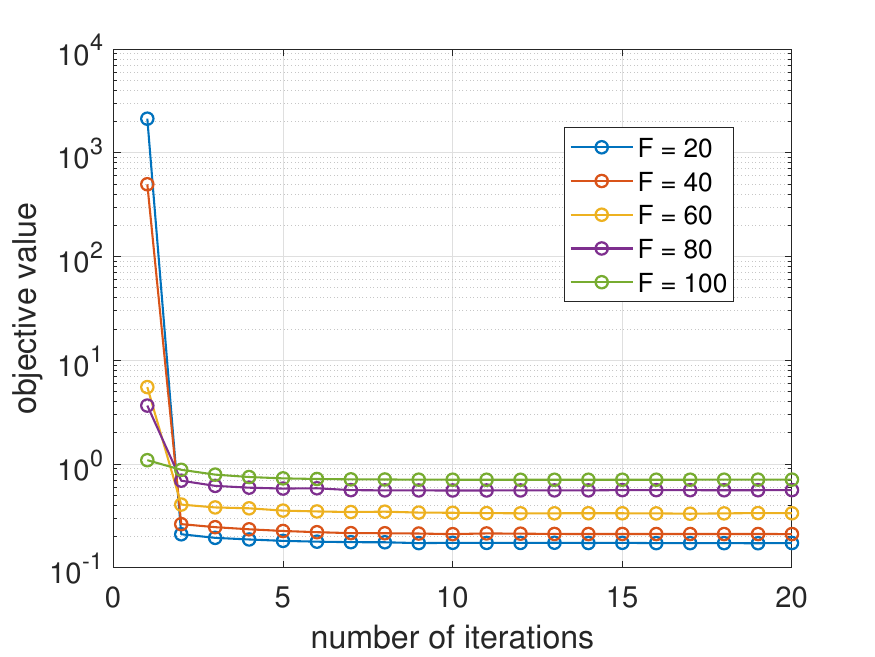}
    \caption{Convergence performance of the proposed algorithm.}
    \label{Re_Conver}
    \vspace{-0.6cm}
\end{figure}

In this section, we evaluate the convergence performance of the proposed algorithm. Fig.~\ref{Re_Conver} shows the objective values of problem ${\rm P}1.1$ versus the number of iterations under different numbers of faulty elements. We can observe that the objective values in different scenarios all stabilize within five iterations. Thus, the convergence of the proposed algorithm is numerically verified, supporting the theoretical convergence proof provided in Section~\ref{Converanalysis}.

\subsection{Beampattern Performance}

\begin{figure*}[!t]
\centering
\begin{subfigure}[b]{0.245\textwidth}
    \centering
    \includegraphics[width=0.95\linewidth]{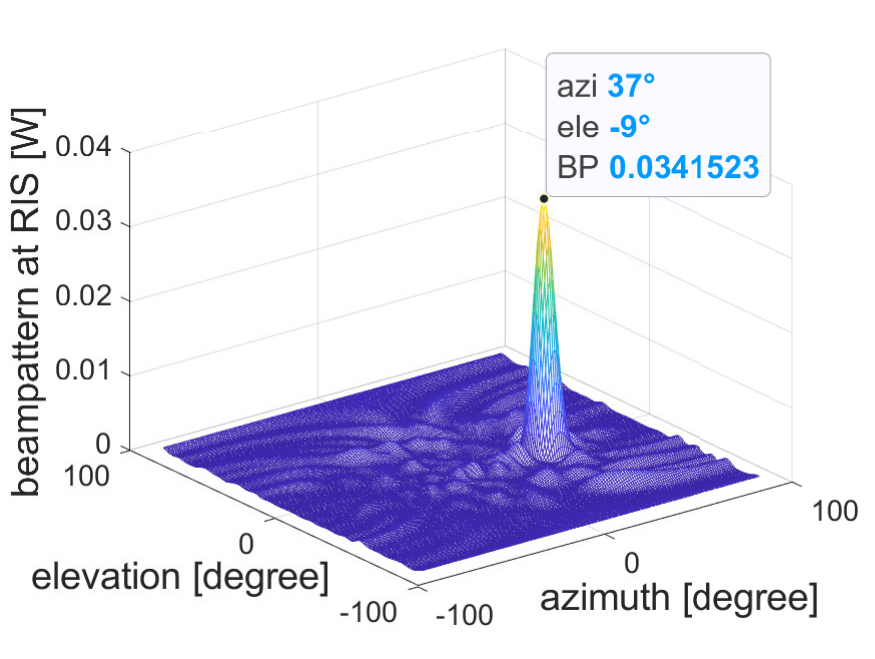}  
    \caption{Ideal case, $F$=0, $R$=150.}
    \label{Re_bpCRB}
\end{subfigure}
\begin{subfigure}[b]{0.245\textwidth}
    \centering
    \includegraphics[width=0.95\linewidth]{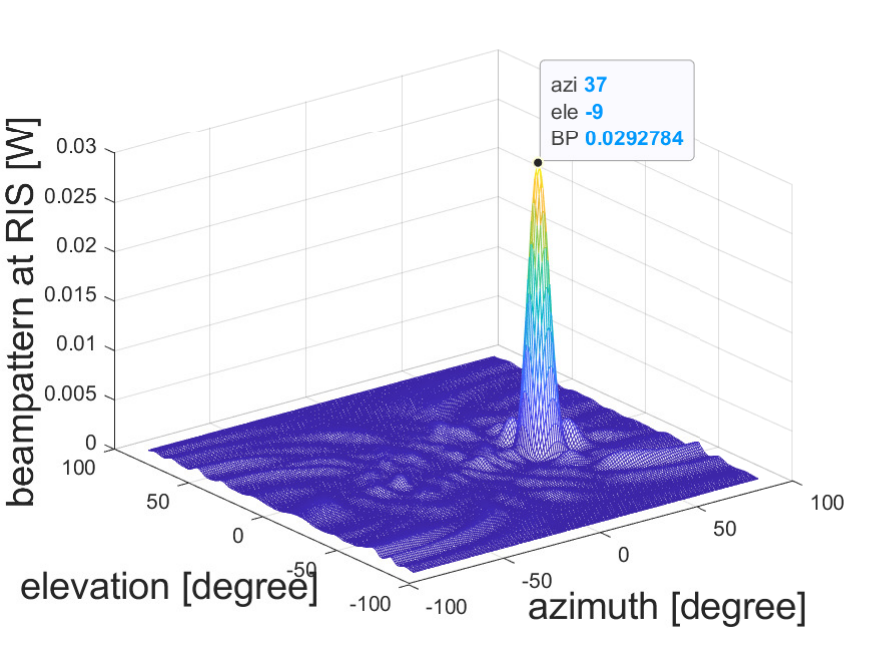}  
    \caption{$F$=20, $R$=150.}
    \label{Re_bpMCRB_F20}
\end{subfigure}
\begin{subfigure}[b]{0.245\textwidth}
    \centering
    \includegraphics[width=0.95\linewidth]{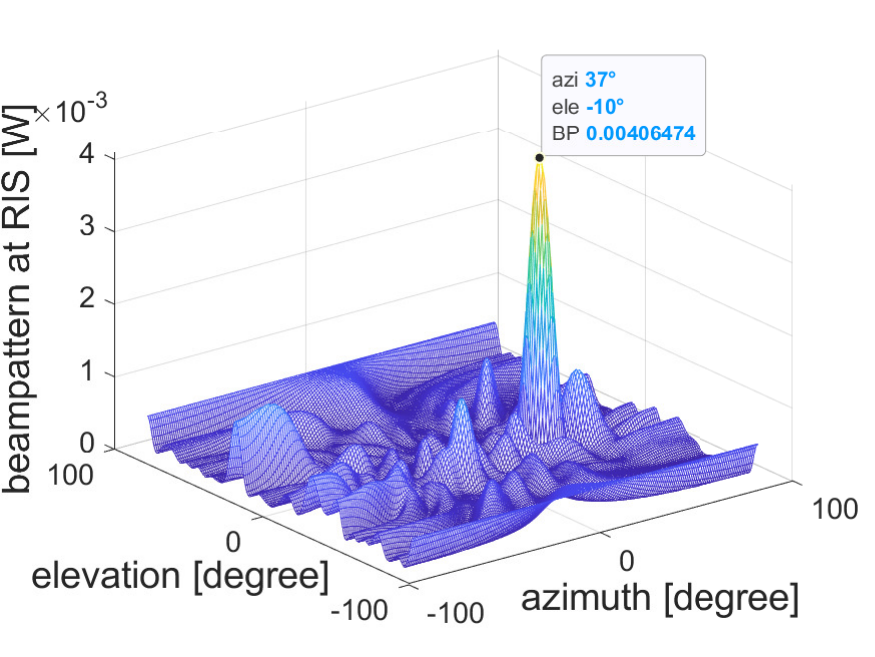}  
    \caption{$F$=40, $R$=150.}
    \label{Re_bpMCRB_F40}
\end{subfigure}
\begin{subfigure}[b]{0.245\textwidth}
    \centering
    \includegraphics[width=0.95\linewidth]{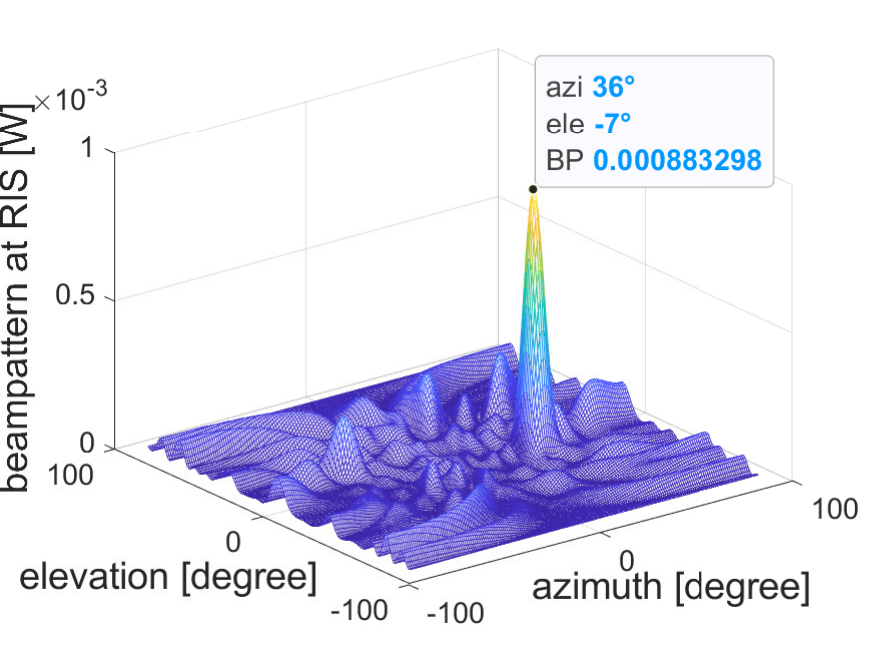}  
    \caption{$F$=60, $R$=150.}
    \label{Re_bpMCRB_F60}
\end{subfigure}
\begin{subfigure}[b]{0.245\textwidth}
    \centering
    \includegraphics[width=0.95\linewidth]{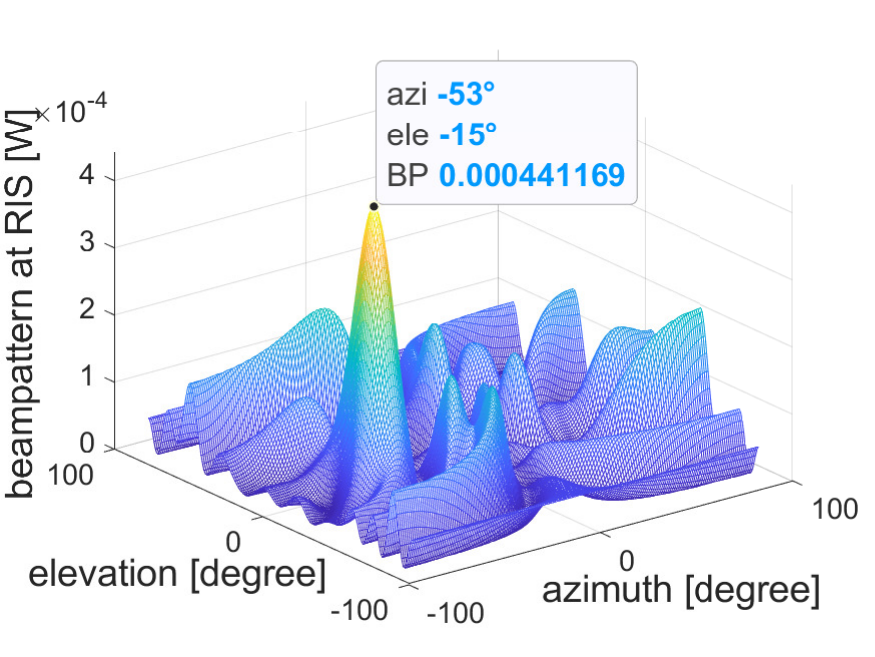}  
    \caption{$F$=40, $R$=50.}
    \label{Re_bpMCRB_R50}
\end{subfigure}
\begin{subfigure}[b]{0.245\textwidth}
    \centering
    \includegraphics[width=0.95\linewidth]{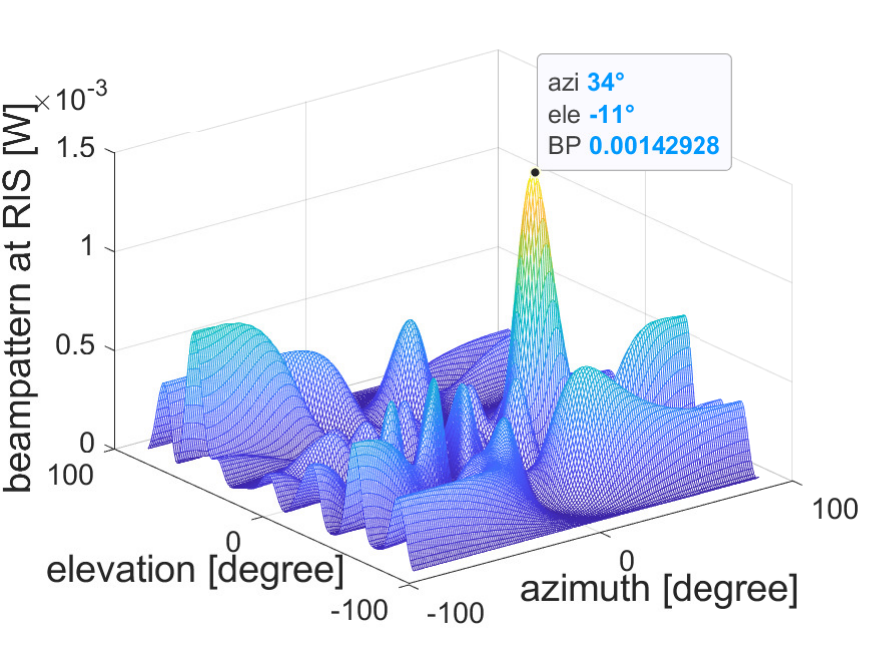}  
    \caption{$F$=40, $R$=80.}
    \label{Re_bpMCRB_R80}
\end{subfigure}
\begin{subfigure}[b]{0.245\textwidth}
    \centering
    \includegraphics[width=0.95\linewidth]{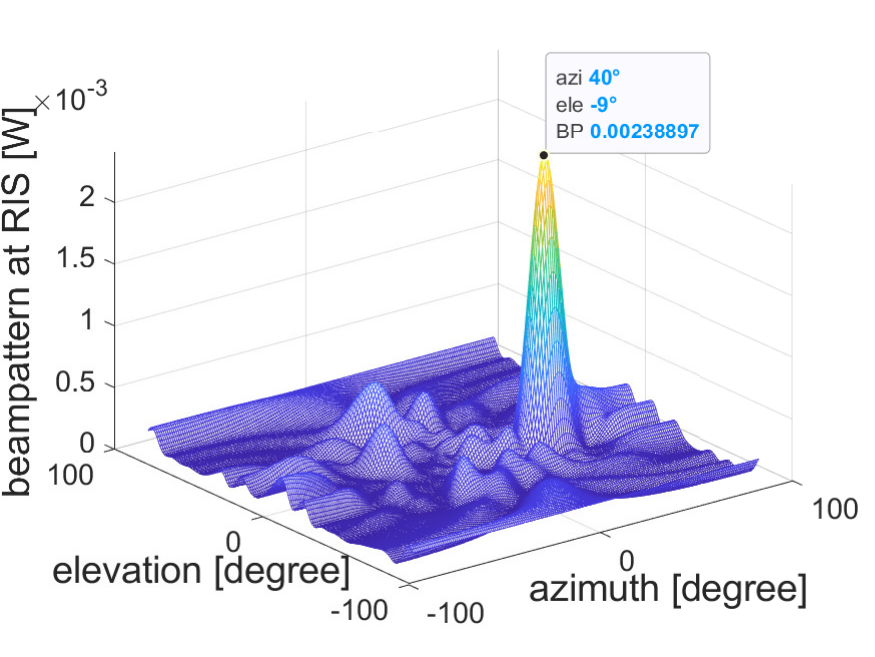}  
    \caption{$F$=40, $R$=100.}
    \label{Re_bpMCRB_R100}
\end{subfigure}
\begin{subfigure}[b]{0.245\textwidth}
    \centering
    \includegraphics[width=0.95\linewidth]{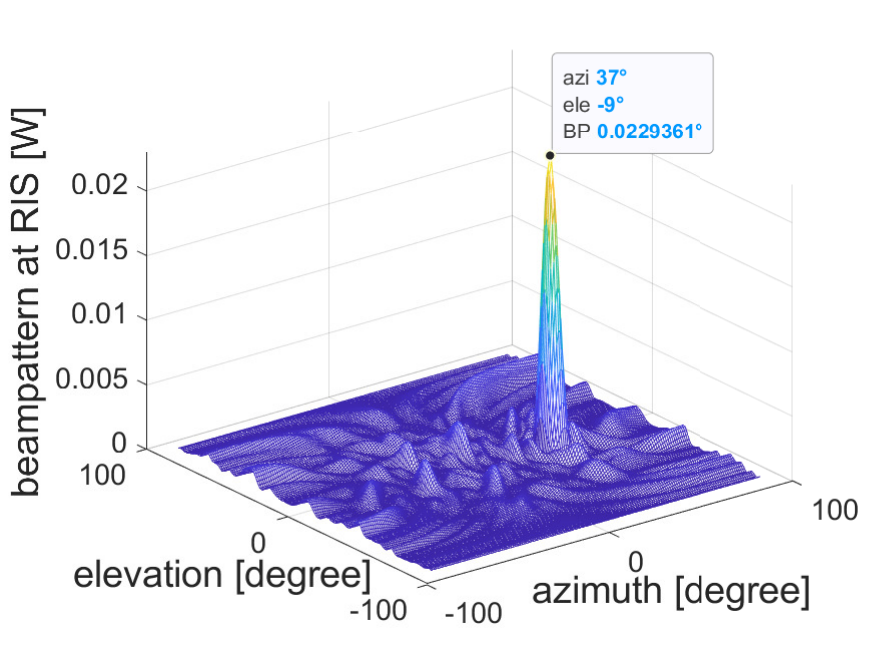}  
    \caption{$F$=40, $R$=200.}
    \label{Re_bpMCRB_R200}
\end{subfigure}
\caption{RIS Beampattern towards target under different numbers of faulty RIS elements and RIS sizes ($R_z=10$, $R_y=5-20$).}
\label{Re_RISbeam}
\vspace{-0.5cm}
\end{figure*}

In this section, we evaluate the RIS beampattern performance under varying numbers of faulty elements and different RIS sizes. The beampatterns in Fig.~\ref{Re_RISbeam}, generated using our proposed algorithm, demonstrate the system's directional response. As a reference, the target’s true 2D AoD relative to the RIS is ${\boldsymbol{\phi}} = {\left[ {{\phi _e},{\phi _a}} \right]^{\rm{T}}} = {\left[{-9.09^ \circ},{36.87^ \circ} \right]^{\rm{T}}}$. As depicted in Fig.~(\ref{Re_bpCRB}), the ideal beampattern aligns precisely with the 2D AoD (the angles are not percentile-accurate due to resolution limits) and achieves a gain of 0.034 W. When the number of faulty elements increases from 0 to 60 (out of 150 elements), the beams exhibit reduced accuracy, increased sidelobe levels, and decreased gain, as illustrated from Fig.~(\ref{Re_bpMCRB_F20}) to Fig.~(\ref{Re_bpMCRB_F60}). With 20 faulty elements, the beampattern still maintains acceptable performance, accurately pointing toward the target with a 14.33\% reduction in peak power gain in the direction of interest. In contrast, when the number of faulty elements reaches 40, the beam exhibits an azimuth error of approximately $1^ \circ$, a tenfold reduction in peak gain, and noticeable sidelobes. The situation deteriorates further with 60 faulty elements, resulting in a beam deviation of around $1^ \circ$ in elevation and $2^ \circ$ in azimuth, a fortyfold reduction in peak gain, and an increase in sidelobe levels.

Meanwhile, with 40 fixed faulty elements, we examine the impact of the RIS size from 50 to 200 elements on the beampattern performance, with results presented from Fig.~(\ref{Re_bpMCRB_R50}) to Fig.~(\ref{Re_bpMCRB_R200}) and also in Fig.~(\ref{Re_bpMCRB_F40}). We observe that a larger RIS size leads to a narrower beamwidth, higher gain, fewer sidelobes, and more accurate beam directions. Specifically, when 40 out of 50 RIS elements experience failures, the beam fails to form correctly and points in an entirely incorrect direction. As the RIS size increases to 80 or 100 elements, the beam aligns more closely with the target direction, although with an angular error of approximately $3^ \circ$ in the azimuth or elevation plane. With 200 RIS elements, the beam accurately points to the target, exhibiting a narrow main lobe, few sidelobes, and a high beampattern gain.

In summary, the presence of faulty elements degrades the beampattern performance, with more faults leading to reduced beam accuracy and lower gain. When such faults are present, increasing the RIS size serves as an effective means to mitigate performance degradation and maintain the desired beampattern quality.

\begin{figure*}[!t]
\begin{minipage}[t]{0.33\linewidth}
\centering
\includegraphics[width=1\textwidth]{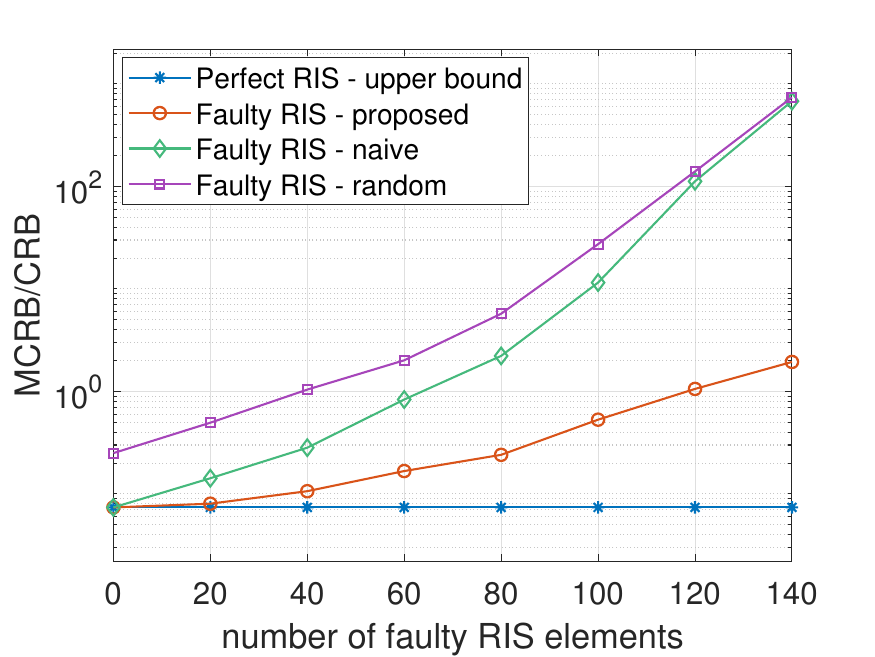}
\caption{MCRB/CRB under varying number of faulty RIS elements.}
\label{Re_RF_sen}
\end{minipage}
\begin{minipage}[t]{0.33\linewidth}
\centering
\includegraphics[width=1\textwidth]{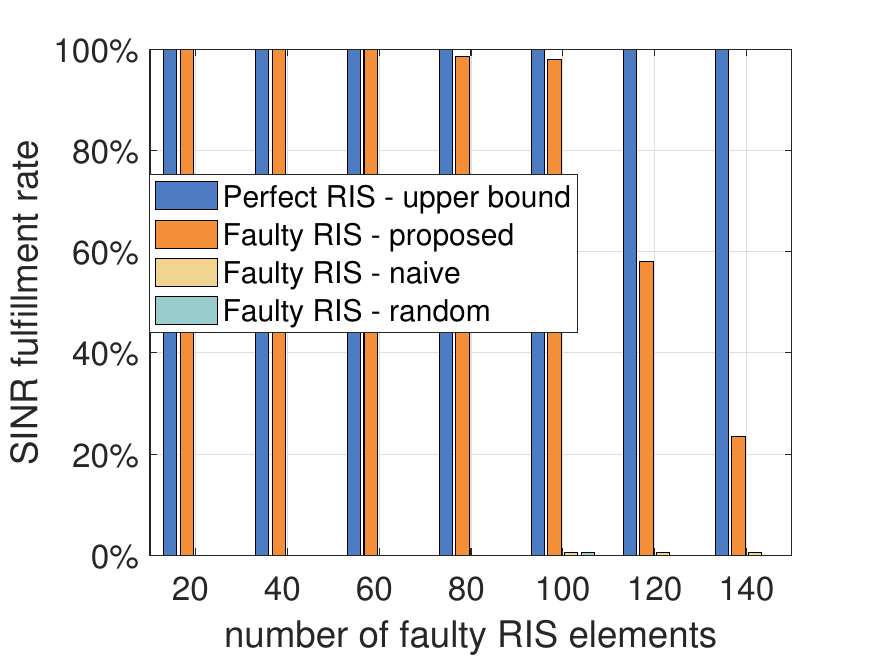}
\caption{SINR fulfillment rate under varying number of faulty RIS elements.}
\label{Re_RF_com}
\end{minipage}
\begin{minipage}[t]{0.33\linewidth}
\centering
\includegraphics[width=1\textwidth]{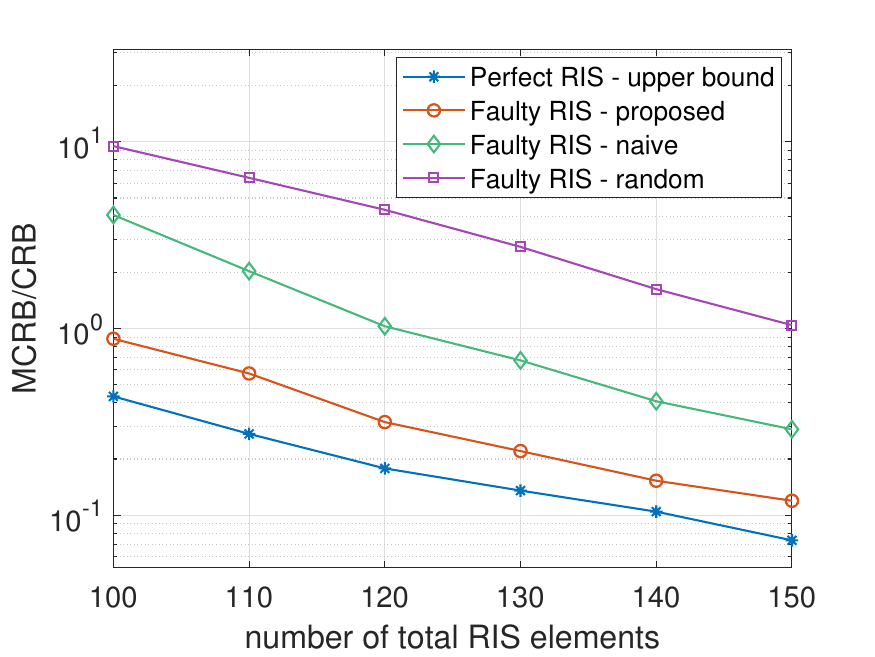}
\caption{\small{MCRB/CRB under varying number of RIS elements ($R_z=10$, $R_y=10/11/12/13/14/15$).}}
\label{Re_R}
\end{minipage}
\vspace{-0.4cm}
\end{figure*}

\subsection{MCRB/SINR Performance under Varying Parameters}
In this section, we evaluate the impact of key parameters on the system performance, including the number of faulty RIS elements, RIS size, and transmit power. The proposed scheme is compared with the three benchmarks introduced earlier.

\subsubsection{Impact of number of faulty RIS elements}

In Fig.~\ref{Re_RF_sen}, the impact of faulty RIS elements on the sensing performance is investigated. When all RIS elements function correctly, the CRB (obtained by the UB scheme) and MCRB (obtained by the proposed scheme) results are identical. However, as faulty behavior emerges, a noticeable performance gap arises between CRB and MCRB due to the mismatch in the RIS model. Specifically, as the number of faulty RIS elements increases from 20 to 140 (out of 150), the gap between the proposed and UB schemes progressively widens from 0.0065 to 1.8658. This observation indicates that more faulty elements result in more pronounced performance degradation. Similarly, as faults increase from 20 to 140, the gap between UB and naive schemes also widens from 0.0684 to 679.0832, which is significantly larger than the performance gap with respect to our approach, highlighting the importance of accounting for faulty elements. These results demonstrate the effectiveness of our method in reducing the negative impact of faulty elements, compared to naively ignoring the faults. Notably, the random scheme performs worst among all, emphasizing the necessity of optimizing the functional RIS elements for maintaining system performance. Additionally, we observe that as the number of faulty elements increases, the performance gap between the naive scheme and the random scheme gradually narrows and converges. This is because when all the elements are faulty, the naive scheme reduces to a configuration in which all element coefficients are randomly set. In other words, under fully faulty conditions, the naive and random schemes are equivalent.


\begin{figure*}[!t]
\begin{minipage}[t]{0.33\linewidth}
\centering
\includegraphics[width=1\textwidth]{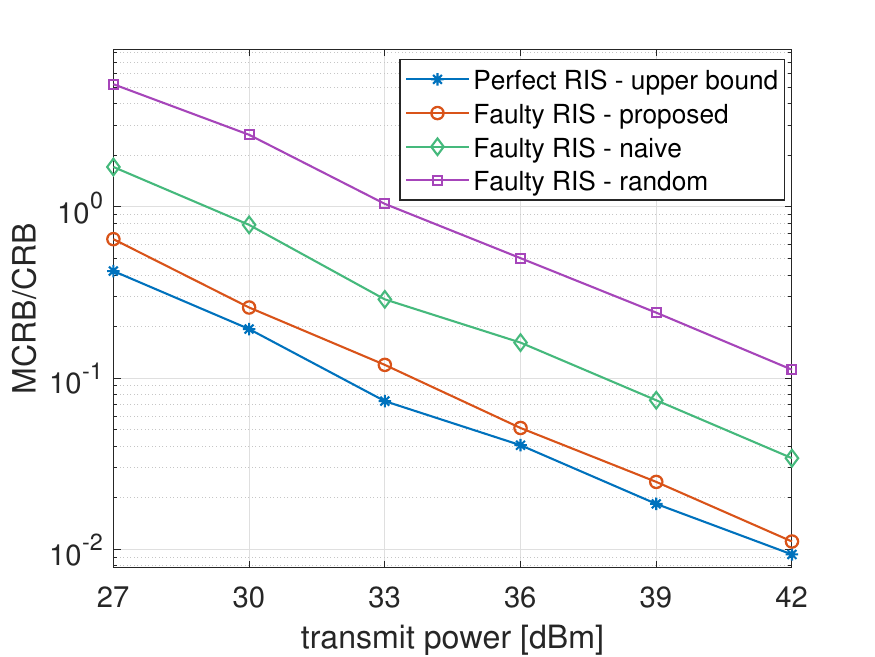}
\caption{\small{MCRB/CRB under varying transmit power.}}
\label{Re_Pwr}
\end{minipage}
\begin{minipage}[t]{0.33\linewidth}
\centering
\includegraphics[width=1\textwidth]{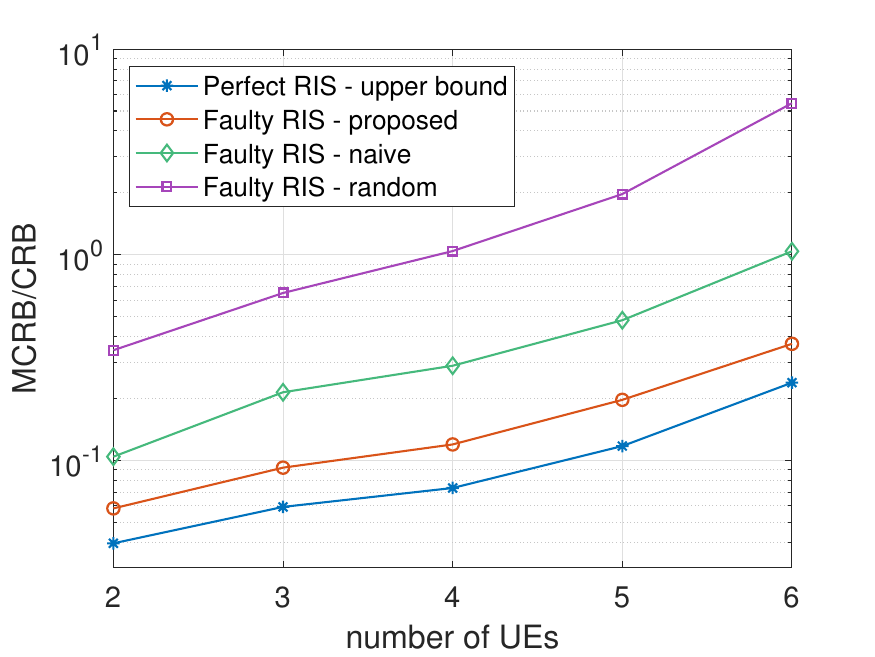}
\caption{\small{MCRB/CRB under varying number of UEs.}}
\label{Re_UE}
\end{minipage}
\begin{minipage}[t]{0.33\linewidth}
\centering
\includegraphics[width=1\textwidth]{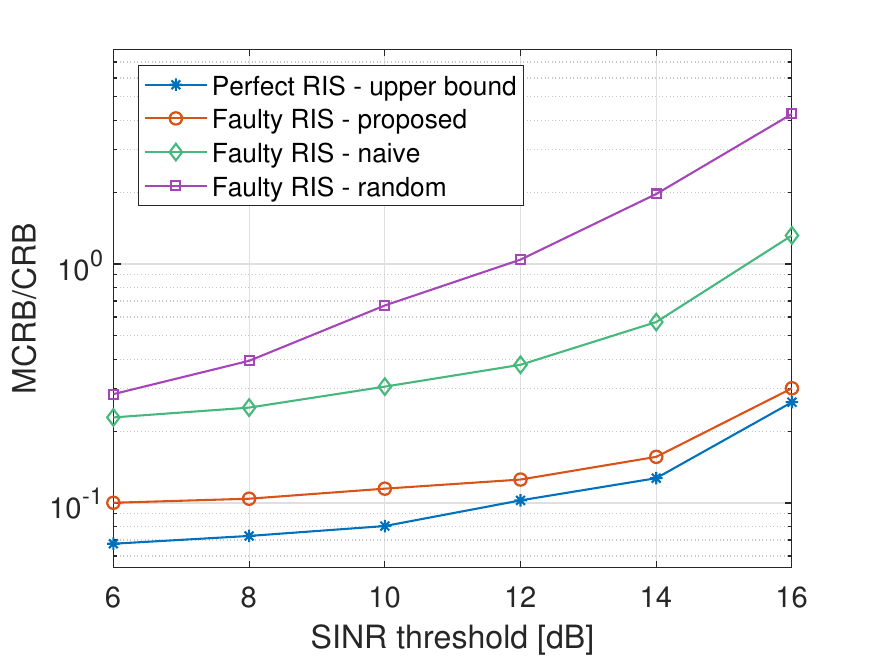}
\caption{\small{MCRB/CRB under varying SINR threshold.}}
\label{Re_Ga}
\end{minipage}
\vspace{-0.4cm}
\end{figure*}

Meanwhile, to validate the preservation of the minimum required communication reliability, we adopt the \textit{SINR constraint fulfillment rate} as a metric, computed as follows.
\begin{equation}\label{SINRrate}
{\gamma _{{\rm{rate}}}} = \tfrac{{\sum\limits_i^{{I_{MC}}} {\sum\limits_k^K {{\rm{Num}}({\gamma _{ki}} \ge {\gamma _{th}})} } }}{{K \times {I_{MC}}}} \times 100\%,
\end{equation}
where ${I_{MC}}$ is the number of Monte Carlo iterations. ${\gamma_{ki}}$ is the SINR of UE $k$ at the $i$-th Monte Carlo iteration, and $\sum\nolimits_i^{{I_{MC}}} {\sum\nolimits_k^K {{\rm{Num}}({\gamma _{ki}} \ge {\gamma _{th}})}}$ counts the total number of UEs whose SINRs satisfy the constraints over all iterations. The SINR constraint fulfillment rates ${\gamma _{{\rm{rate}}}}$ under different numbers of faulty elements are shown in Fig.~\ref{Re_RF_com}. As expected, the UB scheme always meets the SINR requirements, which is reasonable due to the absence of faulty elements. The proposed scheme achieves 100\% fulfillment mostly when the number of faulty elements is at most 100 (out of 150). When the number of faulty elements increases to 120 or 140, the fulfillment rate decreases to 58\% and 23.5\%, respectively. This degradation is reasonable since, with 80\% or 93\% of RIS elements being faulty, the remaining 20\% or 7\% of working elements are insufficient to effectively optimize system performance. In contrast, the naive and random schemes fail to satisfy the SINR requirements. This is because the RIS configurations in both naive and random schemes do not consider the SINR requirements.

Simulations under other system parameters (transmit power, RIS size, number of UEs, and SINR thresholds) are conducted with 40 faulty elements out of 150 RIS elements (26.67\% elements are faulty). These results exhibit similar SINR fulfillment behavior to the 40 faulty elements case in Fig.~\ref{Re_RF_com}, where both the proposed and UB schemes achieve a 100\% SINR constraint fulfillment rate. Due to space limitations, the SINR fulfillment results under other system parameters are omitted, and only the sensing performance measured by MCRB or CRB is presented in the subsequent analysis.

\subsubsection{Impact of total number of RIS elements}

To enhance the beamforming precision and overall system performance, we analyze the effect of increasing the RIS size, i.e., the number of RIS elements, with the corresponding results illustrated in Fig.~\ref{Re_R}. We observe that expanding the RIS size from 100 to 150 elements results in a steady decline in both MCRB (proposed scheme) and CRB (UB scheme) values, highlighting the performance benefits of a larger RIS in ISAC systems. Under a fixed 40 faulty elements, our proposed scheme exhibits an average MCRB increase of 42.63\% compared to the CRB from the fault-free UB baseline, which aligns with expectations due to RIS model mismatch. Meanwhile, compared to the naive and random schemes, our approach achieves significantly better performance, with average improvements of 38.60\% and 52.14\%, respectively. The way of calculating performance improvement is shown as \eqref{PerImprove1} and \eqref{PerImprove2}, which also applies to other parameter settings. This confirms the effectiveness of explicitly modeling RIS faults and optimizing the use of the functional elements.


\begin{equation}\label{PerImprove1}
\frac{({{\rm{CRB}}_{\rm{naive}}}-{{\rm{CRB}}_{{\rm{UB}}}})}{{{\rm{CRB}}_{{\rm{naive}}}}} - \frac{({{\rm{MCRB}}_{\rm{proposed}}}-{{\rm{CRB}}_{{\rm{UB}}}})}{\rm{MCRB}_{\rm{proposed}}},
\end{equation}

\begin{equation}\label{PerImprove2}
\frac{({{{\rm{MCRB}}_{\rm{random}}}-{{\rm{CRB}}_{\rm{UB}}})}}{{\rm{MCRB}}_{\rm{random}}} - \frac{({{{\rm{MCRB}}_{\rm{proposed}}}-{{\rm{CRB}}_{\rm{UB}}})}}{{\rm{MCRB}}_{\rm{proposed}}}.
\end{equation}

\subsubsection{Impact of transmit power}
Fig.~\ref{Re_Pwr} presents the MCRB and CRB performance as a function of transmit power. As expected, increasing transmit power leads to a reduction in both MCRB and CRB. The performance trends of the evaluated approaches align with the observations in Fig.~\ref{Re_R}. Our proposed method exhibits an average performance gap of 26.82\% compared to the UB scheme, which is unavoidable due to the presence of faulty elements. Additionally, our proposed solution consistently outperforms both the naive and random schemes (with average 47.83\% and 65.43\% performance improvement, respectively), validating the effectiveness of our fault-aware optimization strategy.


\subsection{Trade-off Performance}
In this section, we analyze the trade-off between sensing and communication in terms of the impact of the number of UEs and the SINR threshold. As before, the MCRB/CRB results of the proposed and the UB schemes are provided on the premise that all UEs satisfy the communication SINR requirements.

\subsubsection{Impact of number of UEs}
Fig.~\ref{Re_UE} illustrates the relationship between the number of UEs and the sensing performance, measured by MCRB or CRB. When the number of UEs increases from 2 to 6, both the MCRB from the proposed scheme and CRB from the UB scheme worsen. This degradation is attributed to more transmit power being allocated to meet the SINR requirements of additional UEs, leaving less power available for optimizing sensing performance. This indicates the inherent trade-off between sensing accuracy and communication quality, which should be considered in practical system design. Among the evaluated schemes, our proposed method exhibits the smallest average performance gap to the UB scheme (36.44\%), outperforming the naive (72.31\%) and random (92.40\%) approaches.


\begin{figure*}[htbp]
\centering
\begin{subfigure}[t]{0.245\textwidth}
    \centering
    \includegraphics[width=0.95\linewidth]{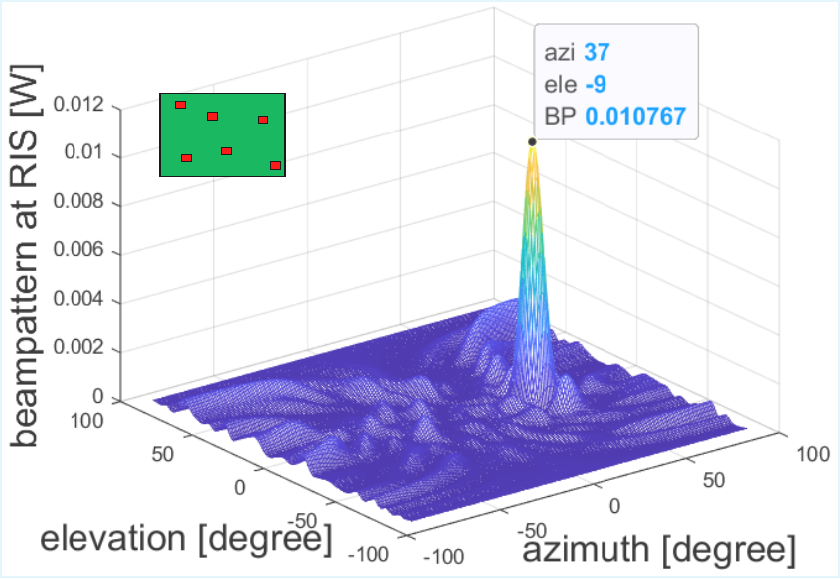}  
    \caption{Random fault distribution.}
    \label{Re_DisRnd}
\end{subfigure}
\begin{subfigure}[t]{0.245\textwidth}
    \centering
    \includegraphics[width=0.95\linewidth]{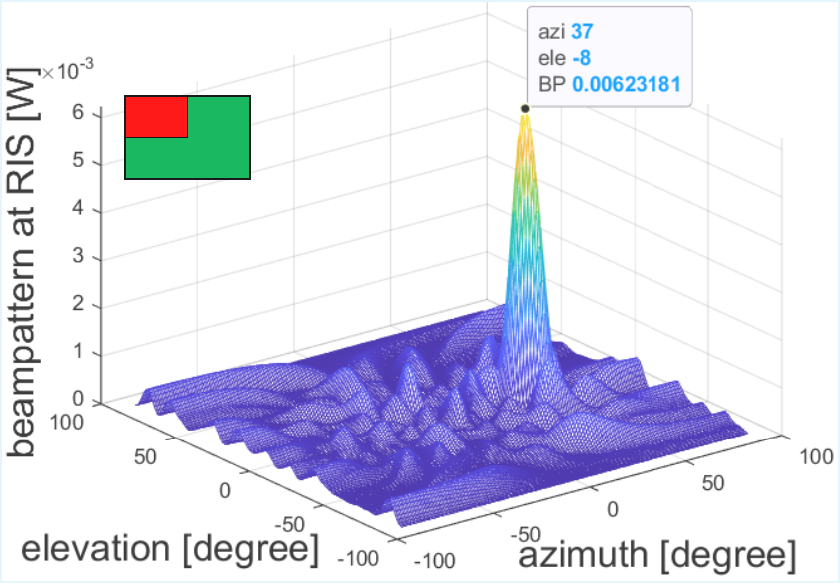}  
    \caption{Clustered upper-left block fault distribution.}
    \label{Re_DisLeftUp}
\end{subfigure}
\begin{subfigure}[t]{0.245\textwidth}
    \centering
    \includegraphics[width=0.95\linewidth]{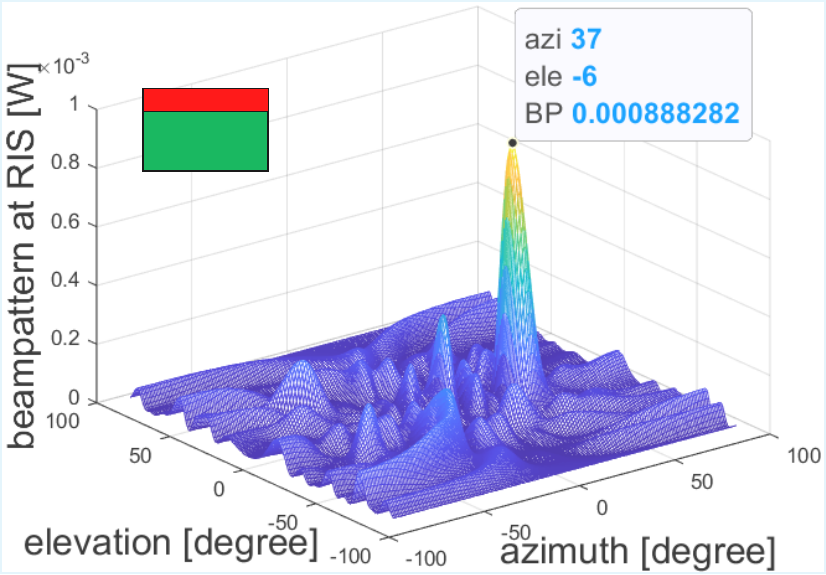}  
    \caption{Top horizontal fault distribution.}
    \label{Re_DisUp}
\end{subfigure}
\begin{subfigure}[t]{0.245\textwidth}
    \centering
    \includegraphics[width=0.95\linewidth]{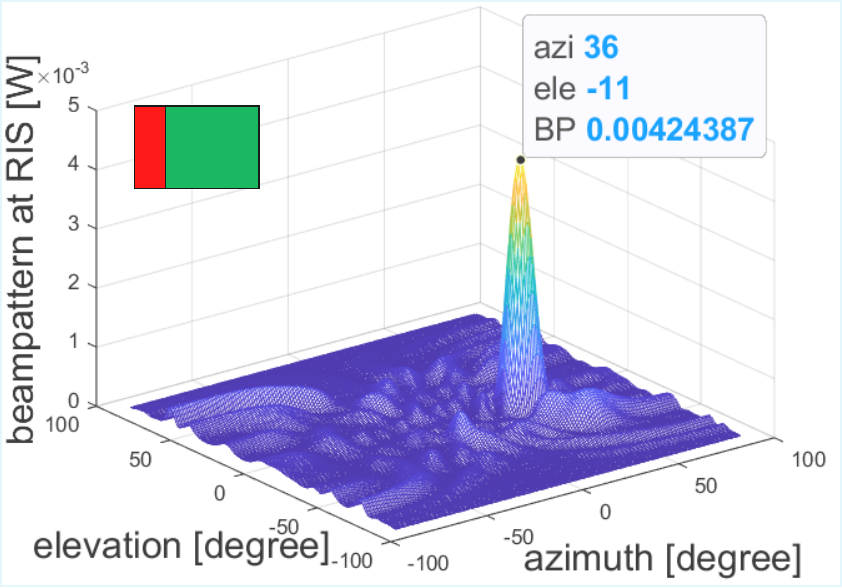}  
    \caption{Left vertical fault distribution.}
    \label{Re_DisLeft}
\end{subfigure}
\caption{Performance under different fault distributions.}
\label{Re_DisFault}
\vspace{-0.4cm}
\end{figure*}

\subsubsection{Impact of SINR thresholds}
In Fig.~\ref{Re_Ga}, the impact of SINR threshold on the communication and sensing performances is shown. When the SINR threshold increases from 6 dB to 16 dB, a noticeable degradation in MCRB (proposed scheme) and CRB (UB scheme) is observed. This is consistent with the trend observed in Fig.~\ref{Re_UE}, where higher SINR requirements reduce the available transmit power for sensing, leading to worse sensing performance. The trade-off between sensing and communication is evident under varying SINR thresholds. This indicates that by appropriately configuring system parameters, such as the number of UEs and SINR thresholds, a balance between sensing accuracy and communication quality can be achieved. In practice, the allocation of resources to sensing and communication users can be tailored based on specific application demands.

\subsection{Performance under Different Fault Distributions}

In the previously conducted simulations, the faulty elements are assumed to be randomly distributed across the RIS, which is a widely adopted assumption to model unpredictable hardware impairments. Nevertheless, other fault distributions also exist in practice. For instance, as illustrated in Fig.~\ref{Re_DisFault}, alternative distributions may include clustered faults or faults concentrated along one boundary. These distributions reflect potential hardware degradation scenarios, such as localized manufacturing defects or edge wear-out of the surface. We extend our simulations to incorporate these fault distributions to provide deeper insights into practical deployment. Given the RIS size $R = R_z \times R_y =10 \times 15$ (namely 10 and 15 elements in the z-axis and y-axis, respectively), we assume 40 out of 150 elements are faulty.

Under the same channel realization, the beampatterns under different fault distributions are illustrated in Fig.~\ref{Re_DisFault}. It can be observed from Fig.~(\ref{Re_DisRnd}) that the most accurate beam direction can be obtained under the random fault distribution among all distributions, with no strong azimuth or elevation bias. This is due to the relatively larger effective aperture compared to the other three fault distributions. In contrast, clustered or contiguous faults cause more severe performance degradation, particularly in beamforming accuracy. Specifically, the clustered upper-left block fault model in Fig.~(\ref{Re_DisLeftUp}) removes a contiguous portion of both the elevation and azimuth apertures. As a result, both dimensions experience degraded beamforming performance, with a broader mainlobe and a pointing bias of $1^ \circ$ mainly in the elevation direction. In Fig.~(\ref{Re_DisUp}), the top horizontal fault reduces the elevation (z-axis) aperture, which impairs elevation beamforming accuracy with a bias of $3^ \circ$ observed in the elevation direction. Finally, Fig.~(\ref{Re_DisLeft}) corresponds to the left vertical fault distribution, which degrades the beamforming accuracy mostly in the azimuth direction with $1^ \circ$ shift. However, a $2^ \circ$ shift in the elevation direction also arises because the number of functional RIS elements is reduced.
\section{Conclusion}

This paper investigated a practical challenge in RIS-aided ISAC systems arising from the presence of faulty RIS elements. By deriving MCRB for sensing and SINR for communication, we quantified the performance degradation due to RIS model mismatch when the BS is unaware of these faults. To counteract this degradation, we proposed an optimization framework that jointly designs the transmit beamforming and the remaining functional RIS elements, solved via the BCD-based algorithm. Simulations under diverse scenarios reveal the following key takeaways:
\begin{itemize}
    \item Faulty RIS elements significantly impair system performance, with severity increasing as faults accumulate. Nonetheless, through strategic design of the functional elements, the performance loss can be minimized compared to approaches that neglect the presence of faults.      
    \item System performance benefits from increased resources, such as a larger RIS array or higher transmit power. Additionally, a desirable sensing-communication trade-off can be achieved by tuning system parameters according to specific application requirements.
    
\end{itemize}
As a next step, we plan to explore a more realistic scenario where only the positions of faulty RIS elements are known, while their amplitudes and phases remain uncertain. A robust RIS configuration design under such conditions is left for future work to maintain performance under uncertainty.


\appendix
\section{Appendix}

\subsection{Calculation of ${{{\bf{A}}_{{{\boldsymbol{\eta }}_0}}}}$ and ${{{\bf{B}}_{{{\boldsymbol{\eta }}_0}}}}$}
\label{AppCalcuAB}
The calculation of MCRB is dependent on the derivation of ${{{\bf{A}}_{{\boldsymbol{\eta }}}}}$ and ${{{\bf{B}}_{{\boldsymbol{\eta }}}}}$ given in (\ref{MCRBA}) and (\ref{MCRBB}), in which the first-order and second-order partial derivatives of ${\boldsymbol{\tilde \mu }}\left( {\boldsymbol{\eta }} \right)$ with respect to ${\boldsymbol{\eta }}$ are needed. To achieve this goal, ${\boldsymbol{\tilde \mu }}( {\boldsymbol{\eta }}) = {\rm{vec}}( {{\bf{\tilde GX}}} )$ is further written as follows based on (\ref{GsensingAssumed}).
\begin{equation}\label{Noisefree_S}
{\boldsymbol{\tilde \mu }}\left( {\boldsymbol{\eta }} \right) \!=\! \alpha{\rm{vec}}\left( {{{\bf{H}}_{BR}}{{\bf{\Theta }}^{\rm{H}}}{\bf{a}}\left( {{\phi _e},{\phi _a}} \right){{\bf{a}}^{\rm{H}}}\left( {{\phi _e},{\phi _a}} \right){\bf{\Theta H}}_{BR}^{\rm{H}}{\bf{X}}} \right).
\end{equation}
Denoting ${\bf{\Omega }} = {\bf{a}}\left( {{\phi _e},{\phi _a}} \right){{\bf{a}}^{\rm{H}}}\left( {{\phi _e},{\phi _a}} \right)$ and ${\boldsymbol{\eta }} = {[ { {{\boldsymbol{\phi}} ^{\rm{T}}},{{{\boldsymbol{\tilde \alpha }}}^{\rm{T}}}}]^{\rm{T}}}$ where ${\boldsymbol{\phi}} = {\left[ {{\phi _e},{\phi _a}} \right]^{\rm{T}}}$ and ${\boldsymbol{\tilde \alpha }} = {\left[ {{\rm{Re}}\left( \alpha  \right),{\rm{Im}}\left( \alpha  \right)} \right]^{\rm{T}}}$, the first-order partial derivative is obtained as follows
\begin{equation}\label{1st_deri_angle}
\frac{{\partial {\boldsymbol{\tilde \mu }}\left( {\boldsymbol{\eta }} \right)}}{{\partial {\boldsymbol{\phi}} }} = \left[ {\begin{array}{*{20}{c}}
{\alpha {\rm{vec}}\left( {{{\bf{H}}_{BR}}{{\boldsymbol{\Theta }}^{\rm{H}}}{{{\bf{\dot \Omega}}}_{{\phi _e}}}{\boldsymbol{\Theta H}}_{BR}^{\rm{H}}{\bf{X}}} \right)}\\
{\alpha {\rm{vec}}\left( {{{\bf{H}}_{BR}}{{\boldsymbol{\Theta }}^{\rm{H}}}{{{\bf{\dot \Omega}}}_{{\phi _a}}}{\boldsymbol{\Theta H}}_{BR}^{\rm{H}}{\bf{X}}} \right)}
\end{array}} \right],
\end{equation}
\begin{equation}\label{1st_deri_alpha}
\frac{{\partial {\boldsymbol{\tilde \mu }}\left( {\boldsymbol{\eta }} \right)}}{{\partial {\boldsymbol{\tilde \alpha }}}} = {\rm{vec}}\left( {{{\bf{H}}_{BR}}{{\boldsymbol{\Theta }}^{\rm{H}}}{\bf{\Omega \Theta H}}_{BR}^{\rm{H}}{\bf{X}}} \right)\left[ {\begin{array}{*{20}{c}}
1\\
j
\end{array}} \right],
\end{equation}
where ${{{\bf{\dot \Omega}}}_{{\phi _e}}}$ and ${{{\bf{\dot \Omega}}}_{{\phi _a}}}$ are given as (\ref{1st_deri_elevation}) and (\ref{1st_deri_azimuth}) at the top of next page.
\begin{figure*}[!t]
\centering
\begin{equation}\label{1st_deri_elevation}
{{{\bf{\dot \Omega}}}_{{\phi _e}}} \!=\! \frac{{\partial {\bf{\Omega}}}}{{\partial {\phi _e}}} \!=\! j\kappa \left[ {{d_y}\sin \left( {{\phi _e}} \right)\sin \left( {{\phi _a}} \right)\left( {{\rm{diag}}\left( {{{\bf{r}}_Y}} \right){\bf{a}}{{\bf{a}}^{\rm{H}}} \!-\! {\bf{a}}{{\bf{a}}^{\rm{H}}}{\rm{diag}}\left( {{{\bf{r}}_Y}} \right)} \right)\left. { \!-\! {d_z}\cos \left( {{\phi _e}} \right)\left( {{\rm{diag}}\left( {{{\bf{r}}_Z}} \right){\bf{a}}{{\bf{a}}^{\rm{H}}} \!-\! {\bf{a}}{{\bf{a}}^{\rm{H}}}{\rm{diag}}\left( {{{\bf{r}}_Z}} \right)} \right)} \right]} \right. ,
\end{equation}
\vspace{-0.8cm}
\end{figure*}
\begin{figure*}[!t]
\centering
\begin{equation}\label{1st_deri_azimuth}
{{{\bf{\dot \Omega}}}_{{\phi _a}}} = \frac{{\partial {\bf{\Omega}}}}{{\partial {\phi _a}}} =  - j\kappa {d_y}\cos \left( {{\phi _e}} \right)\cos \left( {{\phi _a}} \right)\left( {{\rm{diag}}\left( {{{\bf{r}}_Y}} \right){\bf{a}}{{\bf{a}}^{\rm{H}}} - {\bf{a}}{{\bf{a}}^{\rm{H}}}{\rm{diag}}\left( {{{\bf{r}}_Y}} \right)} \right),
\end{equation}
\vspace{-0.7cm}
\end{figure*}
$\bf{a}$ stands for ${\bf{a}}\left( {{\phi _e},{\phi _a}} \right)$ for the sake of notational convenience. ${{\bf{r}}_Y} \in \mathbb{R}^{R\times1}$ and ${{\bf{r}}_Z} \in \mathbb{R}^{R\times1}$ collect the index of each RIS element on the y-axis and z-axis, respectively. Next, we take the second-order partial derivative of ${\boldsymbol{\tilde \mu }}\left( {\boldsymbol{\eta }} \right)$ with respect to ${\boldsymbol{\eta }}$, which can be further partitioned as follows
\begin{equation}\label{2nd_deri_Partition}
\frac{{{\partial ^2}{\boldsymbol{\tilde \mu }}\left( {\boldsymbol{\eta }} \right)}}{{\partial {\boldsymbol{\eta }}\partial {{\boldsymbol{\eta }}^{\rm{H}}}}} = 
\left[ {\begin{array}{*{20}{c}}
{\frac{{{\partial ^2}{\boldsymbol{\tilde \mu }}\left( {\boldsymbol{\eta }} \right)}}{{\partial {\boldsymbol{\phi}} \partial {{\boldsymbol{\phi}} ^{\rm{H}}}}}}&{\frac{{{\partial ^2}{\boldsymbol{\tilde \mu }}\left( {\boldsymbol{\eta }} \right)}}{{\partial {\boldsymbol{\tilde \alpha }}\partial {{\boldsymbol{\phi}} ^{\rm{H}}}}}}\\
{\frac{{{\partial ^2}{\boldsymbol{\tilde \mu }}\left( {\boldsymbol{\eta }} \right)}}{{\partial {\boldsymbol{\phi}} \partial {{{\boldsymbol{\tilde \alpha }}}^{\rm{H}}}}}}&{\frac{{{\partial ^2}{\boldsymbol{\tilde \mu }}\left( {\boldsymbol{\eta }} \right)}}{{\partial {\boldsymbol{\tilde \alpha }}\partial {{{\boldsymbol{\tilde \alpha }}}^{\rm{H}}}}}}
\end{array}} \right],
\end{equation}
with three elements in (\ref{2nd_deri_Partition}) given as follows
\begin{equation}\label{2nd_deri_angalp}
{\frac{{{\partial ^2}{\boldsymbol{\tilde \mu }}\left( {\boldsymbol{\eta }} \right)}}{{\partial {\boldsymbol{\phi}} \partial {{{\boldsymbol{\tilde \alpha }}}^{\rm{H}}}}}} \!=\! {\frac{{{\partial ^2}{\boldsymbol{\tilde \mu }}\left( {\boldsymbol{\eta }} \right)}}{{\partial {\boldsymbol{\tilde \alpha }}\partial {{\boldsymbol{\phi}} ^{\rm{H}}}}}} \!=\! \left[ \!{\begin{array}{*{20}{c}}
\!1\!\\
\!j\!
\end{array}} \!\right]\!{\left[\! {\begin{array}{*{20}{c}}\!
{{\rm{vec}}\left( {{{\bf{H}}_{BR}}{{\bf{\Theta }}^{\rm{H}}}{{{\bf{\dot \Omega}}}_{{\phi _e}}}{\bf{\Theta H}}_{BR}^{\rm{H}}{\bf{X}}} \right)}\!\\\!
{{\rm{vec}}\left( {{{\bf{H}}_{BR}}{{\bf{\Theta }}^{\rm{H}}}{{{\bf{\dot \Omega}}}_{{\phi _a}}}{\bf{\Theta H}}_{BR}^{\rm{H}}{\bf{X}}} \right)}\!
\end{array}} \!\right]^{\rm{T}}},
\end{equation}

\begin{equation}\label{2nd_deri_alpalp}
{\frac{{{\partial ^2}{\boldsymbol{\tilde \mu }}\left( {\boldsymbol{\eta }} \right)}}{{\partial {\boldsymbol{\tilde \alpha }}\partial {{{\boldsymbol{\tilde \alpha }}}^{\rm{H}}}}}} = {{\bf{0}}_2},
\end{equation}
${{\partial ^2}{\boldsymbol{\tilde \mu }}\left( {\boldsymbol{\eta }} \right)}/{{\partial {\boldsymbol{\phi}} \partial {{\boldsymbol{\phi}} ^{\rm{H}}}}}$ is derived as (\ref{2nd_deri_angang}), together with ${{{\bf{\ddot \Omega}}}_{{\phi _e}{\phi _e}}}$, ${{{\bf{\ddot \Omega}}}_{{\phi _e}{\phi _a}}}$, ${{{\bf{\ddot \Omega}}}_{{\phi _a}{\phi _e}}}$, and ${{{\bf{\ddot \Omega}}}_{{\phi _a}{\phi _a}}}$, all shown at the top of next page.
\begin{figure*}[!t]
\centering
\begin{equation}\label{2nd_deri_angang}
{\frac{{{\partial ^2}{\boldsymbol{\tilde \mu }}\left( {\boldsymbol{\eta }} \right)}}{{\partial {\boldsymbol{\phi}} \partial {{\boldsymbol{\phi}} ^{\rm{H}}}}}} = \left[ {\begin{array}{*{20}{c}}
{\frac{{{\partial ^2}{\boldsymbol{\tilde \mu }}\left( {\boldsymbol{\eta }} \right)}}{{\partial \phi _e^2}}}&{\frac{{{\partial ^2}{\boldsymbol{\tilde \mu }}\left( {\boldsymbol{\eta }} \right)}}{{\partial {\phi _a}\partial {\phi _e}}}}\\
{\frac{{{\partial ^2}{\boldsymbol{\tilde \mu }}\left( {\boldsymbol{\eta }} \right)}}{{\partial {\phi _e}\partial {\phi _a}}}}&{\frac{{{\partial ^2}{\boldsymbol{\tilde \mu }}\left( {\boldsymbol{\eta }} \right)}}{{\partial \phi _a^2}}}
\end{array}} \right] = \left[ {\begin{array}{*{20}{c}}
{\alpha {\rm{vec}}\left( {{{\bf{H}}_{BR}}{{\bf{\Theta }}^{\rm{H}}}{{{\bf{\ddot \Omega}}}_{{\phi _e}{\phi _e}}}{\bf{\Theta H}}_{BR}^{\rm{H}}{\bf{X}}} \right)}&{\alpha {\rm{vec}}\left( {{{\bf{H}}_{BR}}{{\bf{\Theta }}^{\rm{H}}}{{{\bf{\ddot \Omega}}}_{{\phi _a}{\phi _e}}}{\bf{\Theta H}}_{BR}^{\rm{H}}{\bf{X}}} \right)}\\
{\alpha {\rm{vec}}\left( {{{\bf{H}}_{BR}}{{\bf{\Theta }}^{\rm{H}}}{{{\bf{\ddot \Omega}}}_{{\phi _e}{\phi _a}}}{\bf{\Theta H}}_{BR}^{\rm{H}}{\bf{X}}} \right)}&{\alpha {\rm{vec}}\left( {{{\bf{H}}_{BR}}{{\bf{\Theta }}^{\rm{H}}}{{{\bf{\ddot \Omega}}}_{{\phi _a}{\phi _a}}}{\bf{\Theta H}}_{BR}^{\rm{H}}{\bf{X}}} \right)}
\end{array}} \right],
\end{equation}
\vspace{-0.5cm}
\end{figure*}
\begin{figure*}[htbp]
\centering
\begin{equation}\label{2nd_deri_eleele}
\begin{array}{l}
{{{\bf{\ddot \Omega}}}_{{\phi _e}{\phi _e}}} = j\kappa \left[ {{d_y}\cos \left( {{\phi _e}} \right)\sin \left( {{\phi _a}} \right)\left( {{\rm{diag}}\left( {{{\bf{r}}_Y}} \right){\bf{a}}{{\bf{a}}^{\rm{H}}} - {\bf{a}}{{\bf{a}}^{\rm{H}}}{\rm{diag}}\left( {{{\bf{r}}_Y}} \right)} \right) + {d_z}\sin \left( {{\phi _e}} \right)\left( {{\rm{diag}}\left( {{{\bf{r}}_Z}} \right){\bf{a}}{{\bf{a}}^{\rm{H}}} - {\bf{a}}{{\bf{a}}^{\rm{H}}}{\rm{diag}}\left( {{{\bf{r}}_Z}} \right)} \right)} \right.\\
 + {d_y}\sin \left( {{\phi _e}} \right)\sin \left( {{\phi _a}} \right)\left( {{\rm{diag}}\left( {{{\bf{r}}_Y}} \right){{{\bf{\dot \Omega}}}_{{\phi _e}}} - {{{\bf{\dot \Omega}}}_{{\phi _e}}}{\rm{diag}}\left( {{{\bf{r}}_Y}} \right)} \right)\left. { - {d_z}\cos \left( {{\phi _e}} \right)\left( {{\rm{diag}}\left( {{{\bf{r}}_Z}} \right){{{\bf{\dot \Omega}}}_{{\phi _e}}} - {{{\bf{\dot \Omega}}}_{{\phi _e}}}{\rm{diag}}\left( {{{\bf{r}}_Z}} \right)} \right)} \right],
\end{array}
\end{equation}
\vspace{-0.6cm}
\end{figure*}
\begin{figure*}[!t]
\centering
\begin{equation}\label{2nd_deri_eleazi}
\begin{array}{l}
{{{\bf{\ddot \Omega}}}_{{\phi _e}{\phi _a}}} = {{{\bf{\ddot \Omega}}}_{{\phi _a}{\phi _e}}} = j\kappa \left[ {{d_y}\sin \left( {{\phi _e}} \right)\cos \left( {{\phi _a}} \right)\left( {{\rm{diag}}\left( {{{\bf{r}}_Y}} \right){\bf{a}}{{\bf{a}}^{\rm{H}}} - {\bf{a}}{{\bf{a}}^{\rm{H}}}{\rm{diag}}\left( {{{\bf{r}}_Y}} \right)} \right)} \right.\\
 + {d_y}\sin \left( {{\phi _e}} \right)\sin \left( {{\phi _a}} \right)\left( {{\rm{diag}}\left( {{{\bf{r}}_Y}} \right){{{\bf{\dot \Omega}}}_{{\phi _a}}} - {{{\bf{\dot \Omega}}}_{{\phi _a}}}{\rm{diag}}\left( {{{\bf{r}}_Y}} \right)} \right)\left. { - {d_z}\cos \left( {{\phi _e}} \right)\left( {{\rm{diag}}\left( {{{\bf{r}}_Z}} \right){{{\bf{\dot \Omega}}}_{{\phi _a}}} - {{{\bf{\dot \Omega}}}_{{\phi _a}}}{\rm{diag}}\left( {{{\bf{r}}_Z}} \right)} \right)} \right],
\end{array}
\end{equation}
\vspace{-0.6cm}
\end{figure*}
\begin{figure*}[!t]
\centering
\begin{equation}\label{2nd_deri_aziazi}
{{{\bf{\ddot \Omega}}}_{{\phi _a}{\phi _a}}} \!=\! j\kappa \left[ {{d_y}\cos \left( {{\phi _e}} \right)\sin \left( {{\phi _a}} \right)\left( {{\rm{diag}}\left( {{{\bf{r}}_Y}} \right){\bf{a}}{{\bf{a}}^{\rm{H}}} \!-\! {\bf{a}}{{\bf{a}}^{\rm{H}}}{\rm{diag}}\left( {{{\bf{r}}_Y}} \right)} \right)} \right.\left. { \!-\! {d_y}\cos \left( {{\phi _e}} \right)\cos \left( {{\phi _a}} \right)\left( {{\rm{diag}}\left( {{{\bf{r}}_Y}} \right){{{\bf{\dot \Omega}}}_{{\phi _a}}} \!-\! {{{\bf{\dot \Omega}}}_{{\phi _a}}}{\rm{diag}}\left( {{{\bf{r}}_Y}} \right)} \right)} \right].
\end{equation}
\vspace{-1.1cm}
\end{figure*}
Then, we partition ${{\bf{A}}_{{\boldsymbol{\eta }}}}$ and ${{{\bf{B}}_{{\boldsymbol{\eta }}}}}$ into 4 blocks, namely
\begin{equation}\label{ABblock}
{{\bf{A}}_{{\boldsymbol{\eta }}}} = \left[ {\begin{array}{*{20}{c}}
{{{\bf{A}}_{{\boldsymbol{\phi}} {\boldsymbol{\phi}} }}}&{{{\bf{A}}_{\alpha {\boldsymbol{\phi}} }}}\\
{{{\bf{A}}_{{\boldsymbol{\phi}} \alpha }}}&{{{\bf{A}}_{\alpha \alpha }}}
\end{array}} \right],
{{\bf{B}}_{{\boldsymbol{\eta }}}} = \left[ {\begin{array}{*{20}{c}}
{{{\bf{B}}_{{\boldsymbol{\phi}} {\boldsymbol{\phi}} }}}&{{{\bf{B}}_{\alpha {\boldsymbol{\phi}} }}}\\
{{{\bf{B}}_{{\boldsymbol{\phi}} \alpha }}}&{{{\bf{B}}_{\alpha \alpha }}}
\end{array}} \right].
\end{equation}
According to (\ref{MCRBA}) and (\ref{MCRBB}), each sub-block matrix in ${{\bf{B}}_{{\boldsymbol{\eta }}}}$ is obtained as (\ref{Bphiphi})-(\ref{Balphaphi}). Each sub-block matrix in ${{\bf{A}}_{{\boldsymbol{\eta }}}}$ is obtained as (\ref{Aphiphi})-(\ref{Aalphaphi}), all shown below and on the next page.

\begin{equation}\label{Bphiphi}
{{\bf{B}}_{{\boldsymbol{\phi}} {\boldsymbol{\phi}} }} \!=\! \frac{{2{{\left| \alpha  \right|}^2}}}{{\sigma _s^2}}{\rm{Re}}\!\left[ \!{\begin{array}{*{20}{c}} 
{{\rm{tr}}\left( {{{{\bf{\bar \Omega }}}_{{\phi _e}}}{{\bf{R}}_x}{\bf{\bar \Omega }}_{{\phi _e}}^{\rm{H}}} \right)}\!&\!{{\rm{tr}}\left( {{{{\bf{\bar \Omega }}}_{{\phi _e}}}{{\bf{R}}_x}{\bf{\bar \Omega }}_{{\phi _a}}^{\rm{H}}} \right)}\!\\
\!{{\rm{tr}}\left( {{{{\bf{\bar \Omega }}}_{{\phi _a}}}{{\bf{R}}_x}{\bf{\bar \Omega }}_{{\phi _e}}^{\rm{H}}} \right)}\!&\!{{\rm{tr}}\left( {{{{\bf{\bar \Omega }}}_{{\phi _a}}}{{\bf{R}}_x}{\bf{\bar \Omega }}_{{\phi _a}}^{\rm{H}}} \right)}\!
\end{array}} \!\right],
\end{equation}
\begin{equation}\label{Bphialpha}
{{\bf{B}}_{{\boldsymbol{\phi}} \alpha }} \!=\! \frac{2}{{\sigma _s^2}}{\rm{Re}}\left[ \!{\begin{array}{*{20}{c}}
{{\alpha ^{\rm{H}}}{\rm{tr}}\left( {{\bf{\bar \Omega }}{{\bf{R}}_x}{\bf{\bar \Omega }}_{{\phi _e}}^{\rm{H}}} \right)}\!&\!{{\alpha ^{\rm{H}}}{\rm{tr}}\left( {{\bf{\bar \Omega }}{{\bf{R}}_x}{\bf{\bar \Omega }}_{{\phi _a}}^{\rm{H}}} \right)}\\
\!{j{\alpha ^{\rm{H}}}{\rm{tr}}\left( {{\bf{\bar \Omega }}{{\bf{R}}_x}{\bf{\bar \Omega }}_{{\phi _e}}^{\rm{H}}} \right)}\!&\!{j{\alpha ^{\rm{H}}}{\rm{tr}}\left( {{\bf{\bar \Omega }}{{\bf{R}}_x}{\bf{\bar \Omega }}_{{\phi _a}}^{\rm{H}}} \right)}\!
\end{array}} \!\right],
\end{equation}

\begin{equation}\label{Balphaphi}
{{\bf{B}}_{\alpha {\boldsymbol{\phi}} }} = {\bf{B}}_{{\boldsymbol{\phi}} \alpha }^{\rm{T}}, {{\bf{B}}_{\alpha \alpha }} = \frac{2}{{\sigma _s^2}}{{\bf{I}}_2}{\rm{tr}}\left( {{\bf{\bar \Omega }}{{\bf{R}}_x}{{{\bf{\bar \Omega }}}^{\rm{H}}}} \right),
\end{equation}

\begin{figure*}[!t]
\begin{small}
\begin{equation}\label{Aphiphi}
\begin{array}{l}
{{\bf{A}}_{{\boldsymbol{\phi}} {\boldsymbol{\phi}} }} = \frac{2}{{\sigma _s^2}}{\rm{Re}}\left[ {\begin{array}{*{20}{c}}
{{\boldsymbol{\varepsilon }}{{\left( {\boldsymbol{\eta }} \right)}^{\rm{H}}}\alpha {\rm{vec}}\left( {{{{\bf{\bar \Omega }}}_{{\phi _e}{\phi _e}}}{\bf{X}}} \right)}&{{\boldsymbol{\varepsilon }}{{\left( {\boldsymbol{\eta }} \right)}^{\rm{H}}}\alpha {\rm{vec}}\left( {{{{\bf{\bar \Omega }}}_{{\phi _a}{\phi _e}}}{\bf{X}}} \right)}\\
{{\boldsymbol{\varepsilon }}{{\left( {\boldsymbol{\eta }} \right)}^{\rm{H}}}\alpha {\rm{vec}}\left( {{{{\bf{\bar \Omega }}}_{{\phi _e}{\phi _a}}}{\bf{X}}} \right)}&{{\boldsymbol{\varepsilon }}{{\left( {\boldsymbol{\eta }} \right)}^{\rm{H}}}\alpha {\rm{vec}}\left( {{{{\bf{\bar \Omega }}}_{{\phi _a}{\phi _a}}}{\bf{X}}} \right)}
\end{array}} \right] - {{\bf{B}}_{\phi \phi }} = \frac{{2{{\left| \alpha  \right|}^2}}}{{\sigma _s^2}} \times \\
{\rm{Re}}\!\left[ {\begin{array}{*{20}{c}}
{{\rm{tr}}\left( {{{{\bf{\bar \Omega }}}_{{\phi _e}{\phi _e}}}{{\bf{R}}_x}{{{\bf{\bar \Omega }}}_W}} \right) \!+\! {\rm{tr}}\left( {{{{\bf{\bar \Omega }}}_{{\phi _e}{\phi _e}}}{{\bf{R}}_x}{{{\bf{\bar \Omega }}}_F}} \right) \!-\!{\rm{tr}}\left( {{{{\bf{\bar \Omega }}}_{{\phi _e}{\phi _e}}}{{\bf{R}}_x}{\bf{\bar \Omega }}} \right)}\!&\!{{\rm{tr}}\left( {{{{\bf{\bar \Omega }}}_{{\phi _a}{\phi _e}}}{{\bf{R}}_x}{{{\bf{\bar \Omega }}}_W}} \right) \!+\! {\rm{tr}}\left( {{{{\bf{\bar \Omega }}}_{{\phi _a}{\phi _e}}}{{\bf{R}}_x}{{{\bf{\bar \Omega }}}_F}} \right) \!-\! {\rm{tr}}\left( {{{{\bf{\bar \Omega }}}_{{\phi _a}{\phi _e}}}{{\bf{R}}_x}{\bf{\bar \Omega }}} \right)}\\
{{\rm{tr}}\left( {{{{\bf{\bar \Omega }}}_{{\phi _e}{\phi _a}}}{{\bf{R}}_x}{{{\bf{\bar \Omega }}}_W}} \right) \!+\! {\rm{tr}}\left( {{{{\bf{\bar \Omega }}}_{{\phi _e}{\phi _a}}}{{\bf{R}}_x}{{{\bf{\bar \Omega }}}_F}} \right) \!-\! {\rm{tr}}\left( {{{{\bf{\bar \Omega }}}_{{\phi _e}{\phi _a}}}{{\bf{R}}_x}{\bf{\bar \Omega }}} \right)}\!&\!{{\rm{tr}}\left( {{{{\bf{\bar \Omega }}}_{{\phi _a}{\phi _a}}}{{\bf{R}}_x}{{{\bf{\bar \Omega }}}_W}} \right) \!+\! {\rm{tr}}\left( {{{{\bf{\bar \Omega }}}_{{\phi _a}{\phi _a}}}{{\bf{R}}_x}{{{\bf{\bar \Omega }}}_F}} \right) \!-\! {\rm{tr}}\left( {{{{\bf{\bar \Omega }}}_{{\phi _a}{\phi _a}}}{{\bf{R}}_x}{\bf{\bar \Omega }}} \right)}
\end{array}} \right] \!-\! {{\bf{B}}_{\phi \phi }}.
\end{array}
\end{equation}
\end{small}
\vspace{-1cm}
\end{figure*}

\begin{figure*}[!t]
\begin{equation}\label{Aphialpha}
\begin{array}{l}
{{\bf{A}}_{{\boldsymbol{\phi}} \alpha }} = \frac{2}{{\sigma _s^2}}{\rm{Re}}\left[ {\begin{array}{*{20}{c}}
{{\boldsymbol{\varepsilon }}{{\left( {\boldsymbol{\eta }} \right)}^{\rm{H}}} {\rm{vec}}\left( {{{{\bf{\bar \Omega }}}_{{\phi _e}}}{\bf{X}}} \right)}&{{\boldsymbol{\varepsilon }}{{\left( {\boldsymbol{\eta }} \right)}^{\rm{H}}} {\rm{vec}}\left( {{{{\bf{\bar \Omega }}}_{{\phi _a}}}{\bf{X}}} \right)}\\
{{\boldsymbol{\varepsilon }}{{\left( {\boldsymbol{\eta }} \right)}^{\rm{H}}}j {\rm{vec}}\left( {{{{\bf{\bar \Omega }}}_{{\phi _e}}}{\bf{X}}} \right)}&{{\boldsymbol{\varepsilon }}{{\left( {\boldsymbol{\eta }} \right)}^{\rm{H}}}j {\rm{vec}}\left( {{{{\bf{\bar \Omega }}}_{{\phi _a}}}{\bf{X}}} \right)}
\end{array}} \right] - {{\bf{B}}_{\phi \alpha }} = \frac{2{\alpha}}{{\sigma _s^2}} \times \\
{\rm{Re}}\!\left[ {\begin{array}{*{20}{c}} \!
{{\rm{tr}}\left( {{{{\bf{\bar \Omega }}}_{{\phi _e}}}{{\bf{R}}_x}{{{\bf{\bar \Omega }}}_W}} \right) \!+\! {\rm{tr}}\left( {{{{\bf{\bar \Omega }}}_{{\phi _e}}}{{\bf{R}}_x}{{{\bf{\bar \Omega }}}_F}} \right) \!-\! {\rm{tr}}\left( {{{{\bf{\bar \Omega }}}_{{\phi _e}}}{{\bf{R}}_x}{\bf{\bar \Omega }}} \right)}\!&\!{{\rm{tr}}\left( {{{{\bf{\bar \Omega }}}_{{\phi _a}}}{{\bf{R}}_x}{{{\bf{\bar \Omega }}}_W}} \right) \!+\! {\rm{tr}}\left( {{{{\bf{\bar \Omega }}}_{{\phi _a}}}{{\bf{R}}_x}{{{\bf{\bar \Omega }}}_F}} \right) \!-\! {\rm{tr}}\left( {{{{\bf{\bar \Omega }}}_{{\phi _a}}}{{\bf{R}}_x}{\bf{\bar \Omega }}} \right)}\!\\
\!{j\left( {{\rm{tr}}\left( {{{{\bf{\bar \Omega }}}_{{\phi _e}}}{{\bf{R}}_x}{{{\bf{\bar \Omega }}}_W}} \right) \!+\! {\rm{tr}}\left( {{{{\bf{\bar \Omega }}}_{{\phi _e}}}{{\bf{R}}_x}{{{\bf{\bar \Omega }}}_F}} \right) \!-\! {\rm{tr}}\left( {{{{\bf{\bar \Omega }}}_{{\phi _e}}}{{\bf{R}}_x}{\bf{\bar \Omega }}} \right)} \right)}\!&\!{j\left( {{\rm{tr}}\left( {{{{\bf{\bar \Omega }}}_{{\phi _a}}}{{\bf{R}}_x}{{{\bf{\bar \Omega }}}_W}} \right) \!+\! {\rm{tr}}\left( {{{{\bf{\bar \Omega }}}_{{\phi _a}}}{{\bf{R}}_x}{{{\bf{\bar \Omega }}}_F}} \right) \!-\! {\rm{tr}}\left( {{{{\bf{\bar \Omega }}}_{{\phi _a}}}{{\bf{R}}_x}{\bf{\bar \Omega }}} \right)} \right)}
\end{array}} \!\right] \!-\! {{\bf{B}}_{\phi \alpha }}.
\end{array}
\end{equation}
\vspace{-1.1cm}
\end{figure*}

\begin{equation}\label{Aalphaphi}
{{\bf{A}}_{\alpha {\boldsymbol{\phi}} }} = {\bf{A}}_{{\boldsymbol{\phi}} \alpha }^{\rm{T}},{{\bf{A}}_{\alpha \alpha }} =  - {{\bf{B}}_{\alpha \alpha }},
\end{equation}
where
\begin{equation}\label{Omegabar}
{\bf{\bar \Omega }} = {{\bf{H}}_{BR}}{{\bf{\Theta }}^{\rm{H}}}{\bf{\Omega \Theta H}}_{BR}^{\rm{H}},
\end{equation}
\begin{equation}\label{OmegaPhiebar}
{{{\bf{\bar \Omega }}}_{{\phi _e}}} = {{\bf{H}}_{BR}}{{\bf{\Theta }}^{\rm{H}}}{{{\bf{\dot \Omega }}}_{{\phi _e}}}{\bf{\Theta H}}_{BR}^{\rm{H}},
\end{equation}
\begin{equation}\label{OmegaPhiabar}
{{{\bf{\bar \Omega }}}_{{\phi _a}}} = {{\bf{H}}_{BR}}{{\bf{\Theta }}^{\rm{H}}}{{{\bf{\dot \Omega }}}_{{\phi _a}}}{\bf{\Theta H}}_{BR}^{\rm{H}},
\end{equation}
\begin{equation}\label{OmegaPhiePhiebar}
{{{\bf{\bar \Omega }}}_{{\phi _e}{\phi _e}}} = {{\bf{H}}_{BR}}{{\bf{\Theta }}^{\rm{H}}}{{{\bf{\ddot \Omega }}}_{{\phi _e}{\phi _e}}}{\bf{\Theta H}}_{BR}^{\rm{H}},
\end{equation}
\begin{equation}\label{OmegaPhiePhiabar}
{{{\bf{\bar \Omega }}}_{{\phi _e}{\phi _a}}} = {{\bf{H}}_{BR}}{{\bf{\Theta }}^{\rm{H}}}{{{\bf{\ddot \Omega }}}_{{\phi _e}{\phi _a}}}{\bf{\Theta H}}_{BR}^{\rm{H}},
\end{equation}
\begin{equation}\label{OmegaPhiaPhiebar}
{{{\bf{\bar \Omega }}}_{{\phi _a}{\phi _e}}} = {{\bf{H}}_{BR}}{{\bf{\Theta }}^{\rm{H}}}{{{\bf{\ddot \Omega }}}_{{\phi _a}{\phi _e}}}{\bf{\Theta H}}_{BR}^{\rm{H}},
\end{equation}
\begin{equation}\label{OmegaPhiaPhiabar}
{{{\bf{\bar \Omega }}}_{{\phi _a}{\phi _a}}} = {{\bf{H}}_{BR}}{{\bf{\Theta }}^{\rm{H}}}{{{\bf{\ddot \Omega }}}_{{\phi _a}{\phi _a}}}{\bf{\Theta H}}_{BR}^{\rm{H}},
\end{equation}
\begin{equation}\label{OmegaWbar}
{{{\bf{\bar \Omega }}}_W} = {{\bf{H}}_{BR,W}}{\bf{\Theta }}_W^{\rm{H}}{{\bf{H}}_{{\rm{TRM}},W}}{{\bf{\Theta }}_W}{\bf{H}}_{BR,W}^{\rm{H}},
\end{equation}
\begin{equation}\label{OmegaFbar}
{{{\bf{\bar \Omega }}}_F} = {{\bf{H}}_{BR,F}}{\bf{\Theta }}_F^{\rm{H}}{{\bf{H}}_{{\rm{TRM}},F}}{{\bf{\Theta }}_F}{\bf{H}}_{BR,F}^{\rm{H}}.
\end{equation}

With all the above derivation, we can obtain both ${{\bf{A}}_{{{\boldsymbol{\eta }}_0}}}$ and ${{\bf{B}}_{{{\boldsymbol{\eta }}_0}}}$ by letting ${{\boldsymbol{\eta }} = {{\boldsymbol{\eta }}_0}}$.

\subsection{Proof of definiteness of ${{\bf{Z}}^{ - 1}}$ and ${{\bf{Z}}^{-1}}{{\bf{Z}}^{-1}}$}
\label{AppZZ}
The eigendecomposition of ${\bf{Z}}$ is ${\bf{Z}} = {{\bf{Q}}_z}{\bf{\Lambda Q}}_z^{ - 1}$, where ${\bf{Q}}_z$ and ${\bf{\Lambda}}$ store the eigenvectors and eigenvalues of ${\bf{Z}}$. For ${{\bf{Z}}^{ - 1}}$ and ${{\bf{Z}}^{-1}}{{\bf{Z}}^{-1}}$, we have
\begin{equation}\label{ZinvEVD}
{{\bf{Z}}^{ - 1}} = {\left( {{{\bf{Q}}_z}{\bf{\Lambda Q}}_z^{ - 1}} \right)^{ - 1}} = {{\bf{Q}}_z}{{\bf{\Lambda }}^{ - 1}}{\bf{Q}}_z^{\rm{T}},
\end{equation}
\begin{equation}\label{Zinv2EVD}
{{\bf{Z}}^{ - 1}}{{\bf{Z}}^{ - 1}} = {{\bf{Q}}_z}{\bf{\Lambda Q}}_z^{ - 1}{{\bf{Q}}_z}{\bf{\Lambda Q}}_z^{ - 1} = {{\bf{Q}}_z}{{\bf{\Lambda }}^2}{\bf{Q}}_z^{ - 1}.
\end{equation}
Given that ${\bf{Z}}$ is negative definite, we have ${{\bf{\Lambda }}_{ii}}={\lambda _{z,i}} < 0,\forall i$, where ${\lambda _{z,i}}$ is the $i$-th eigenvalue of ${\bf{Z}}$. Thus, we have
\begin{equation}\label{ZinvEigVa}
{\left( {{{\bf{\Lambda }}^{ - 1}}} \right)_{ii}} = \frac{1}{{{\lambda _{z,i}}}} < 0,\forall i,
\end{equation}
\begin{equation}\label{Zinv2EigVa}
{\left( {{{\bf{\Lambda }}^2}} \right)_{ii}} = \lambda _{z,i}^2 > 0,\forall i.
\end{equation}
We can infer that ${{\bf{Z}}^{ - 1}}$ is also negative definite, while ${{\bf{Z}}^{ - 1}}{{\bf{Z}}^{ - 1}}$ is positive definite.

\subsection{Proof of equivalence for (\ref{DAschur})}
\label{DAproof}
Denoting ${{{\bf{\tilde I}}}_2} = \left[ {{{\bf{I}}_2}\;{\bf{0}}} \right] \in \mathbb{R}^{2\times 4}$, we can rewrite (\ref{DAschur}) as follows
\begin{equation}\label{DAschur1}
 - \left[ {\begin{array}{*{20}{c}}
{{\bf{\tilde D}}}&{{{{\bf{\tilde I}}}_2}}\\
{{\bf{\tilde I}}_2^{\rm{T}}}&{\bf{A}}
\end{array}} \right] \succeq 0 .
\end{equation}
Based on Schur complement, we have 
\begin{equation}\label{DAschur2}
- \left( {{\bf{\tilde D}} - {{{\bf{\tilde I}}}_2}{{\bf{A}}^{ - 1}}{\bf{\tilde I}}_2^{\rm{T}}} \right)\succeq 0.
\end{equation}
Subsituting ${{\bf{A}}^{ - 1}}$, namely (\ref{Ainv}), into (\ref{DAschur2}), we can obtain ${{{\bf{\tilde I}}}_2}{{\bf{A}}^{ - 1}}{\bf{\tilde I}}_2^{\rm{T}} = {{\bf{Z}}^{ - 1}}$. Therefore, $- ( {{\bf{\tilde D}} - {{\bf{Z}}^{ - 1}}} ) \succeq 0$, namely ${{\bf{Z}}^{ - 1}}\succeq {\bf{\tilde D}}$. The proof is completed.

\bibliographystyle{IEEEtran}

\end{document}